\RequirePackage{ifpdf}
\documentclass[letterpaper]{article}
\pdfoutput=1

\usepackage[usenames,dvipsnames]{pstricks}
\usepackage{./auxi/jheppub}
\usepackage[T1]{fontenc}
\usepackage[utf8]{inputenc}
\usepackage[english]{babel}
\usepackage[babel]{csquotes}
\usepackage{amsmath}
\usepackage{amsfonts}
\usepackage{amssymb}
\usepackage{mathtools,colonequals}
\usepackage{mathrsfs}
\usepackage{bm}
\usepackage{bbold}
\usepackage{braket}
\usepackage{graphicx}
\usepackage{graphics,multicol}
\usepackage{tikz}
\usepackage{cleveref}
\usetikzlibrary{decorations.pathmorphing}
\usetikzlibrary{decorations.markings}
\usepgflibrary{shapes.geometric}
\usepackage{booktabs}
\usepackage{array}
\usepackage{color   }
\usepackage{subcaption}
\usepackage[percent]{overpic}
\usepackage{multirow}
\usepackage{soul}
\usepackage{comment}
\usepackage{cleveref}
\usepackage{pict2e}
\usepackage{transparent}
\usepackage{cancel}

\newcommand{\Hy}{H_\textup{bry}}
\newcommand{\Hk}{H_\textup{bulk}}

\newcommand{\Dotwo}{\frac{\Delta}{2}}
\newcommand{\hgapa}{h_{\text{gap}}^{\text{ann}}}
\newcommand{\hgaptp}{h_{\text{gap}}^{\text{2pt}}}

\def\zb{\bar{z}}
\def\hb{\bar{h}}

\def\qaq{\quad \text{and} \quad}

\def\wh{\widehat}

\def\lra{\leftrightarrow}

\def\dm{\Delta_{\text{min}}}
\def\hm{h_{\text{min}}}

\def\ldef{\mathrel{\mathop:}=}

\newcommand{\reef}[1]{(\ref{#1})}

\def\mca{\mathcal}
\def\mfr{\mathfrak}
\def\mrm{\mathrm}

\def\th{\tfrac{1}{2}}

\def\eps{\epsilon}

\newcommand*\circled[1]{\tikz[baseline=(char.base)]{
            \node[shape=circle,draw,inner sep=2pt, scale=0.8] (char) {#1};}}

\newcommand{\D}{\Delta}

\renewcommand{\th}{\theta}

\newcommand{\n}{\nu}

\newcommand{\beq}{\begin{equation}}
\newcommand{\eeq}{\end{equation}}

\newcommand{\bz}{\bar{z}}
\newcommand{\bh}{\bar{h}}

\def\eps{\epsilon}

\def\eps{\epsilon}

\def\D{\Delta}

\def\sutwok{\mathfrak{su}(2)_k}

\author[a]{Marco Meineri,}
\author[b,c]{Bharathkumar Radhakrishnan}

\affiliation[a]{Istituto Nazionale di Fisica Nucleare, Sezione di Torino, and Department of Physics, University of Turin, Via P. Giuria 1, 10125, Turin, Italy}
\affiliation[b]{Department of Theoretical Physics, University of Geneva,
24 quai Ernest-Ansermet, 1211 Genève 4, Suisse}
\affiliation[c]{Walter Burke Institute for Theoretical Physics, Caltech, Pasadena, California 91125, USA}

\emailAdd{marco.meineri@unito.it}
\emailAdd{r.bharathkumar@outlook.com}

\bigskip
\abstract{
We study two-dimensional conformal field theories (CFTs) with boundaries via the conformal bootstrap. We derive a positive semi-definite program from crossing symmetry of three observables: the annulus partition function, the two-point function of identical operators in the presence of a boundary, and the four-point function of the same operators on the infinite plane. The mixed-correlator system allows the numerical bootstrap to access new data, like the bulk-to-boundary Operator Product Expansion coefficients, and to strengthen the bounds on observables already contained in the partition function on the annulus, such as the boundary entropy. We test the method on the free boson CFT; then, as a first application, we produce new non-perturbative bounds on the entropy and the gaps in boundary CFTs with central charge $c=3/2,$ with special emphasis on the $\mathfrak{su}(2)_2$ WZW model.
}

\title{The bootstrap of points and lines}

\keywords{Conformal field theory, bootstrap, boundaries and defects, convex optimization}

\begin{document}

\maketitle

\section{Introduction}

The numerical bootstrap program \cite{Poland:2018epd,Rychkov:2023wsd} is an extremely successful and versatile tool, which lends itself to precision physics \cite{Kos:2016ysd,Simmons_Duffin:2017,Chester:2019ifh} and to answering rigorously qualitative questions, such as the existence and the nature of phase transitions in specific models \cite{Nakayama:2016jhq,Chester:2020iyt}. The list of observables amenable to a convex optimisation formulation through self consistency constraints is ever growing. A few notable examples include four-point functions of local operators in a conformal field theory (CFT) \cite{Rattazzi:2008pe}, two-to-two scattering amplitudes in a quantum field theory (QFT) \cite{Paulos:2016fap}, expectation values in quantum mechanics, lattice models and matrix models \cite{Anderson:2016rcw,Han:2020bkb,Lin:2020mme,Cho:2022lcj}. Explorations of higher-point correlators and amplitudes have also been initiated \cite{Poland:2023vpn,Antunes:2023kyz,Guerrieri:2024ckc}. 

This paper introduces an entry to this growing list, focusing on the bootstrap of boundary conditions in two dimensional CFTs. This is a long standing topic of research, and, as with correlation functions of local operators, boundary CFTs (BCFTs) have been studied within the bootstrap framework since its inception \cite{Cardy:1984bb,Cardy:1989ir,Cardy:1991tv}---see \cite{Recknagel:2013uja} for a comprehensive review. In \cite{Lewellen:1991tb}, the full set of crossing constraints defining a conformal boundary condition was established. In addition to the usual bulk constraints on the infinite plane and on the torus \cite{SONODA1988417}, the relevant observables are the annulus partition function, the four-point functions of boundary operators, and the correlators of two bulk operators and one boundary operator, including the case where the boundary operator is the identity. In order to obtain information on boundary condition changing operators as well, one needs to also check consistency of the correlators of one bulk operator and two boundary operators. In cases where the crossing equations involve only a finite number of unknowns associated to these observables, one can find the complete set of solutions compatible with unitarity \cite{Runkel:1998he,Behrend:1999bn}. This happens in rational CFTs for boundary conditions that preserve the largest possible chiral algebra. Beyond rational boundary conditions, the realm of the numerical bootstrap begins.

The crossing equation obeyed by the annulus partition function was first cast into a linear programming problem in \cite{Friedan:2012jk}, and the first rigorous bounds were obtained on the boundary entropy as a function of the central charge and the lightest scalar operator in the bulk. This was significantly extended in \cite{Collier:2021ngi}, where state-of-the-art numerical \cite{Simmons-Duffin:2015qma} and analytic \cite{Mazac:2016qev} bootstrap techniques were applied, yielding a wealth of results, including compelling conjectures on the existence of a unique boundary state for certain CFTs, and analytic bounds on the ground state degeneracy when the central charge $c=8$ or $24$.

In this paper, we demonstrate how to bootstrap a set of correlators involving both bulk local operators and boundaries. The setup includes the four-point function of primaries in infinite space, the annulus partition function, and the two-point function with a single boundary. The resulting system enjoys the required positivity properties necessary to be turned into a polynomial matrix program (PMP) and solved using SDPB. At a technical level, the bootstrap of points and lines is analogous to the mixed correlators bootstrap for the four-point function \cite{Kos:2014bka}. The enlarged set of constraints allows access to new CFT data, most notably the bulk-to-boundary OPE coefficients. It also entails a bigger parameter space, which we begin to explore in this work but do not fully exhaust. It should be noted that the two-point function of local operators with a boundary has been a central focus of bootstrap methods in the presence of defects---see \cite{McAvity1995,Liendo2013,Gliozzi2015,Billo:2016cpy,Liendo:2016ymz,deLeeuw:2017dkd,Lemos:2017vnx,Mazac:2018biw,Kaviraj:2018tfd,Bissi:2018mcq,Liendo:2019jpu,Padayasi:2021sik,Barrat:2021yvp,Bianchi:2022sbz,Gimenez-Grau:2023fcy} and many others---as it is the simplest correlation function incorporating both bulk and defect data while being subject to crossing constraints in arbitrary dimensions. This paper provides the first demonstration that rigorous bounds can be derived from the crossing properties of this observable.\footnote{See also \cite{Lanzetta:2024fmy, LanzettaLiuMetlitski} for a different approach, available when considering defects that can end, which happens in two dimensions for a class of interfaces.}

It is useful to classify the targets of the mixed correlator bootstrap in two classes. In the first class lie quantities which are already accessible via the annulus partition function: the ground state degeneracy and the expectation value of specific bulk operators are the main representatives explored in this work. For these observables, we find that in many situations, although not always, the bootstrap of points and lines tightens the bounds obtained from crossing symmetry on the annulus. The second class comprises observables which are only accessible via the two-point function. Here, we focus on the gap above the identity in the boundary operator product expansion (OPE) of the lightest bulk primary. This is a natural quantity to bound via the bootstrap, and has a transparent physical interpretation when the boundary condition preserves a subgroup of the bulk symmetry, under which the lightest bulk operator is charged.

We tackle boundary conditions in three classes of CFTs of growing complexity, labeled by central charges $c=1/2,1,\text{and }3/2$ respectively. Conformal boundary conditions in the Ising model ($c=1/2$) and in the free boson ($c=1$) have been classified, and this allows us to test the bootstrap of points and lines in a controlled setting. On the contrary, no complete classification exists for conformal boundary of $c=3/2$ CFTs. In this case, we explore the moduli space by using crossing symmetry on the annulus---see \cref{fig:su(2)2_hann_delta} and \cref{fig:su2-annulus}---and we focus on the $\mathfrak{su}(2)_2$ WZW model at level 2 for the mixed correlator bootstrap. There, we find the non-perturbative bounds reported in \cref{fig:su(2)2_hann_h2pt}, \cref{fig:su2-whale-0.1875}, \cref{fig:su2-whale-0.5} and \cref{fig:su2-asq}.

The structure of the paper is as follows. \Cref{sec:full-system} details the components of each crossing equation and formulates the numerical optimisation problem. \Cref{sec:numerical} is dedicated to the numerical explorations highlighted above.
Finally, \cref{sec:outlook} offers a more detailed summary of the main results and an outlook. A few appendices contain details about the numerical implementation and about the CFTs whose boundary conditions we bootstrap. All the plots in this paper can be reproduced using the data provided in the Zenodo archive at the following DOI: \url{https://doi.org/10.5281/zenodo.15704687}.

\section{The bootstrap program for points and lines}\label{sec:full-system}

The data which define a $2d$ CFT on the infinite plane are the scaling dimensions $\Delta$ and spins $\ell$ of local operators, and the coefficients in their operator product expansions (OPE). We will only be interested in the fusion of identical external scalar Virasoro primary operators,
\begin{equation}\label{bulkOPE}
    \phi(z,\bar{z})\phi(0) \sim\frac{1}{(z\bar{z})^{\D_\phi}}  \sum_{\Delta,\ell} \frac{c_{\D\ell}}{1+\delta_{\ell,0}} \left(z^h \bar{z}^{\bar{h}}\,O_{h,\bar{h}}(0)+z^{\bar{h}} \bar{z}^{h}\,O_{\bar{h},h}(0)\right) +\dots~,
\end{equation}
where we suppressed the $\phi$ dependence from the OPE coefficients $c_{\D\ell}$, and defined the quantities
\begin{equation}\label{hhb}
    h=\frac{\Delta+\ell}{2}~, \quad \bh=\frac{\Delta-\ell}{2}~, \qquad \ell \in 2\mathbb{N}~.
\end{equation}
In writing eq. \eqref{bulkOPE} we assumed parity invariance. The sum runs over quasi-primary operators, and the dots hide $sl(2,\mathbb{C})$ descendants. This means that the $c_{\D \ell}$'s associated to operators belonging to the same Verma module are not independent. In this paper, we systematically enforce $sl(2,\mathbb{C})$ rather than the full Virasoro symmetry. There will be some exceptions: when bootstrapping the annulus partition function Virasoro characters will often be used, and in a few occasions the larger symmetry will be reintroduced by hand in the low lying spectrum and three-point coefficients $c_{\D\ell}$. These exceptions are always mentioned explicitly in text and figures. We take $\phi$ to be real, by which we mean that the OPE \eqref{bulkOPE} contains the identity operator.

A CFT on the upper half-plane is further characterized by a new set of scaling dimensions $h$ of boundary local operators, together with their three-point coefficients. Bulk local operators can be expanded in terms of boundary operators weighted by new OPE coefficients $b_{\Delta_O h}$ as
\begin{equation}\label{BOPE}
    O(z,\bar{z}) \sim \sum_{h} b_{\Delta_O h} \left(2\text{Im} z\right)^{h-\Delta_O}\wh{O}_h(\text{Re}z)+\dots~.
\end{equation}
Here, we placed the boundary along the real axis and we considered a scalar bulk operator for simplicity. Extensive information about the OPE channels with a defect can be found, for instance, in \cite{Billo:2016cpy}. In two dimensions, scale invariance of the boundary, together with unitarity and the existence of a local stress tensor, implies invariance under a full copy of the conformal algebra \cite{Nakayama:2012ed} (see also \cite{Meineri:2019ycm}). This is captured in the Cardy gluing conditions,
\begin{equation}\label{CardyCond}
    T(x)=\bar{T}(x)~, \qquad x \in \mathbb{R}~.
\end{equation}
The boundary spectrum is organised into Virasoro multiplets, and the coefficients $b_{\Delta_O h}$ are not all independent. In this work, numerical bounds are obtained without imposing this constraint on the bulk to boundary OPE coefficients. Furthermore, if the CFT has an extended symmetry, the boundary may either preserve it or violate it. Our primary focus is on boundary conditions which only preserve conformal symmetry. This gives access to the enticing setup where the CFT is rational, where the bulk data is known, while the boundary conditions may be irrational.

In eq. \eqref{BOPE}, the coefficient of the boundary identity has a special role, and we denote it by
\begin{equation}\label{eq:bd-id}
    b_{\Delta\, h=0} \equiv a_\Delta~.
\end{equation} 
Conformal symmetry implies that only scalar bulk operators can have a non-vanishing coupling to the boundary identity.
One way to see the importance of the coefficients $a_\Delta$ is to map the boundary to a circle, or equivalently look at the BCFTs on the semi-infinite cylinder. Then, a circlular boundary of radius $R$ defines a state in the radial quantization Hilbert space of the CFT, which we denote as $\ket{\bigcirc,R}$. The components of the boundary state, when expressed in the basis of eigenstates of the dilatation operator, are precisely the coefficients $a_O$:
\begin{equation}\label{bulkOPE_circ}
    \ket{\bigcirc,R}=\sum_{\Delta}  a_\Delta R^\D |O \rangle\! \rangle~.
\end{equation}
The Ishibashi states $|O \rangle\! \rangle$ resum the contribution of an $sl(2,\mathbb{C})$ family.\footnote{The name of Ishibashi is in general associated to states which are invariant under the action of a prescribed symmetry. In this case, we use it for states invariant under the $sl(2,\mathbb{R})$ subalgebra of $sl(2,\mathbb{C})$ which leaves a plane boundary invariant. The appropriate invariance ($sl(2,\mathbb{R})$, Virasoro, or an extension thereof) will be clear from the context in the rest of the paper.}
Let us reiterate that the one-point functions of Virasoro descendants are fixed in terms of the ones of the corresponding primary.

        \begin{figure}[t]
            \captionsetup[subfigure]{labelformat=empty} 
            \centering
            \subfloat[$\bra{0}\phi(-1) \phi(1) \phi(-\rho) \phi(\rho)\ket{0}$]{%
                \begin{tikzpicture}
                    \draw[gray,dashed] (2,2) circle (1.75cm);
                    \node at (3.75,2) [circle,fill,inner sep=1.5pt]{};
                    \node at (3.75,2) [anchor=west]{$\phi$};
                    \node at (0.25,2) [circle,fill,inner sep=1.5pt]{};
                    \node at (1.45,1.45) [circle,fill,inner sep=1.5pt]{};
                    \node at (2.55,2.55) [circle,fill,inner sep=1.5pt]{};
                    \node at (2.55,2.55) [anchor=south west]{$\phi$};
                    \node[gray] at (2,2) {$\times$};
                    \draw[gray,->] (2.1,2.1) -- (2.5,2.5);
                    \node at (2.6,2.2) {$\rho$};
                \end{tikzpicture}
            }\hfill
            \subfloat[$\bra{\bigcirc,1}\phi(-\rho_{\text{bulk}}) \phi(\rho_{\text{bulk}}) \ket{0}$]{%
                \begin{tikzpicture}
                    \draw[red] (2,2) circle (1.75cm);
                    \node at (1.45,1.45) [circle,fill,inner sep=1.5pt]{};
                    \node at (2.55,2.55) [circle,fill,inner sep=1.5pt]{};
                    \node[gray] at (2,2) {$\times$};
                    \draw[gray,->] (2.1,2.1) -- (2.5,2.5);
                    \node at (2.8,2.2) {$\rho_{\text{bulk}}$};
                    \node at (0.5, 4) [anchor=west]{$\ket{\bigcirc,1}$};
                    \node at (2.55,2.55) [anchor=south west]{$\phi$};
                \end{tikzpicture}
            } \hfill
            \subfloat[$\braket{\bigcirc,R^{-1} | \bigcirc,R}$]{%
                \begin{tikzpicture}
                    \draw[red] (2,2) circle (1.75cm);
                    \draw[gray, dashed] (2,2) circle (1cm);
                    \draw[red] (2,2) circle (0.57cm);
                    \node[gray] at (2,2) {$\times$};
                    \node at (0.5, 4) [anchor=west]{$\ket{\bigcirc,R^{-1}}$};
                    \draw[gray,->] (2.1,2.1) -- (2.4,2.4);
                    \draw[gray,->] (2.1,1.9) -- (3.2,0.8);
                     \node at (2.5,2.6) {$R$};
                     \node at (2.7,0.7) {$R^{-1}$};
                \end{tikzpicture}
            }
        \caption{Representation of the observables involved in the bootstrap of points and lines, together with a choice of cross ratios. See \Cref{sec:block-imp} for the relations of $\rho$ and $\rho_{\text{bulk}}$ to other useful cross ratios.}
        \label{fig:rho-coords}
        \end{figure}
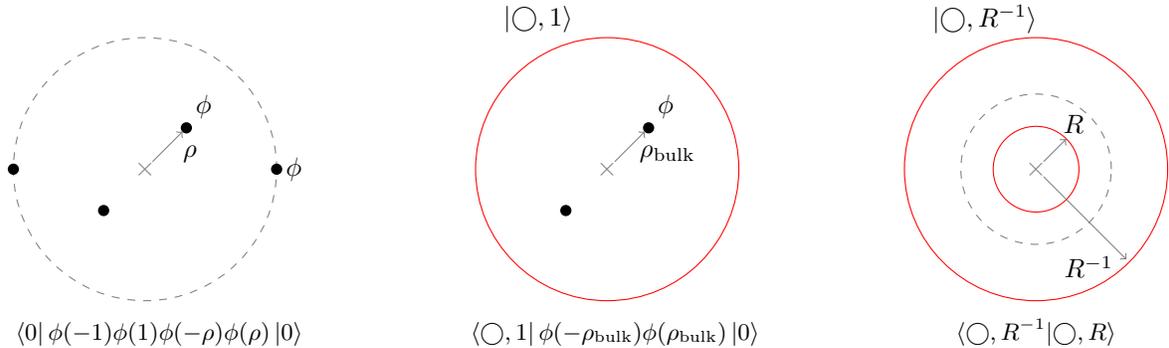

If we further denote the state created by two identical bulk primaries as $\ket{\phi(x) \phi(y)}$, the correlation functions we will bootstrap can be organized into the following Kac matrix:
\begin{equation}\label{eq:2by2-schema}
    \begin{pmatrix}
        \braket{\phi(\mathcal{R}x) \phi(\mathcal{R}y) | \phi(x) \phi(y)} & \braket{\phi(\mathcal{R}x) \phi(\mathcal{R}y) | \bigcirc,R} \\
        \braket{\bigcirc, R^{-1} | \phi(x) \phi(y)} &\braket{\bigcirc,R^{-1} | \bigcirc,R}
    \end{pmatrix}~,
\end{equation}
where the overlaps are evaluated in the Hilbert space at unit radius, $\mathcal{R}$ denotes an inversion, and we assume $R,\,|x|,\,|y|<1$.\footnote{We use the notation $\bra{\mathcal{R}x}$ and $\bra{\bigcirc, R^{-1} }$ in order to emphasise the location of the corresponding operators in the path-integral. A more principled notation would be $\bra{\phi(x)}=\bra{0}\phi(\mathcal{R}x)$.}
Each of the entries in the matrix satisfies at least one crossing equation. One side of each equation is obtained by inserting within each overlap a resolution of the identity in terms of the eigenstates of the dilatation operator. Reflection positivity then implies that the resulting sums in the diagonal terms consists solely of positive terms, each being a modulus squared. This expansion is equivalent to plugging the bulk channel OPEs \eqref{bulkOPE} and \eqref{bulkOPE_circ} in \eqref{eq:2by2-schema}, and we conclude that
\begin{equation}\label{positivityBulkOPE}
    c^2_{\D \ell}>0,\qquad a_\Delta^2>0~,
\end{equation}
as expected.\footnote{More precisely, one can choose a basis $\{\ket{\Delta,\ell}\}$ in the Hilbert space so that $c_{\Delta \ell}\in \mathbb{R}$, but then $a_\Delta$ could be complex. In fact, $a_\Delta \in \mathbb{R}$ is enforced for the operators appearing in the $\phi\times\phi$ fusion thanks to the crossing equation \eqref{2pCrossing}, where reflection positivity of the \emph{boundary channel} imposes that the two-point function is real and positive for any value of $\xi$.}
As is common in the bootstrap approach, constraints on the four-point function arise from different kinematical regions beyond those in the matrix \eqref{eq:2by2-schema}. The generic configuration is depicted in \cref{fig:rho-coords}. However, the Kac matrix \eqref{eq:2by2-schema} highlights the property \eqref{positivityBulkOPE}, which is instrumental in overcoming the lack of positivity of the off-diagonal terms. This property further constrains the solutions to crossing for the mixed correlator system to lie in a convex region in the $(c^2_{\D \ell},a_\Delta^2)$ space, which can be bound numerically.

In the following subsection, we outline the crossing symmetry properties of the annulus partition function, the two-point function, and the four-point function. 
In subsection \ref{ssec:SDPs}, we describe the semi-definite system that we will solve numerically in the rest of the paper. Finally, \cref{ssec:elementary} is dedicated to linear combinations of boundary states and to their imprint on the bootstrap bounds.

\subsection{The crossing equations}\label{ssec:mixed-correlator}

In this subsection, we define the conformal blocks appearing in the OPE decompositions of the correlators discussed so far. It is important to emphasise that, while the explicit form of the blocks depends on the symmetry algebra whose multiplets they resum, the general structure of the crossing equations, and in particular the relations between four-point blocks and two-point blocks dictated by the method of images, are valid both for $sl(2,\mathbb{R})$ and for Virasoro blocks.

\subsubsection{Four-point function}

The first diagonal element of the matrix \eqref{eq:2by2-schema} computes the four point function of identical scalar operators. Using the OPE \eqref{bulkOPE}, one can expand the four-point function in conformal blocks as follows
\begin{equation}\label{4pCrossing}
    \braket{\phi(z_1)\phi(z_2)\phi(z_3)\phi(z_4)}=\frac{1}{\left(z_{12}z_{34}\bz_{12}\bz_{34}\right)^{\Delta_\phi}}
    \sum_{\D,\ell} c^2_{\D\ell}\, g_{\D\ell}(z,\bar{z})~.
\end{equation}
The cross ratios are
\begin{equation}
    z= \frac{z_{12}z_{34}}{z_{13}z_{24}}~, \quad
    \bz= \frac{\bz_{12}\bz_{34}}{\bz_{13}\bz_{24}}~, \qquad z_{ij}=z_i-z_j~.
\end{equation}
The conformal blocks factorise as follows\footnote{the definition of $\kappa$ differs from the usual one, which includes the power law prefactor. This is done for later convenience.}
\begin{equation}\label{gkappa}
    g_{\D\ell}(z,\bar{z}) = z^h {\bz}^{\bh}\kappa_h(z) \kappa_{\bh}(\bz)+(z \leftrightarrow \bar{z})~,
\end{equation}
where we used the definitions \eqref{hhb} of $h$ and $\bar{h}$. If we focus on the global conformal group, the function $\kappa_h(z)$ takes the form
\begin{equation}\label{eq:kappa-defn}
    \kappa_h(z)= {}_2 F_1(h,h,2h,z)~.
\end{equation}
Since the operators are identical, crossing is equivalent to permutation invariance. The cross ratios transform under permutations as follows,
\begin{align}
    &1 \leftrightarrow 2   \qquad z \to \frac{z}{z-1}~, \\
    &1 \leftrightarrow 3   \qquad z \to 1-z~,\\
    &1 \leftrightarrow 4   \qquad z \to \frac{1}{z}~.
\end{align}
As usual, since the permutation $(12)(34)$ leaves $z$ invariant, the first two permutations generate the full group action. The $(12)$ permutation constrains $\ell$ to be even.
Then, the $(13)$ permutation gives the crossing equation
\begin{equation}\label{4ptSumRule}
    \sum_{\Delta, \ell}  c^2_{\Delta \ell} F_{\Delta \ell}(z,\bz)=0~, \qquad c_{00}=1~,
\end{equation}
with
\begin{equation}
    F_{\Delta \ell}(z,\bz)=v^{\Delta_\phi}g_{\Delta \ell}(z,\bz)-u^{\Delta_\phi}g_{\Delta \ell}(1-z,1-\bz)~,
\end{equation}
and
\begin{equation}
    u=z\bz~, \qquad v=(1-z)(1-\bz)~.
\end{equation}

\subsubsection{Annulus partition function}
\label{ssec:annulus}

The second diagonal element of the matrix \eqref{eq:2by2-schema} computes the partition function on the annulus. If we map it to a portion of cylinder of width $r$ and circumference $\beta$\footnote{The precise map is $R=\exp \frac{\pi r}{\beta}$.} we find the following crossing equation:
\begin{equation}\label{StripCrossing}
    \textup{Tr}\, e^{-\beta \Hy}=\braket{B|e^{-r\Hk}|B}~,
\end{equation}
where the trace is taken over the Hilbert space of the boundary operators, and the boundary state $\ket{B}$ is related to the one defined in eq. \eqref{bulkOPE_circ} as 
\begin{equation}\label{BtoCirc}
    \ket{B}=g\ket{\bigcirc,R=1}~.
\end{equation}
The factor $g$ is necessary because the state $\ket{\bigcirc,R}$ was normalized so that $a_0=1$---see eq. \eqref{eq:bd-id}---while it is the left hand side of eq. \eqref{StripCrossing} that normalizes $\ket{B}$. In particular, one has that 
\begin{equation}\label{gdef}
    \braket{0|B}=g
\end{equation}
which is a measurable quantity called the ground state degeneracy. It equals the partition function of the theory on a disk with boundary conditions dictated by $\ket{B}$. Since the partition function is symmetric under the involution generated by the reflection across a diameter, reflection positivity implies $g>0$.

The Hamiltonians $\Hy$ and $\Hk$ in eq. \eqref{StripCrossing} generate translations in the two directions of the finite cylinder. Due to the anomaly, they are shifted with respect to the generators of dilatations on the UHP and on the plane, respectively, as follows:
\begin{subequations}
\begin{align}
\Hy &=\frac{\pi}{r}\left(\wh{L}_0-\frac{c}{24}\right)~, \\
\Hk &=\frac{2\pi}{\beta}\left( L_0+\bar{L}_0-\frac{c}{12}\right)~.
\end{align}
\label{Hamiltonians}    
\end{subequations}
The eigenvalues of $\wh{L}_0$ are the scaling dimensions $h$ of the boundary operators, while the eigenvalues of  $L_0$ and $\bar{L}_0$ are respectively the $h$ and $\bar{h}$ of eq. \eqref{hhb}. 
Each side of eq. \eqref{StripCrossing} only depends on the ratio $r/\beta$, and it is useful to define the modular parameters
\begin{equation}
q(t)=e^{-2\pi t}~, ~~ \tilde{q}(t)=q(1/t)~, \qquad t=\frac{2r}{\beta}~.
\label{eq:t-defn}
\end{equation}
When expanded in $sl(2)$ characters, eq. \eqref{StripCrossing} becomes
\begin{equation}
\sum_{h} n_{h} \chi_{h}(1/t)=g^2\sum_{\Delta} a^2_{\Delta} \chi_{\frac{\Delta}{2}}(t)~,
\label{annulus_sl2}
\end{equation}
where
\begin{equation}\label{sl2_char}
    \chi_h(t) =
        \begin{cases}
            \frac{q^{h-c/24}}{(1-q)} & h>0, \\
            q^{-c/24} & h = 0,
        \end{cases}
\end{equation}
and we used the fact that among the bulk quasi-primaries, only scalars appear in the decomposition of the boundary state $\ket{B}$.
Equivalently, if we expand the crossing equation in terms of Virasoro modules, the characters in eq. \eqref{annulus_sl2} are replaced by
\begin{equation}\label{vir_char}
    \chi^{\textup{Vir}}_h(t) = 
    \begin{cases}
        \frac{q^{h-\frac{c-1}{24}}}{\eta(\mathrm{i}t)} & h > 0, \\
        \frac{q^{-\frac{c-1}{24}}}{\eta(\mathrm{i}t)} (1-q) & h = 0, \\
    \end{cases}
    \qquad
    \eta(\mathrm{i}t)=q^{1/24}\prod_{n=1}^\infty(1-q^n)~,
\end{equation}
if the modules are non-degenerate. For $c\leq1$, the representations of the Virasoro algebra become degenerate at special values of $h$. In particular, when $c=1$, these states lie at $h=\frac{n^2}{4}$, with $n$ a non-negative integer, and their degenerate Virasoro characters are given by
\begin{equation}\label{Vir_char_deg}
    \chi^{\textup{Vir, deg}}_n(t)=\frac{q^{\frac{n^2}{4}}}{\eta(\mathrm{i}t)} \left( 1 - q^{n+1} \right).
\end{equation}

 Notice that, even when using the $sl(2)$ characters \eqref{sl2_char} in the bootstrap, we always impose the specific form of the Casimir energies on the central charge fixed by eqs. \eqref{Hamiltonians}.

\subsubsection{Two-point function} \label{ssec:twopcross}

Let us consider the two-point function of identical operators on the upper half-plane, with the boundary condition described by the boundary state $\ket{\bigcirc,R}$ as before. The two point function can be written as
\begin{equation} \label{eq:2pt-fn}
    \braket{\phi(z_1)\phi(z_2)}_{\textup{UHP}} = \frac{1}{(4\text{Im} z_1 \text{Im} z_2)^{\Delta_\phi}} H(\xi),
\end{equation}
where we have defined the cross ratio
\begin{equation}
    \xi = \frac{z_{12}\bar{z}_{12}}{4 \text{Im} z_1 \text{Im} z_2}~.
\end{equation}
The function $H(\xi)$ satisfies the crossing relation \cite{Liendo:2012hy}
\begin{equation}\label{2pCrossing}
    H(\xi)=\sum_{h} b^2_{h} \xi^{-h} \kappa_{h}(-1/\xi)
    =\sum_{\substack{\Delta \\ \ell=0}} c_{\Delta}a_\Delta \xi^{-\D_\phi+\Dotwo}\kappa_{\Dotwo}(-\xi)~,
\end{equation}
where $b_h \equiv b_{\Delta_\phi h}$ (see eq. \eqref{BOPE}). The sum on the right-hand side only includes scalar (quasi)primaries, and $c_{\Delta}\equiv c_{\D, \ell=0}$. $\kappa_h(z)$ is the same function defined by eq. \eqref{gkappa}---explicitly eq. \eqref{eq:kappa-defn} in the case of $sl(2)$ blocks.

The coincidence of the two-point function blocks with the chiral blocks of the four-point function can be deduced by the method of images, which implies the replacement $z \to -\xi$ or $z \to -1/\xi$, depending on the channel.
While the method of images yields the conformal blocks, it has nothing to say about the OPE coefficients in eq. \eqref{2pCrossing}, which are instead determined by eqs. \eqref{bulkOPE} and \eqref{BOPE}. 

Finally, notice that one should be careful in setting $b_0=a_{\Delta_\phi}$ in eq. \eqref{2pCrossing}. Indeed, when the $\Delta=\Delta_\phi$ subspace of the Hilbert space is degenerate, this equality is consistent with eq. \eqref{annulus_sl2} only when at most one operator, $\phi$, gets a one-point function. In this work, whenever we enforce the relation between $b_0$ and $a_{\Delta_\phi}$ in SDPB, we always find ourselves in precisely this situation, so we have adopted the loose notation $a_\Delta$ rather than the unambiguous one $a_{O_\Delta}$. A different way to violate the equation $b_0=a_{\Delta_\phi}$ is to consider boundary conditions which do not respect the clustering property, \emph{i.e.} support more than one copy of the identity operator. These are called non-elementary branes, and we discuss them in \cref{ssec:elementary}.

\subsection{The full system and semi-definite programs}\label{ssec:SDPs}
We now have the following three crossing relations from the above discussion
\begin{align}
    \sum_{\Delta, \ell=0}  c^2_{\Delta} F_{\Delta, \ell=0}(z,\bz) + \sum_{\Delta, \ell\neq0}  c^2_{\Delta, \ell\neq0} F_{\Delta, \ell\neq0}(z,\bz)&=0~, \label{eq:four-point-unnormal}\\
         \sum_{\Delta, \ell=0} c_{\Delta}a_\Delta \xi^{-\D_\phi+\Dotwo}\kappa_{\Dotwo}(-\xi) - \sum_{h} b^2_{h} \xi^{-h} \kappa_{h}(-1/\xi) &= 0~, \label{eq:two-point-unnormal} \\
          \sum_{\Delta, \ell=0} a^2_{\Delta} \chi_{\frac{\Delta}{2}}(t) - \frac{1}{{g^2}}\sum_{h} n_{h} \chi_{h}(1/t)  &= 0~, \label{eq:annulus-unnormal}
\end{align}
where we have separated out the scalar contribution to the four point function. 
Adding the three equations together and re-arranging, we get the crossing relation
\begin{multline}\label{full-system}
    \begin{pmatrix}
        1~~ & 1
        \end{pmatrix}
        \begin{pmatrix}
        F_{\Delta=0,\ell=0}(z,\bz)  & \frac{1}{2} \mathcal{G}^{\text{bulk}}_{\Delta=0}(\xi) \\
        \frac{1}{2}\mathcal{G}^{\text{bulk}}_{\Delta=0}(\xi) & 
        \chi_{0}(t)
    \end{pmatrix}
    \begin{pmatrix}
        1 \\
        1
        \end{pmatrix} +
        \sum_{\Delta\neq0,\ell=0}
        \begin{pmatrix}
        c_{\Delta} & a_{\Delta}
    \end{pmatrix}
    \begin{pmatrix}
        F_{\Delta,\ell=0}(z,\bz)  & \frac{1}{2} \mathcal{G}^{\text{bulk}}_{\Delta}(\xi) \\
        \frac{1}{2}\mathcal{G}^{\text{bulk}}_{\Delta}(\xi) & \chi_{\frac{\Delta}{2}}(t)
    \end{pmatrix}
    \begin{pmatrix}
        c_{\Delta} \\
        a_{\Delta}
    \end{pmatrix} \\
    + \sum_{\Delta,\ell\neq 0}  c^2_{\Delta, \ell\neq0} F_{\Delta, \ell\neq0}(z,\bz)
    +\sum_{h} \left[ \frac{n_{h}}{{g^2}} (-\chi_{h}(1/t))+
    b^2_{h} \big(-\mathcal{G}^{\text{bry}}_{h}(\xi) \big)\right] = 0~.
\end{multline}
Here, we have defined the two-point blocks $\mathcal{G}^{\text{bulk}}(\xi)$ and $\mathcal{G}^{\text{bry}}(\xi)$ as
\begin{align}
    \mathcal{G}^{\text{bulk}}_{\Delta}(\xi) &= \xi^{-\D_\phi+\Dotwo}\kappa_{\Dotwo}(-\xi), \\
    \mathcal{G}^{\text{bry}}_{h}(\xi) &= \xi^{-h} \kappa_{h}(-1/\xi)~.
\end{align}
We have also separated the contribution of the bulk channel identity, and used the normalizations
\begin{equation}
    c_0=1~, \qquad a_0=1~.
\end{equation}
Eq. \eqref{full-system} holds for both global block and Virasoro block decompositions of the involved correlators, with the obvious modification in the latter case that the full Verma module of the identity can be separated from the bulk channel sum. In the $sl(2)$ case, $\chi_h(t)$ is defined in eq. \eqref{sl2_char} and $\kappa_h(z)$ in \eqref{eq:kappa-defn}. In the Virasoro case, $\chi_h(t)$ is defined in eq. \eqref{vir_char}. In this work, when bootstrapping all three correlators, we will only use global blocks.

Now, following the general mantra of numerical bootstrap \cite{Rattazzi:2008pe}, we can argue for the non-existence of a CFT by looking for appropriate functionals $\alpha : \mathcal{F} \rightarrow \mathbb{R}$, which act on the space $\mathcal{F}$ of functions of a single variable.
To this end, we define the vector of functionals
\begin{equation}
    \vec{\alpha} = (\alpha_4,\alpha_2,\alpha_a),
\end{equation}
where we use $a$ as subscript to denote the annulus. Further, we define the vectors 
\begin{subequations}
\begin{align}
    \vec{V}_{\Delta,\ell=0} &= 
    \begin{pmatrix}
        \begin{pmatrix}
            F_{\Delta,\ell=0}(z,\bz)  & 0 \\
            0 & 0
        \end{pmatrix},
        &
        \begin{pmatrix}
            0 & \frac{1}{2} \mathcal{G}^{\text{bulk}}_{\Delta}(\xi) \\
            \frac{1}{2}\mathcal{G}^{\text{bulk}}_{\Delta}(\xi) & 0
        \end{pmatrix},
        &
        \begin{pmatrix}
            0 & 0 \\
            0 & \chi_{\frac{\Delta}{2}}(t)
        \end{pmatrix}
    \end{pmatrix}~, \label{vec_Vscalars}\\
    \vec{V}_{\Delta,\ell\neq0} &= \begin{pmatrix}
        F_{\Delta, \ell}(z,\bz), &  0, & 0
    \end{pmatrix}~, \\
    \vec{V}^{\text{2pt}}_{h} &= \begin{pmatrix}
        0, & -\mathcal{G}^{\text{bry}}_{h}(\xi), & 0
    \end{pmatrix}~, \\
    \vec{V}^{a}_{h} &= \begin{pmatrix}
        0, & 0, & -\chi_{h}(1/t)
    \end{pmatrix}~.
\end{align}
\label{blockVectors}
\end{subequations}
Thus, the action of the functionals on \eqref{full-system} can be written as 
\begin{multline}\label{eq:alpha-full-system}
    \begin{pmatrix}
        1~~ & 1
    \end{pmatrix}
    \vec{\alpha}\cdot\vec{V}_{\Delta=0,\ell=0}
    \begin{pmatrix}
        1 \\
        1
    \end{pmatrix} + \sum_{\Delta\neq0,\ell=0}
    \begin{pmatrix}
        c_{\Delta} & a_{\Delta}
    \end{pmatrix}
    \vec{\alpha}\cdot\vec{V}_{\Delta,\ell=0}
    \begin{pmatrix}
        c_{\Delta} \\
        a_{\Delta}
    \end{pmatrix} \\
    + \sum_{\Delta, \ell\neq0}  c^2_{\Delta, \ell} ~ \vec{\alpha}\cdot\vec{V}_{\Delta,\ell\neq0} 
    + \frac{1}{g^2} ~ \vec{\alpha}\cdot\vec{V}^{a}_{h=0}+
    a^2_{\D_\phi} ~ \vec{\alpha}\cdot\vec{V}^{\text{2pt}}_{h=0}
    +\sum_{h\neq0} \left[ \frac{n_{h}}{g^2} ~ \vec{\alpha}\cdot\vec{V}^{a}_{h}+
    b^2_{h} ~ \vec{\alpha}\cdot\vec{V}^{\text{2pt}}_{h}\right] = 0~,
\end{multline}
where we isolated the identity contribution to the boundary blocks, and further used $b_0 = a_{\D_\phi}$.

It will be useful to define the following vector of parameters, which we will dial in our numerical explorations, together with the central charge $c$:
\begin{equation}\label{pdef}
    \vec{p} = (\Delta_\phi| \Delta_{\text{gap}}; \hgapa, \hgaptp |b_{\Delta_\phi 0}).
\end{equation}
Here $\Delta_\phi$ is the scaling dimension of the external operator, $b_{\Delta_\phi 0}$ is non-vanishing when the two-point function exchanges the boundary identity---see also eq. \eqref{eq:bd-id}---while $\Delta_{\text{gap}},\, \hgapa$ and $\hgaptp$ denote the gaps above which we will allow continuous spectra to contribute to the crossing equations. Often, one is looking for constraints on boundary conditions of known CFTs. In this case, the boundary spectrum is unknown, and $\hgapa$ is the dimension of the first operator in the boundary spectrum above the identity, while $\hgaptp$ is the dimension of the first operator in the boundary spectrum above the identity that appears in the bulk-to-boundary OPE \eqref{BOPE}. The two might differ, for accidental or symmetry reasons. 
Similarly, $\Delta_{\text{gap}}$ is the scaling dimension above which \emph{non-isolated} scalar bulk operators may contribute to both the four-point function and the annulus partition functions, \emph{i.e.} both $c_\D$ and $a_\D$ can be different from zero. In practice, above $\Delta_{\text{gap}}$ SDPB will assume that the spectrum can be continuous---see eq. \eqref{eq:SDP-bulk-sc} below. $\Delta_{\text{gap}}$ need not be equal to the minimal bulk scaling dimension. Sometimes, we will input in the bootstrap a set of low lying bulk states. The set of their scaling dimension is denoted as $\mathcal{I}_{2\times2}$. There might also be a known discrete\footnote{Continuous sets can also be accounted for via a fractional linear transform of $\Delta$, to get additional sets of positivity constraints. We do not explore this situation in the current work. We thank Andrea Cavaglià for discussions on this.}  set of operators $\mathcal{I}_a$ whose one-point function is not assumed to vanish, but which do not appear in the $\phi \times \phi$ OPE. Analogously, there might be a second discrete set $\mathcal{I}_4$, which appear in the bulk channel OPE of the external operators but whose coupling to the boundary vanishes, for instance by symmetry. These two sets are accounted for explicitly by including additional positivity conditions on the following vectors, 
\begin{align}
     &\vec{V}^{a}_{\Delta} = \begin{pmatrix}
        0, & 0, & \chi_{\frac{\Delta}{2}}(t)
    \end{pmatrix}~, \quad \D \in \mathcal{I}_a~, \\
    &\vec{V}^{4}_{\Delta} = \begin{pmatrix}
        F_{\Delta,\ell=0}(z,\bz), & 0, & 0
    \end{pmatrix}~, \quad \D \in \mathcal{I}_4~.
\end{align}
In a few cases, we will assume the knowledge of a discrete set of low-lying operators appearing in the boundary OPE of $\phi$. We will call this set $\mathcal{I}_2$.
Finally, the gap in the spinning sector is taken to be the unitarity bound, unless otherwise stated.

We are now able to formulate our first numerical bootstrap problem with points and lines. If we
can find a functional $\vec{\alpha}$ such that
\begin{subequations}\label{eqs:SDP}
    \begin{alignat}{2}
       \begin{pmatrix}
            1~~ & 1
        \end{pmatrix} 
        \vec{\alpha}\cdot\vec{V}_{\Delta=0,\ell=0}
        \begin{pmatrix}
            1 \\ 1
        \end{pmatrix} &= 1, \label{eq:SDP-bulk-id}\\
        \vec{\alpha}\cdot\vec{V}^{a}_{h=0} &\geq 0, \label{eq:SDP-bd-ann-id}\\
        \vec{\alpha}\cdot\vec{V}^{\text{2pt}}_{h=0} &\geq 0 , \label{eq:SDP-bd-2pt-id}\\
        \vec{\alpha}\cdot\vec{V}^{a}_{\Delta} &\geq 0 ~~ \forall \Delta \in \mathcal{I}_a, \label{eq:SDP-bulk-disc}\\
        \vec{\alpha}\cdot\vec{V}^{4}_{\Delta} &\geq 0 ~~ \forall \Delta \in \mathcal{I}_4, \label{eq:SDP-bulk-disc-4pt}\\
        \vec{\alpha}\cdot\vec{V}_{\Delta,\ell=0} &\succeq 0 ~~ \forall \Delta \in \mathcal{I}_{2\times 2} \cup [\Delta_{\text{gap}}, \infty), \label{eq:SDP-bulk-sc}\\
        \vec{\alpha}\cdot\vec{V}_{\Delta,\ell\neq0} &\geq 0 ~~ \forall \Delta-\ell \geq 0, \label{eq:SDP-bulk-l}\\
        \vec{\alpha}\cdot\vec{V}^{\text{2pt}}_{h} &\geq 0 ~~ \forall h \in \mathcal{I}_2 \cup [\hgaptp, \infty), \label{eq:SDP-bd-2pt}\\
        \vec{\alpha}\cdot\vec{V}^{a}_{h} &\geq 0 ~~ \forall h \geq h_{\text{gap}}^{\text{ann}}, \label{eq:SDP-bd-ann}
    \end{alignat}
\end{subequations}
then we have excluded the existence of a unitary theory at this specific choice of $\vec{p}~$, $\mathcal{I}_a$, $\mathcal{I}_{2}$, $\mathcal{I}_4$, and $\mathcal{I}_{2\times 2}.$\footnote{We will only explicitly write the sets $\mathcal{I}_a,\,\mathcal{I}_2,\,\mathcal{I}_4,\,\mathcal{I}_{2 \times 2}$ when they are not empty.} The above bootstrap problem is utilised to place bounds on gaps in the spectrum (see \cref{fig:ising_hann_h2pt} and \cref{fig:su(2)2_hann_h2pt}). We will discuss the bootstrap problem for OPE coefficients in more detail as it becomes relevant to the specific situations under consideration. 

The inequality \eqref{eq:SDP-bulk-sc} makes it clear why the inclusion of the four-point function and of the annulus partition function are necessary to get rigorous bounds out of crossing symmetry of the two-point function. Indeed, a symmetric matrix with a zero on the diagonal has one negative eigenvalue unless it is diagonal, and so eq.  \eqref{eq:SDP-bulk-sc}, without either of the two diagonal crossing equations, would require $\alpha_2$ to vanish when acting on $\mathcal{G}^{\text{bulk}}_{\Delta}(\xi)$, for all $\Delta$'s in the specified range. This implies $\alpha_2=0$ identically, at least within the finite dimensional space of functionals explored by SDPB.

In our implementation of the bootstrap problem, we adopt the standard convention of choosing the derivative basis for the functionals,
\begin{eqnarray}
    \alpha_4 &=& \sum_{\substack{m \geq n \\ m+n \leq N}} \frac{1}{m!\,n!} \partial_z^m \partial_{\bar{z}}^{n} |_{z=\bar{z}=\frac{1}{2}}, \label{eq:alphaz} \\
    \alpha_2 &=& \sum_{m \leq N} \partial_{\xi}^m |_{\xi=\xi^*}, \label{eq:alphaxi} \\
    \alpha_a &=& \sum_{m \leq N} \partial_{\tau}^m |_{\tau = 1}. \label{eq:alphatau}
\end{eqnarray}
In principle, one can choose a different $N$ for each of the three $\alpha_i$'s. However, in practice we observed that numerical stability requires them to be equal. Further, requiring a common prefactor in all elements of $\vec{V}_{\Delta,\ell=0}$ forces a choice of $\xi$ dependent on the choice of $z$, $\bar{z}$, and $\tau$. For the specific choice made in \eqref{eq:alphaz} and \eqref{eq:alphatau}, $\xi^*$ takes the value \eqref{eq:reln-b/w-crossratio},
\begin{equation}
    \xi^* \simeq 0.03010. 
\end{equation}
This choice of $\xi$ also leads to an adjustment of the one point function by a $\Delta$ dependent factor \eqref{eq:allBlocks}
\begin{equation}
    a_{\Delta} \rightarrow a_{\Delta} \nu^{-\Delta}, \qquad \nu \simeq 3.98513,
\end{equation}
and needs to be accounted for in OPE optimisation problems involving the one-point coefficient. For additional information on the different choices made in this paper in order to ensure a well defined and numerically stable semi-definite problem, we refer the reader to \cref{sec:block-imp}. 

\subsection{Non-elementary boundary conditions and bootstrap bounds}
\label{ssec:elementary}

Boundary states that contain only a single copy of the identity representation are commonly referred to as elementary. These states satisfy equation \eqref{annulus_sl2} with $n_0=1$. In what follows, we examine how non-elementary boundary states influence the resulting bootstrap bounds.

The bootstrap is agnostic about integrality of the degeneracies $n_h$, and the crossing equation \eqref{full-system} depends on $g^2$ only through the ratio $g^2/n_0$, which is the one we will bound. For simplicity, we will set $n_0=1$ in all formulae after this subsection, but it is important to notice that, at first sight, the bootstrap is blind to whether or not the boundary condition is elementary. 
In fact, one can sometimes exclude non-elementary branes by imposing certain assumptions.

Before discussing the latter, let us consider the value of $g^2/n_0$, $\Delta_\text{gap}$, $\hgapa$ and $\hgaptp$ for non-elementary boundary states.
A linear combination of elementary boundary states
\begin{equation}
    \ket{B}=\sum_{i=1}^nk_i \ket{B_i}~, \quad k_i>0~,
    \label{nonelem}
\end{equation} 
with ground state degeneracies $g_1,\,g_2,\,\dots,\,g_n$, is located in our plots at the following value of $g^2/n_0$:
\begin{equation}
    \frac{g^2}{n_0}=\frac{\left(\sum_{i=1}^nk_ig_i\right)^2}{\sum_{i=1}^nk_i^2}~.
    \label{gaverage}
\end{equation}
Eq. \eqref{gaverage} requires that the open string spectrum in $\bra{B_i} \exp (-r H_{bulk})\ket{B_j}$ is not supported at $h=0$ when $i\neq j$. The veracity of this statement can be seen from the following argument: if there was a topological boundary changing operator $\varphi_{top}$, all one-point functions of bulk operators would obey $\braket{O}_i=\nu \braket{O}_j$, with $\nu$ an overall constant, hence $\ket{B_i}=\nu \ket{B_j}$. Indeed, one can start from the correlator $\braket{O \varphi_{top}\varphi_{top}}$ on the unit disk, which is independent of the position of the defect operators, and notice that their fusion can only contain boundary operators of dimension $h=0$ in either theory $i$ or theory $j$. Since $\ket{B_i}$ and $\ket{B_j}$ are elementary, the only such operator is the identity, and the statement follows.

With the redefinition $v_i=k_i \bigg/ \sqrt{\sum_{j=1}^nk_j^2} $, eq. \eqref{gaverage} can be written as
\begin{equation}
    \frac{g^2}{n_0} = \left(\vec{v}\cdot \vec{g}\right)^2~, \qquad\quad
    \vec{v}^2=1~,\ \vec{g}=(g_1,\,\dots,g_n)~,\ v_i>0~,\ g_i>0~.
\end{equation}
It is easy to see that 
\begin{equation}
    \min g^2/n_0=\min_i \{g_i^2\}~,
    \label{gmincomb}
\end{equation}
obtained by choosing $v_i=\delta_{i i_0}$, with $g_{i_0}=\min_i \{g_i\}$. Similarly, 
\begin{equation}
    \max g^2/n_0=\vec{g}^2~,
    \label{gmaxcomb}
\end{equation}
is reached when $\vec{v} \parallel \vec{g}$. Notice that the maximum is generically not reached by a state with integer degeneracies, but any value between $\min g^2$ and $\max g^2$ can be approximated arbitrarily well by choosing (sufficiently large) integers $k_i$, because any real unit vector $\vec{v}$ can be correspondingly approximated. Furthermore, one can always choose all $v_i$'s to be non-vanishing (perhaps small) and so any allowed value of $g^2$ can be reached by boundary states with the same gap $\hgapa$, equal to the scaling dimension of the lightest boundary condition changing operator among the elementary branes. 

Regarding the value of $\Delta_\text{gap}$ for $\ket{B}$, while generically it is simply the smallest among the gaps in the elementary branes $\ket{B_i}$, this is not always the case: cancellations might happen, and we will see examples of this phenomenon when discussing \cref{fig:su(2)2_hann_delta}.

The lower bound on $g^2/n_0$ obtained from the annulus bootstrap as a function of $\hgapa$ is not especially affected by non-elementary branes. The bound is by definition a non-decreasing function of $\hgapa$, and the value \eqref{gmincomb} is simply compatible with this feature. 

On the other hand, the maximum \eqref{gmaxcomb} can be made larger by increasing the number of branes in the linear combination \eqref{nonelem}. In many cases, there exists a continuum of elementary boundary conditions, labeled for instance by an element of a Lie group. In these cases, for any value of $\Delta_\text{gap}$, one can make \eqref{gmaxcomb} arbitrarily large, at the price of making $\hgapa$ arbitrarily small.\footnote{This at least is the case when the moduli space of the boundary CFT is compact.} Therefore, in these cases the upper bound on $g^2/n_0$ from crossing of the annulus partition function must diverge as $\hgapa \to 0$, for any value of $\Delta_\text{gap}$.  

Finally, let us comment on the value of $\hgaptp$ for the boundary condition \eqref{nonelem}. The two-point function $\braket{\phi\phi|B}$ is the sum of the two-point functions with the elementary branes. Therefore, the boundary OPE does not include boundary changing operators, and $\hgaptp=\min_i\{{h_{\text{gap},i}^{\text{2pt}}}\}.$

As mentioned above, (some) non-elementary branes might be excluded by special spectral assumptions. For instance, in the annulus, one might either fix $\hgapa$, impose a specific relation between $\hgapa$ and $\Delta_\text{gap}$, or between $\hgapa$ and the dimension of an isolated bulk operator having a one-point function. Similarly, in the mixed-correlator bootstrap one might be interested in studying boundaries which satisfy a relation between $\hgapa$ and the dimension of the external operator $\Delta_\phi$.

Contrary to the annulus bootstrap, the bootstrap of points and lines also provides a model-independent way of excluding non-elementary branes from the space of solutions, which is available when the external operator $\phi$ acquires a one-point function and is isolated. For an elementary brane, one has the relation $b_0^2=a_{\Delta_\phi}^2$ in eq. \eqref{full-system}, which relates boundary OPE data of the two-point function and the annulus. This relation is easily imposed in SDPB as an objective---see \emph{e.g.} eq. \eqref{dbranes_obj_a} below---or as a positivity condition for the functionals. Instead, the boundary condition \eqref{nonelem} has $a_{\Delta_\phi}^2=(\sum_i k_i g_i a_i)^2/(\sum_i k_i g_i)^2$ while $b_0^2=(\sum_i k_i g_i a_i^2)/\sum_i k_i g_i.$ The two do not match unless all $k_i$'s vanish except one, \emph{i.e.} the brane is elementary.

Let us make a comment about cases where not all the $k_i's$ in \eqref{nonelem} are positive. The resulting boundary state is unphysical, as it can be seen by computing the annulus partition function between boundary conditions $\ket{B}$ and any of the $\ket{B_j}$ appearing with negative multiplicity. The only contribution to the $h=0$ eigenspace comes from the term $k_j\bra{B_j} \exp (-r H_{bulk})\ket{B_j}$ and the degeneracy is therefore negative. This off-diagonal term is, however, missing from the mixed-correlator system that we bootstrap, and in fact there are examples where linear combinations of this kind lead to positive degeneracies in the partition function with identical boundary conditions. If one drops the positivity requirement, then obviously eq. \eqref{gmincomb} is replaced by $\min g^2=0$, and cancellations might also lead to larger values of $\hgapa$. However, the cancellations required to keep all degeneracies non-negative are constraining, and bootstrap bounds mostly exclude $g^2=0$, the one exception being \cref{fig:c=1-dirichlet}. It would be nice to determine in which cases, if any, extremal solutions to the numerical bootstrap can be populated by linear combinations of elementary states with negative coefficients.

In practice, we will not try to systematically keep track of all non-negative linear combinations of known boundary states, but we will plot some of them to show how they compare to the bounds. It should be noted that since their values of $\hgapa$ and $\hgaptp$ are always bounded by the ones of elementary boundary conditions, non-elementary branes are never extremal in these directions in parameter space. 

As a final remark, it is useful to notice that a class of boundary conditions in theories obtained by discrete gauging leave similar imprints in the bootstrap bounds as non-elementary branes in the parent (un-gauged) CFT. Discrete gauging, or orbifolding, consists in restricting a theory to its sector invariant under a discrete symmetry, with the addition of twisted states to preserve modular invariance---see \emph{e.g.} \cite{Ginsparg:1988ui} for an introduction. Boundary conditions in the orbifold split into two classes: regular and fractional branes. Regular branes simply consist in gauge invariant linear combinations of boundary states of the parent theory. Fractional branes appear in the image under gauging of a boundary condition which is invariant under a subgroup of the discrete symmetry. Regular and fractional branes are reviewed in \cite{Recknagel:2013uja,Collier:2021ngi}. While fractional branes involve genuinely new Ishibashi states, because they overlap with the twisted sector, regular branes do not, hence considerations about $g^2$, $\hgapa$ and $\hgaptp$ made above hold for the regular branes. In particular, using the fact that $g$ is constant under application of global group-like symmetries, one easily obtains $g^2_\text{reg}=|G| g^2$ where $g$ is the entropy of the boundary condition in the parent theory and $|G|$ is the order of the discrete symmetry group.

\section{Numerical results}\label{sec:numerical}
We now apply the framework developed in the previous section to systems with increasingly richer space of boundary conditions: the Ising model, the $c=1$ CFTs and the $c=3/2$ CFTs, where we mainly focus on the $\mathfrak{su}(2)_2$ WZW model. While boundary states are classified in the first two cases, no such classification exists in the latter. In this regime, our numerical bounds yield genuinely novel information.

\subsection{The Ising model}
\label{subsec:ising}

We begin with the simplest case of a two dimensional CFT that admits boundaries - the Ising model. Since it is a rational theory under Virasoro algebra, the boundary conditions that preserve Virasoro symmetry have been classified \cite{Cardy:1989ir}. We use the Ising model as a benchmark to understand how different parts of the semi-definite program (SDP) fit together, and to set the expectations for more complicated models we consider. 

The model has three Virasoro primary operators, all scalars---the identity, $\sigma$ and $\epsilon$---of dimensions $\Delta=0,\,1/8,\,1$ respectively. It has a $\mathbb{Z}_2$ symmetry which flips the sign of $\sigma$.
There are three conformal boundary states
\begin{subequations}
    \begin{align}
        \ket{\mathbb{1}}&=\frac{1}{\sqrt{2}}\left(|\mathbb{1}\rangle\rangle+2^{1/4}|\mathbb{\sigma}\rangle\rangle+|\mathbb{\epsilon}\rangle\rangle\right)~, \\
        \ket{\epsilon}&=\frac{1}{\sqrt{2}}\left(|\mathbb{1}\rangle\rangle-2^{1/4}|\mathbb{\sigma}\rangle\rangle+|\mathbb{\epsilon}\rangle\rangle~\right), \\
        \ket{\sigma}&=|\mathbb{1}\rangle\rangle-|\mathbb{\epsilon}\rangle\rangle~.
    \end{align}
    \label{IsingCardy}
\end{subequations}
Each equation above expresses the boundary state $|B\rangle$ defined by \eqref{StripCrossing} in the basis of local bulk operators. In particular, each Ishibashi state $|O\rangle\rangle$ resums the contribution of a Verma module. 
$\ket{\mathbb{1}}$ and $\ket{\epsilon}$ break the $\mathbb{Z}_2$ symmetry of the model, and are mapped to each other by a $\mathbb{Z}_2$ transformation. The third Cardy state, $\ket{\sigma}$, is $\mathbb{Z}_2$ symmetric, and corresponds to free boundary conditions in the microscopic model. The Verma modules supported on each boundary condition are those for which
\begin{equation}
    n_{ii}{}^k \neq 0~,
\end{equation}
where $n$ is the fusion matrix of the model, $i=\mathbb{1},\epsilon,\sigma$ labels the boundary condition, and $k$ labels the boundary primary operator. An analogous rule determines the boundary changing operators, and these statements are valid for any rational boundary condition \cite{Cardy:1989ir,Behrend:1999bn}. In particular, $\ket{\mathbb{1}}$ and $\ket{\epsilon}$ both support only the boundary Verma module of the identity, while the $\ket{\sigma}$ boundary condition contains a primary of dimension $h=1/2.$ Notice that this boundary operator is $\mathbb{Z}_2$ even, as it can be deduced from the fact that the boundary OPE of the bulk operator $\sigma$ does not contain the identity and cannot be empty.

We now explore the following question: if we only had knowledge of the CFT on the infinite plane, what information could we extract numerically about the boundary states? Let us focus on the $\mathbb{Z}_2$ preserving boundary conditions. Then, we might choose $\sigma$ as the external operator, impose that it does not get a one-point function, and that the next quasi-primary in the bulk OPE is $\epsilon$. This leads to the following choice of gaps
\begin{equation}\label{eq:ising-gaps}
    \vec{p}=\left(\frac{1}{8} ~\bigg| 1;~h_{\text{gap}}^{\text{ann}},~\hgaptp ~ \bigg| 0\right), ~~ 
\end{equation}
with the notation defined in eq. \eqref{pdef}. 

\begin{figure}[!ht]
    \centering
    \begin{tikzpicture}
        \node at (0,0) {\includegraphics[width=\textwidth]{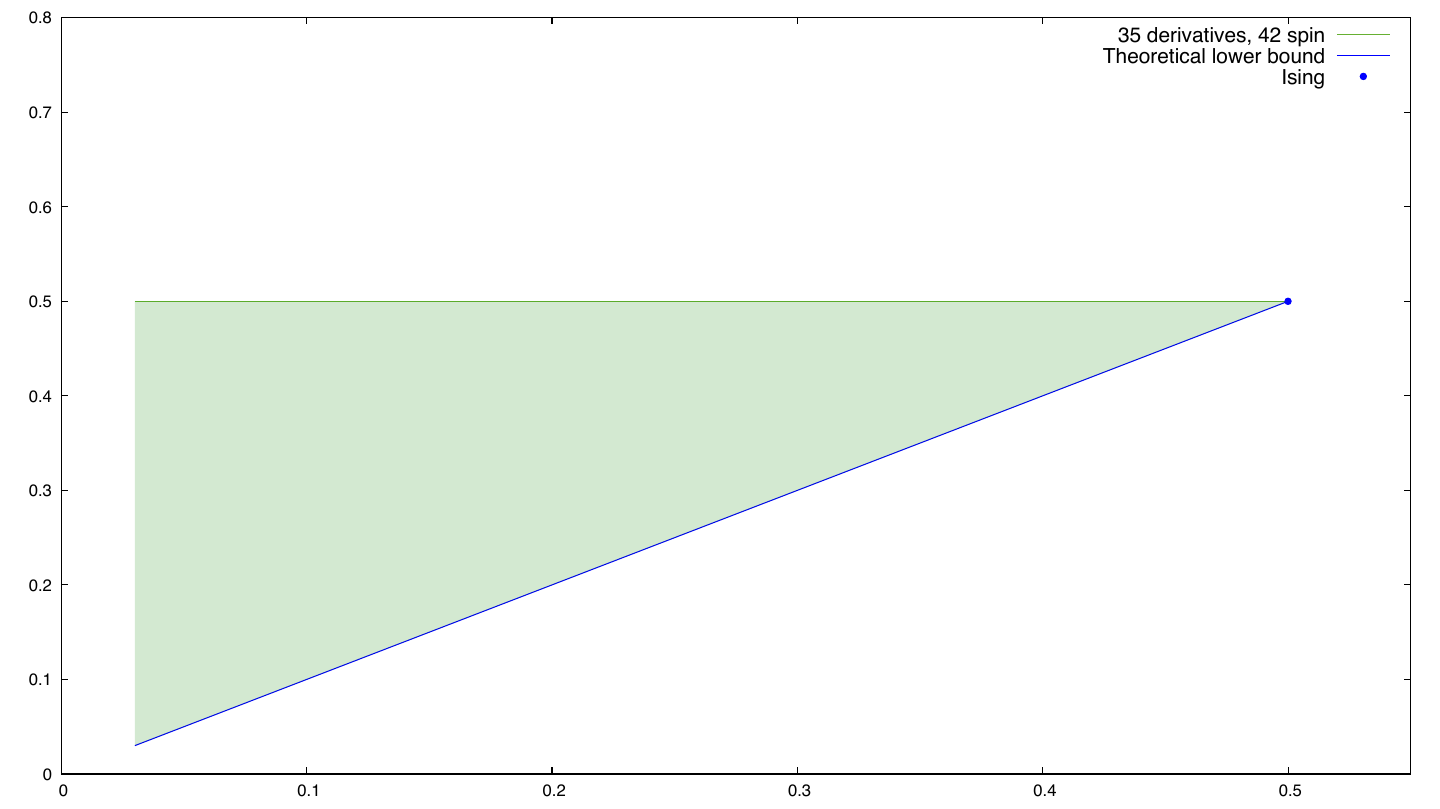}};
        
    
        \node[scale=0.7, anchor=east] at (-7.5,4.25) {$h^{\text{2pt}}_{\text{gap}}$}; 
        \node[scale=0.7, anchor=east] at (8,-4.2) {$h^{\text{ann}}_{\text{gap}}$}; 
        
    \end{tikzpicture}
    \caption{Bounds from the mixed-correlator system on the boundary channel gaps of theories with $c=\frac{1}{2}$, for $\vec{p}=(\frac{1}{8}| 1; h^{\text{ann}}_{\text{gap}}, h^{\text{2pt}}_{\text{gap}}|0)$ at $N=35$ derivatives and including operators with spin up to 42. Here and in the plots below, either the legend or the caption will contain both the number of derivatives and the number of spinning operators.}
    \label{fig:ising_hann_h2pt}
\end{figure}

The natural problem then is to carve the allowed region in the $(h_{\text{gap}}^{\text{ann}},~\hgaptp)$ plane. The result is shown in \cref{fig:ising_hann_h2pt}. Let us first describe a few general features of these kinds of plots. Given an allowed point, all points with smaller $\hgapa$ and $\hgaptp$ are allowed as well, and vice versa, given an excluded point, all points with larger gaps are excluded. However, since $\hgapa \leq \hgaptp$ by definition, points below the 45 degree blue line are unphysical and we exclude that region by hand. 

All in all, \cref{fig:ising_hann_h2pt} appears to be the intersection of two independent bounds: $\hgapa<0.5$ and $\hgaptp<0.5$. Precisely at the tip of the allowed region lies the boundary condition labeled by $\sigma$. It is important to notice that the vertical bound---$\hgaptp<0.5$---is a genuine new prediction from the two-point function of local operators with a boundary, and the first rigorous bootstrap result about this observable.

\begin{figure}[!ht]
    \centering
    \begin{tikzpicture}
        \node at (0,0) {\includegraphics[width=\textwidth]{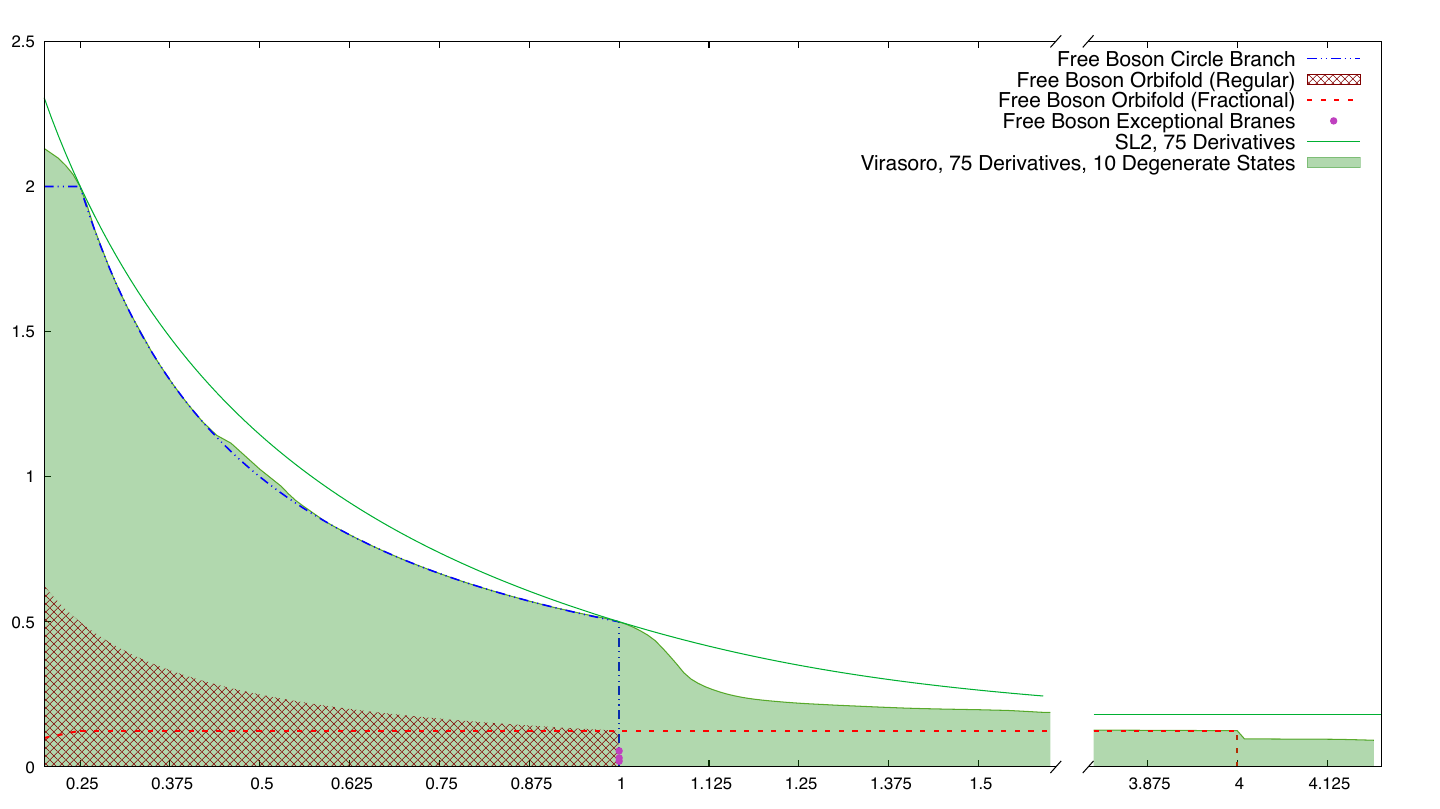}};
    
        \node[scale=0.7, anchor=east] at (-7.5,4.25) {$\Delta_{\text{gap}}$}; 
        \node[scale=0.7, anchor=east] at (7.8,-4.2) {$h_{\text{gap}}$}; 
        
    \end{tikzpicture}
    \caption{Bounds on the bulk gap as a function of boundary gap for $c=1$ theories, from the annulus bootstrap. We compare the bounds obtained from Virasoro characters with $N=75$ and with $10$ degenerate states added (see \cref{Vir_char_deg}), and $sl(2)$ characters with $N=75$. See figure legend for colour scheme. Due to the presence of a quasi-primary with $h=2$ in the Virasoro vacuum character, the largest possible gap in the $sl(2)$ sector is $h_\text{gap} = 2$. To allow a meaninful comparison to the Virasoro bounds, the $sl(2)$ bounds for $h_\text{gap}>2$ have been set to the value at $h_\text{gap} = 2$.}
    \label{fig:c=1_virasoro_ann_sub}
\end{figure}

\subsection{$c=1$}\label{ssec:free-boson}
We now consider the rich landscape of free-boson theories, whose action reads
\begin{equation}
    S = \frac{1}{8\pi} \int d^2z \,\partial \varphi(z, \bar{z}) \bar{\partial} \varphi(z, \bar{z})~.
    \label{SfreeBoson}
\end{equation}
The spectrum of the theory depends on the chosen target space for the field $\varphi$, and the full set of CFTs includes a conformal manifold and three isolated theories \cite{Ginsparg:1987eb}. The conformal manifold has two branches. On the circle branch, the target space is a circle with radius $2\pi R$.\footnote{$R$ is only physical once the normalization of the action \eqref{SfreeBoson} has been chosen. Our choice corresponds to setting $\alpha'=2$ in the string theory language, and different choices are easily distinguished by comparing the spectrum of vertex operators \eqref{vertexOps}.} On the orbifold branch, the target space is a segment of length $\pi R$. This branch is obtained from the circle branch by gauging the $\mathbb{Z_2}$ symmetry that sends  $\varphi \to -\varphi$. The two branches meet at a point. Finally, the three isolated CFTs are obtained by gauging discrete subgroups of an $SU(2)$ global symmetry, which only exists at the special radius $R=\sqrt{2}$. 

At a generic point on the circle branch, the model has a $U(1)_m\times U(1)_w$ continuous symmetry, where the subscripts stand for momentum and winding. The two $U(1)$'s are generated by the currents $j=(\partial \varphi,\bar{\partial} \varphi)$ and $\tilde{j}=(\partial \varphi,-\bar{\partial} \varphi)$ respectively. Taking the $\mathbb{Z}_2$ orbifold breaks both of these symmetries, since neither current is invariant under $\varphi \to -\varphi$. 

On the circle branch, the spectrum of Virasoro primaries comprises the vertex operators, which are labeled by their $U(1)$ charges $m,\,w,$ and an additional tower of states belonging to the identity module of the $U(1)$ chiral algebra. The scaling dimensions of the vertex operators are
\begin{equation}
    h = \frac{1}{2} \left( \frac{m}{R} + w\frac{R}{2} \right)^2, \quad \bar{h} = \frac{1}{2} \left( \frac{m}{R} - w\frac{R}{2} \right)^2,
    \qquad m,\,w \in \mathbb{Z}~.
    \label{vertexOps}
\end{equation}
The Virasoro primaries which are $U(1)$ singlets have scaling dimensions
\begin{equation}\label{eq:bulk-currents}
    h = n^2, \quad \bar{h} = \bar{n}^2, \qquad n,\,\bar{n} \in \mathbb{N}~.
\end{equation}
The spectrum is invariant under the involution $R \to 2/R$: this is $T-$duality, which exchanges winding and momentum modes. In this work, we will use conventions in which $R\geq \sqrt{2}$, the self-dual radius. Therefore, the lightest scalar operator in the theory will always be the first momentum mode, whose scaling dimension $\Delta_{(m=1,w=0)} \in (0,1/2]$. On the other hand, the lightest scalar winding mode has dimension $\Delta_{(m=0,w=1)} \in [1/2,\infty)$. This will be useful in the following.

The boundary states of the free boson theory are classified \cite{Gaberdiel:2001xm,Gaberdiel:2001zq,Cappelli:2002wq,Yamaguchi:2003yq,Janik:2001hb,Gaberdiel:2008zb}, and are nicely reviewed in \cite{Recknagel:2013uja,Collier:2021ngi}, to which we refer for details. Our aim in this section is to check the performance of the bootstrap in a scenario where the solutions to crossing are known. In particular, we will take as input some knowledge of the bulk spectrum, while for the boundary conditions we will enforce a choice of symmetry, and be as agnostic as possible about the gaps. Exploring the boundary conditions at $c=1$ will also be the occasion to compare the bootstrap of the annulus partition function to the one of the mixed-correlator system of points and lines.

We will mainly focus on the Dirichlet and Neumann boundaries for the theories on the circle branch, whose Cardy states are respectively
\begin{subequations}\label{DNCardy}
\begin{eqnarray}\label{eq:circle-states}
    \ket{D(\theta)} &=& \frac{1}{\sqrt{R}} \left( \sum_{n\in\mathbb{Z}^+} \ket{n;n}\rangle + \sum_{m\in\mathbb{Z}} e^{im\theta} \ket{(m,0)}\rangle \right), \label{DCardy} \\
    \ket{N(\theta)} &=& \sqrt{\frac{R}{2}} \left( \sum_{n\in\mathbb{Z}^+} (-1)^n \ket{n;n}\rangle + \sum_{w\in\mathbb{Z}} e^{iw\theta} \ket{(0,w)}\rangle \right).
    \label{NCardy}
\end{eqnarray}
\end{subequations}
The Cardy states show that momentum (winding) modes acquire one-point functions with a $D$-brane (Neumann brane), respectively. Hence, each boundary condition preserves one copy of the $U(1)$ chiral algebra. By comparing the states in \eqref{DNCardy} with \eqref{vertexOps} and \eqref{eq:bulk-currents}, we see that the lightest operator with an expectation value has $\Delta=1/R^2$ for a $D$-brane, and $\Delta=\min(2,R^2/4)$ for an $N$-brane.  From eqs. \eqref{DNCardy}, one can calculate the annulus partition functions, and extract the boundary spectrum by crossing. In the Dirichlet case, one finds (see for example, \cite{Oshikawa:1996dj})
\begin{equation}\label{eq:Z-D}
    Z_D(\theta, \theta') = \sum_{n\in\mathbb{Z}} \frac{ \tilde{q}^{ \frac{1}{2} \left( nR + \frac{(\theta - \theta')R}{2\pi} \right)^2} }{\eta(\tilde{q})}, 
\end{equation}
where $\tilde{q}= q(1/t)$. Thus for $R\geq\sqrt{2}$, setting $\theta = \theta'$, we obtain the boundary gap $h_{\text{gap}}^{\text{ann}} = 1$. This is due to the fact that the $U(1)$ vacuum character $\eta(\tilde{q})^{-1}$ is different from both the $SL(2)$ and the Virasoro vacuum characters. The mismatch is given by the current, while all winding modes have values of $h > 1$. 

Similarly, we can use the state $\ket{N(\phi)}$ to calculate the partition function for Neumann branes as an expansion in the boundary characters:
\begin{equation}
    Z_N(\theta,\theta') = \sum_{n\in\mathbb{Z}} \frac{ \tilde{q}^{ \frac{1}{2} \left( \frac{2n}{R} + \frac{(\theta - \theta')}{\pi R} \right)^2} }{\eta(\tilde{q})}, 
    \label{ZN}
\end{equation}
Thus, the boundary gap is in this case $h_{\text{gap}}^{\text{ann}} = \frac{2}{R^2}$. 
The information about Dirichlet and Neumann boundaries is summarized in \cref{tab:state-bdry-FB}.

The circle branch admits additional branes at rational multiples of the self-dual radius---$ R = \frac{M}{N}\sqrt{2}$. These branes are parametrised by elements in $SU(2)/\mathbb{Z}^N\times\mathbb{Z}^M$. They are labelled by an $SU(2)$ element $g$. Their bulk gaps generically are always equal to $1/2$, while their boundary gap depends on the angle of $g$ on the maximal torus of $SU(2)$. For all values of this angle, $\hgapa < 1/4$, and a vast majority of them have the boundary gap $\hgapa < 1/10$.\footnote{The exception to this is states labelled by a diagonal matrix $\begin{pmatrix}
    e^{i\theta} & 0 \\
    0 & e^{-i\theta} \\
\end{pmatrix}$, which have a $\hgapa$ of $1$.} 

In addition, the $c=1$ space of boundary CFTs includes regular and fractional branes, which are the images of the previous boundary states under the $\mathbb{Z}_2$ orbifold. Their annulus partition functions at generic radius can be found in eqs. 5.55 and 5.56 of \cite{Recknagel:2013uja}. At rational multiples of $R=\sqrt{2}$ one finds regular branes whose $\hgapa$ is bounded by the ones in the circle branch, in accordance with the discussion in \cref{ssec:elementary}.

Finally, the three T-O-I exceptional theories obtained by gauging non-abelian discrete symmetries of $SU(2)$ admit two classes of boundary conditions. The first class includes finitely many Cardy states with respect to the extended chiral algebras of each of the three theories. The second class includes the images under orbifolding of the Cardy states of the $\mathfrak{su}(2)_1$ WZW model. These exceptional branes are reviewed in more detail in \cite{Collier:2021ngi}. 

We start our bootstrap exploration of the free boson theory by placing numerical bounds on $\Delta_{\text{gap}}$ and $h_\text{gap}$. To do so, we restrict the system \eqref{full-system} to the annulus subproblem. The result is shown in  \cref{fig:c=1_virasoro_ann_sub}, together with some of the boundary conditions discussed above  (to maintain visual clarity, a few classes of boundary conditions have been omitted from the plot if they are in the convex hull of the boundary conditions showcased, or if they have values of $\hgapa$ outside the range showed).

This plot nicely complements the bounds on the $g$ function presented in \cite{Collier:2021ngi}, and shows that when Virasoro symmetry is implemented, the bounds are close to optimal. This is remarkable since the only assumptions entering the bootstrap are the value of the central charge and the associated representation theory, which includes a tower of degenerate modules---see eq. \eqref{Vir_char_deg}. The origin of the kinks in the envelope of the boundary conditions is in fact easy to understand from the point of view of Virasoro representation theory: the bumps appear when a new degenerate Verma module becomes available in the bulk or on the boundary. While the $sl(2)$ bootstrap misses a few important features of the space of solutions, the $sl(2)$ bound does capture an important property of branes in $c=1$ theories: \emph{if the boundary condition is stable} ($h_{\text{gap}}\geq 1$), \emph{a bulk operator with $\Delta\leq 1/2$ must acquire a one-point function}. All such scalars are charged under either of the two $U(1)$'s, hence one concludes that stable boundary conditions either belong to orbifold models or to symmetry breaking boundary conditions on the circle branch. One can make a further distinction by using symmetry and the OPE. 
Indeed, if there is a current in the bulk, its boundary OPE is non-singular and always generates a boundary operator of dimension $h=1$ as a consequence of holomorphy. Therefore, all the boundary conditions for the circle branch must lie in the region $h_\text{gap}\leq 1$.
In our $T-duality$ frame, the winding modes all have dimension $\Delta \geq 1/2$, hence the annulus bootstrap alone allows us to conclude that \emph{boundary conditions which preserve $U(1)_m$ are unstable for $R> \sqrt{2}$}. 

For completeness, let us mention that there cannot be a boundary condition which preserves the full chiral algebra $U(1)_m\times U(1)_w$. One way to see this is to notice that the boundary value of a current implements the variation of the boundary condition under the symmetry, and it vanishes if the boundary does not break the symmetry. While one linear combination of holomorphic and antiholomorphic currents can be set to zero via a Cardy gluing condition, holomorphy does not allow both operators to vanish.\footnote{These protected boundary operators are called the \emph{tilt} operators, and in two dimensions they generate \emph{chiral deformations}, in the language of \cite{Recknagel:1998ih}. Cardy conditions for the currents read 
\begin{equation}
j^a=\Omega^{ab}\bar{j}^b~.
\end{equation}
Here $\Omega$ is an automorphism of the symmetry algebra, which in the $U(1)$ case reduces to $\pm 1$. Tilt operators will feature more prominently in \cref{subsec:wzw}.}

Finally, we note that for values of $\hgapa$ below $0.2$, we observe numerical instabilities that prevent the solver from detecting a dual/primal feasible jump. We leave the resolution of this instability to future work.

\bgroup
    \def\arraystretch{1.5}
    \begin{table}
        \centering
        \begin{tabular}{|c|c|c|c|c|c|c|c|}
            \hline
            brane & $\Delta_\phi$ & $\Delta_{\text{min}}^{\text{ann}}$ & $\Delta_{\text{min}}^{\text{4pt}}$ & $h_{\text{min}}$ & $h^{\text{2pt}}_{\text{min}}$ & $g^2$ & $a_\phi^2$\\
            \hline
            $D(\theta)$ & $\frac{1}{R^2}$ & $\frac{1}{R^2}$ & $\frac{4}{R^2}$ & $1$ & $1$ & $\frac{1}{R}$ & 2 \\
            \hline
            $N(\theta)$ & $\frac{1}{R^2}$ & $\min(2,\frac{R^2}{4})$ & $\frac{4}{R^2}$ & $\frac{2}{R^2}$ & $\frac{2}{R^2}$  & $\frac{R}{2}$ & 0 \\
            \hline
        \end{tabular}
        \caption{BCFT data for free bosons on the circle branch. $\Delta_\phi$ denotes the lightest operator in the bulk spectrum.
        The subscript `min' denotes the smallest scaling dimension appearing above the identity  in the corresponding channel.}
        \label{tab:state-bdry-FB}
    \end{table}
\egroup
\smallbreak

We now move to the full system. We pick our external operator to be the following linear combination of the two momentum mode vertex operators,
\begin{equation}\label{eq:phi-c-1}
    \phi = \sqrt{2}~\text{Re}\left( \mathcal{V}_{(1,0)} \right) = \frac{1}{\sqrt{2}}\left( \mathcal{V}_{(-1,0)} + \mathcal{V}_{(1,0)} \right),
\end{equation}
which is the lowest lying real state above the identity in the bulk for $R \geq \sqrt{2}$, and has scaling dimension $\Delta_{\phi} = \frac{1}{R^2}$. The operator $\phi$ in \eqref{eq:phi-c-1} should not be confused with the elementary field $\varphi$. 
The Virasoro fusion rule reads \cite{Nemkov:2021huu}
\begin{equation}\label{eq:4pt-fusion}
    \phi \times \phi \sim 1+ \mathcal{V}_{(-2,0)} + \mathcal{V}_{(2,0)}+\sum_{\substack{n,\bar{n}\\n-\bar{n}\,\text{even}}}\mathcal{O}_{n,\bar{n}},
\end{equation}
where $\mathcal{O}_{n,\bar{n}}$ are the primaries with scaling dimensions as in eq. \eqref{eq:bulk-currents}.
Since $\phi$ is charged under $U(1)_m$ and neutral under $U(1)_w$, its one-point function vanishes with Neumann boundary conditions, while it is non-zero in the presence of a $D$-brane. Since $R>\sqrt{2}$, the first scalar operator in the bulk OPE above the identity is $\mathcal{V}_{2,0}$, with scaling dimension $\Delta=\frac{4}{R^2} \in (0,2]$. 

Our aim is to constrain the space of boundary conditions by inputting knowledge of the bulk spectrum and symmetries. In the context of the free boson, we shall focus on the circle branch and use bootstrap assumptions to input whether $U(1)_m$ is preserved or broken by the boundary.

\subsubsection{$U(1)_m$-preserving boundary conditions}

If $U(1)_m$ is preserved by the boundary condition, as in the case of the Neumann brane, only the winding modes and the Virasoro primaries $\mathcal{O}_{n,n}$ can acquire a one-point function. Of course, we know from the explicit boundary state \eqref{NCardy} that all of them do. The absence of a one-point function for the momentum modes will be our main tool in trying to isolate the Neumann boundary condition. The most natural piece of boundary data to bound with the bootstrap is the value of the ground state degeneracy $g$, defined in eq. \eqref{gdef}, and this will indeed be the focus of this subsection.

Let us discuss more precisely our choice of gaps.
The spectrum of $sl(2)$ primaries in the four-point function does not depend on the boundary conditions, and thus is populated by the states in \eqref{eq:4pt-fusion}. Thus, in view of the discussion above, the bulk gap in the annulus partition function is 
\begin{equation}
    \Delta_\text{gap}=\text{min}\left(2, \frac{R^2}{4}\right)=\text{min}\left(2, \frac{1}{4\Delta_\phi}\right)~.
    \label{deltagapN}
\end{equation}
Here and in the following, we trade $R$ for the scaling dimension of the external operator \eqref{eq:phi-c-1}.

As previously discussed, the annulus bootstrap results show that the boundary gap must necessarily be smaller than one (except at the self-dual radius). This renders our exercise in agnosticism about the boundary data harder, because there is no a priori way of setting $h_\text{gap}$. One option that might seem natural is to set it to the largest possible value allowed by the $sl(2)$ annulus subproblem---see \cref{fig:c=1_virasoro_ann_sub}. However, the extremal solution is expected to be unique, the value of $g$ fixed, and the spectrum is incompatible with Virasoro symmetry, except at most at a couple of points in \cref{fig:c=1_virasoro_ann_sub}. Instead, the Virasoro annulus bootstrap bound falls on top of the Neumann line for a large portion of the plot.\footnote{The small bump in \cref{fig:c=1_virasoro_ann_sub} around $h=0.5$ is a feature in the data, for which we do not have an immediate explanation.} 
At the corresponding values of $(h_\text{gap},\Delta_\text{gap})$, the $sl(2)$ problem is not extremal, and it makes sense to ask whether the mixed-correlator bootstrap can improve the bounds on $g$ with respect to the annulus bootstrap. Hence, we will set $h_{\text{gap}}=2\Delta_{\phi}$---see 
\cref{tab:state-bdry-FB}---in the region where the annulus bulk gap is between 0.5 and 2, \emph{i.e.} for $\Delta_\phi \in [1/8,1/2]$. 

\begin{figure}[!ht]
    \centering
    \begin{tikzpicture}
        \node at (0,0) {\includegraphics[width=\textwidth]{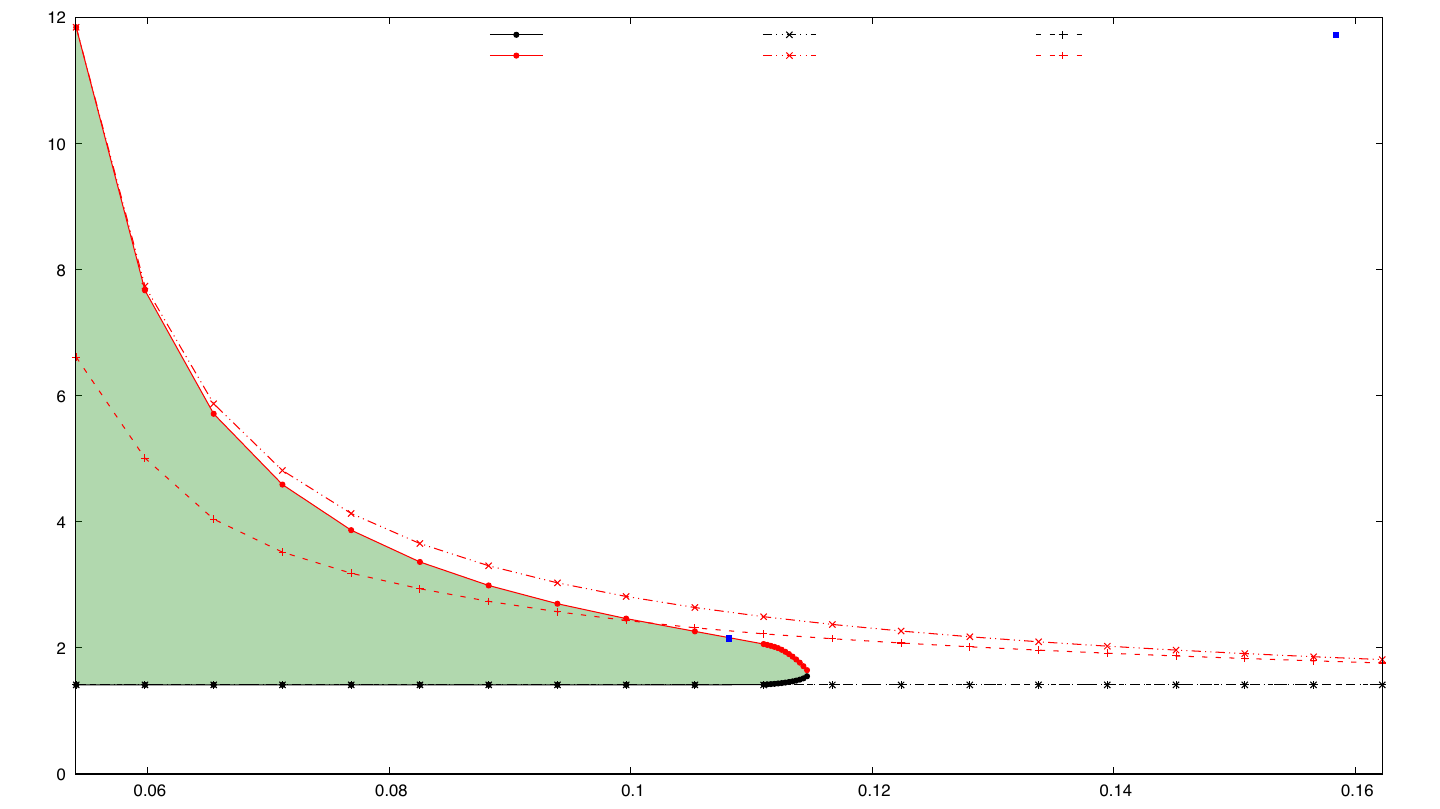}};
        
    
        \node[scale=0.525, anchor=east] at (0.45,4) {$g^2$-min $sl(2)$ Annulus}; 
        \node[scale=0.525, anchor=east] at (0.45,3.775) {$g^2$-max $sl(2)$ Annulus};
        
        \node[scale=0.525, anchor=east] at (-2.5,4) {$g^2$-min Mixed System}; 
        \node[scale=0.525, anchor=east] at (-2.5,3.775) {$g^2$-max Mixed System};
    
        \node[scale=0.525, anchor=east] at (3.4,4) {$g^2$-min Virasoro Annulus}; 
        \node[scale=0.525, anchor=east] at (3.4,3.775) {$g^2$-max Virasoro Annulus};
    
        \node[scale=0.525, anchor=east] at (6.3,4) {Neumann Brane}; 
    
        \node[scale=0.7, anchor=east] at (-7.15,4.25) {$g^2$}; 
        \node[scale=0.7, anchor=east] at (7.8,-4.2) {$h_{\text{gap}}$}; 
        
    \end{tikzpicture}
    \caption{Bounds on $g^2$ for $c=1$ theories as a function of $ h_{\text{gap}}$, with $\vec{p}=(\Delta_\phi| 2; h_{\text{gap}}, h_{\text{gap}})$, and $\mathcal{I}_4 = \{ 4\Delta_\phi \}$. We set $\Delta_\phi = 0.0540833$ ($R=4.3$). These gap assumptions  correspond to the Neumann brane. We compare the bounds obtained from the annulus subproblem with $sl(2)$ characters, annulus subproblem with Virasoro characters, and the full system (solid curves), all computed with $N=35$. The red curves denote the upper bound on $g^2$ and the black curves denote the lower bound. The blue square at $(0.108167, 2.15)$ is the Neumann brane.}
    \label{fig:c=1-neumann-h2pt}
\end{figure}
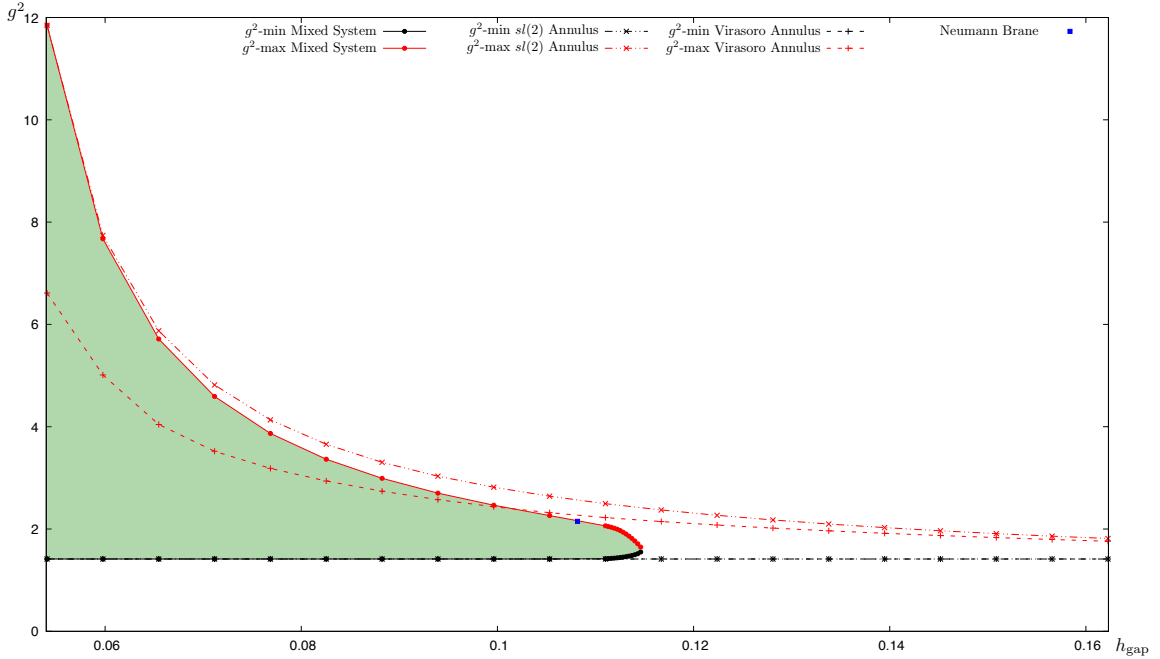

Finally, something different happens in \cref{fig:c=1_virasoro_ann_sub} when $h_\text{gap}<1/4$. Modular invariance of the annulus and Virasoro symmetry allow for $\Delta_\text{gap}>2$, and correspondingly the Neumann boundary state is no longer extremal. In view of our bulk gap assumptions, we will now set $\Delta_\text{gap}=2$ and leave the value of $h_\text{gap}$ as a parameter, to explore the allowed region. Such value of $\Delta_\text{gap}$ is compatible with eq. \eqref{deltagapN} if $\Delta_\phi<1/8$. In \cref{fig:c=1-neumann-h2pt}, we find the allowed region in the $(h_\text{gap},g^2)$ plane, for one such value of the external dimension, corresponding to $R=4.3$. We do not impose selection rules in the boundary OPE of $\phi$, and so, in the notation of \eqref{pdef} and \eqref{eqs:SDP}, for this plot we have 
\begin{equation}
\vec{p} = \left( \Delta_\phi \bigg| 2;h_{\text{gap}},h_{\text{gap}} \bigg| 0 \right), \quad
\mathcal{I}_4 = \{ 4\Delta_\phi \}, \qquad \Delta_\phi=0.0540833. 
\end{equation}

\Cref{fig:c=1-neumann-h2pt} showcases the added constraining power of the mixed system of points and lines. Firstly, the mixed-correlator bootstrap sets a more stringent upper bound on $h_\text{gap}$ than the $sl(2)$ annulus constraint.\footnote{Note that, for the $sl(2)$ annulus bootstrap, there is no noticeable difference between 35 derivatives and 75 derivatives.} In fact, the largest value of $h_\text{gap}$ compatible with modular invariance alone is 1/4, as visible in \cref{fig:c=1_virasoro_ann_sub}. This is the value of the boundary gap for the Neumann brane when $R=2\sqrt{2}$, which is extremal for $\Delta_\text{gap}=2$. The two- and the four-point functions are sensitive to the radius through $\Delta_\phi,$ and allow to refine the bound. Furthermore, while the Neumann brane lies well within the annulus bounds, it saturates the mixed-correlator bound on $g^2$. It would be interesting to input more of the low lying bulk spectrum, or to bootstrap the mixed system with Virasoro blocks, to see if the Neumann boundary condition can be made to extremize $h_\text{gap}$. We do not pursue this direction in the current study.

The lower bound on $g^2$ in \cref{fig:c=1-neumann-h2pt} is independent of $h_\text{gap}$. This feature was already noticed in \cite{Collier:2021ngi}, and it is seen here to extend to the mixed-correlator bootstrap. Analogous statements can be made for \cref{fig:c=1-neumann} and \cref{fig:su2-annulus} below. 

Before moving on, let us comment on the fact that, for small values of $h_\text{gap}$, the mixed system stops improving over the annulus in \cref{fig:c=1-neumann-h2pt}. This can be explained as follows. When boundary operators of dimension $h\leq \Delta_\phi$ are allowed in the spectrum, there is a unitary solution to crossing for the two-point function, where only the identity is exchanged in the bulk channel. This is the trivial defect, whose boundary spectrum is simply obtained by Taylor-expanding the bulk operators. One can choose the CFT data of any crossing symmetric annulus partition function, and add the trivial defect to it so that the mixed-system crossing equation \eqref{full-system} is satisfied. The correlators are decoupled and in particular the value of $g$ is unaffected.

We now proceed to bound $g^2$ as a function of the radius.
To keep things simple, we set $h_\text{gap}=2\Delta_\phi$, the Neumann value, also for $\Delta_\phi<1/8$. The summary of the gap choices made in \cref{fig:c=1-neumann} is then given by the following table:
\begin{subequations}\label{eq:neumann-gaps}
    \begin{alignat}{2}
        \circled{1} \qquad \vec{p} &= \left( \Delta_\phi \bigg| 2;2\Delta_\phi,2\Delta_\phi \bigg| 0 \right), ~~ \mathcal{I}_4 = \{ 4\Delta_\phi \},\quad \Delta_\phi<1/8 ~~ (2\sqrt{2}\leq R), \label{eq:neumann-r<2} \\
        \circled{2} \qquad \vec{p} &= \left( \Delta_\phi \bigg| \frac{1}{4\Delta_\phi};2\Delta_\phi,2\Delta_\phi \bigg| 0 \right), ~~ \mathcal{I}_4 = \{ 4\Delta_\phi \}, \quad 1/8\leq \Delta_\phi<1/4 ~~ (2\leq R \leq 2\sqrt{2}), \\
        \circled{3} \qquad \vec{p} &= \left( \Delta_\phi \bigg| 4\Delta_\phi; 2\Delta_\phi,2\Delta_\phi \bigg| 0 \right), ~~ \mathcal{I}_a = \left\{ \frac{1}{4\Delta_\phi} \right\}, \quad 1/4\leq \Delta_\phi\leq 1/2 ~~ (\sqrt{2}\leq R \leq 2).
    \end{alignat}
\end{subequations}
As we already emphasized, the transition between the gaps \circled{1} and \circled{2} is due to eq. \eqref{deltagapN}. Instead, the transition from \circled{2} to \circled{3} is due to the fact that at small enough radius the second momentum mode $\mathcal{V}_{2,0}$, which appears in the four-point function OPE, becomes heavier than the first winding mode $\mathcal{V}_{0,1}$, which appears in the closed string channel expansion of the annulus partition function.

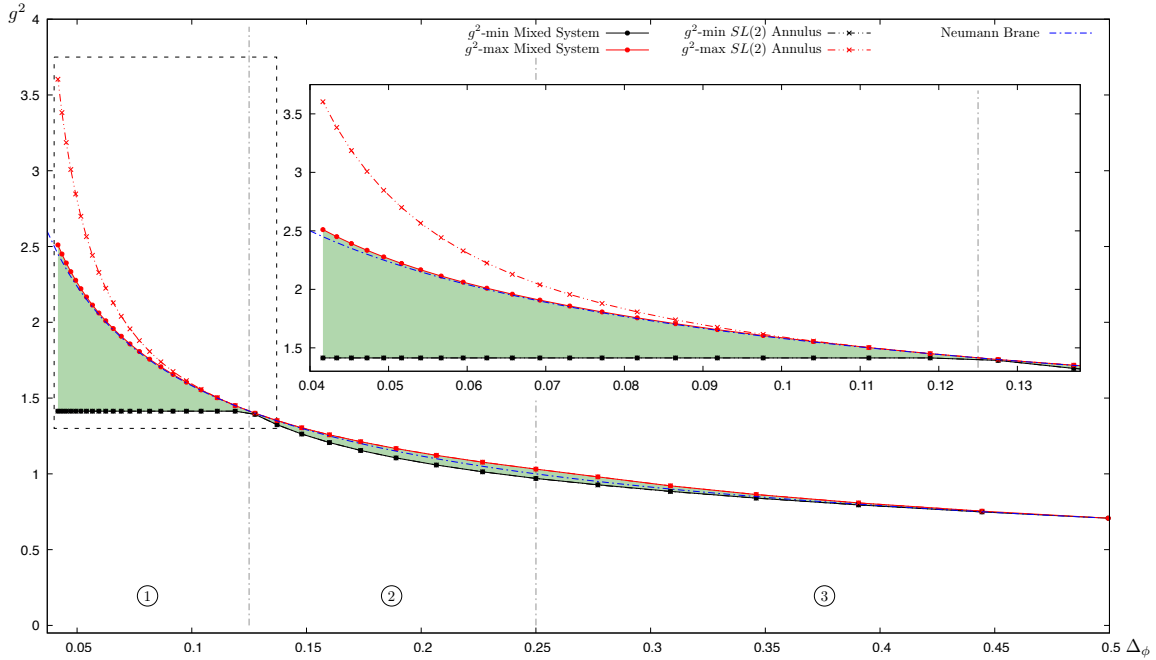
\begin{figure}[ht!]
    \centering
    \begin{tikzpicture}
        \node at (0,0) {\includegraphics[width=\textwidth]{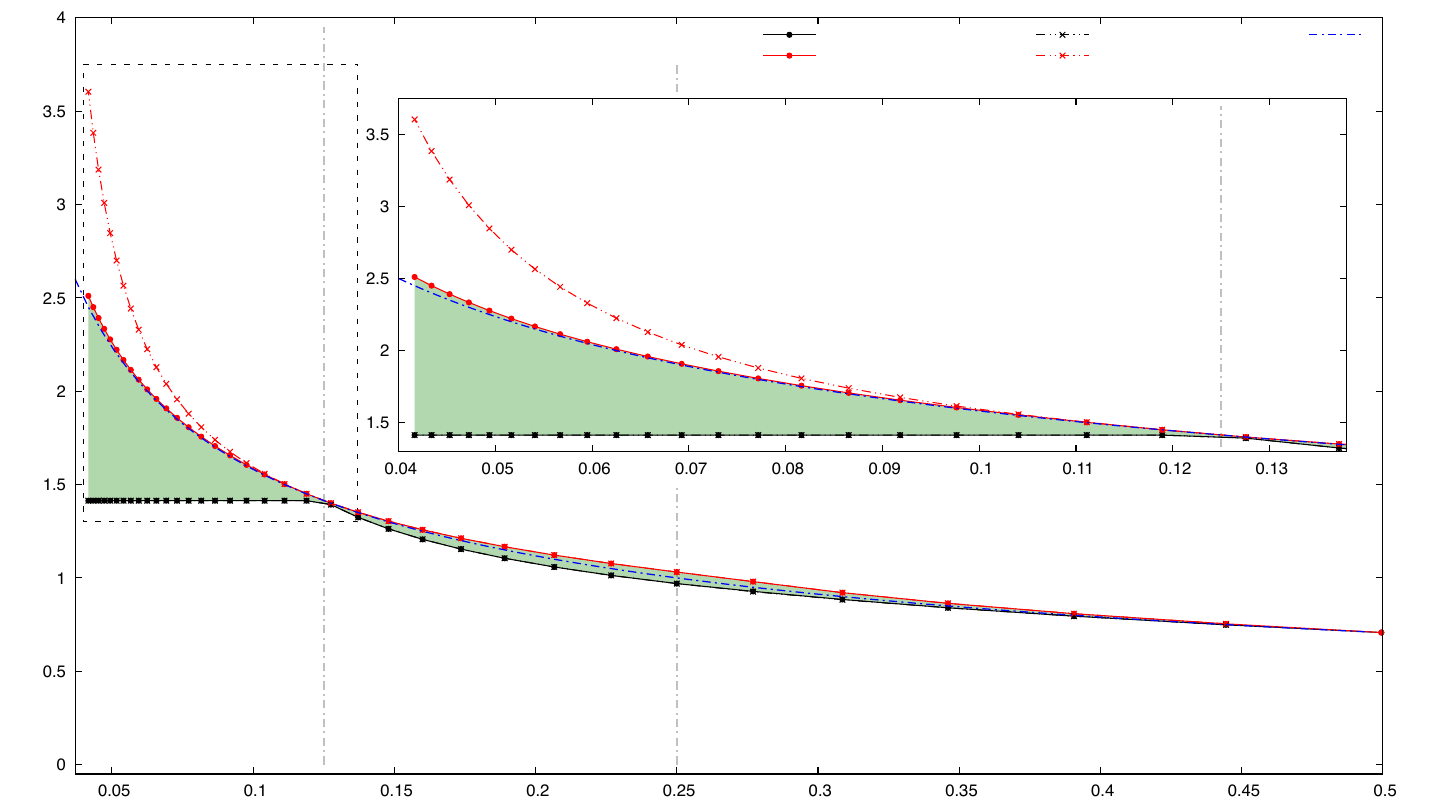}};
        
    
        \node[scale=0.525, anchor=east] at (0.45,4) {$g^2$-min Mixed System}; 
        \node[scale=0.525, anchor=east] at (0.45,3.775) {$g^2$-max Mixed System};
    
        \node[scale=0.525, anchor=east] at (3.4,4) {$g^2$-min $SL(2)$ Annulus}; 
        \node[scale=0.525, anchor=east] at (3.4,3.775) {$g^2$-max $SL(2)$ Annulus};
    
        \node[scale=0.525, anchor=east] at (6.3,4) {Neumann Brane}; 
    
        \node[scale=0.7, anchor=east] at (-7.15,4.25) {$g^2$}; 
        \node[scale=0.7, anchor=east] at (7.8,-4.2) {$\Delta_{\phi}$}; 

        \node[scale=0.7, anchor=east] at (-5.4,-3.5) {$\circled{1}$};
        \node[scale=0.7, anchor=east] at (-2.155,-3.5) {$\circled{2}$};
        \node[scale=0.7, anchor=east] at (3.6,-3.5) {$\circled{3}$};
        
    \end{tikzpicture}
    \caption{Bounds on $g^2$ as a function of $\Delta_\phi$ for $c=1$ theories for the choice of gaps described in \eqref{eq:neumann-gaps}, corresponding to Neumann branes. We compare the bounds obtained from the annulus subproblem with $sl(2)$ characters for $N=35$, and the full system with $N=35$ (solid curves). The red curves denote the upper bound on $g^2$ and the black curves denote the lower bound. The blue dotted curve is the Neumann brane. 
    }
    \label{fig:c=1-neumann}
\end{figure}
The resulting bounds are shown in \cref{fig:c=1-neumann}.  The benefits of the extra crossing equations are confirmed throughout region \circled{1}.\footnote{In region \circled{1}, the annulus bootstrap is sensitive to $\Delta_\phi$ via the boundary gap $h_\text{gap}=2\Delta_\phi$.} In contrast, in regions \circled{2} and \circled{3} the 
mixed-correlator system does not show any improvement over the annulus bootstrap.

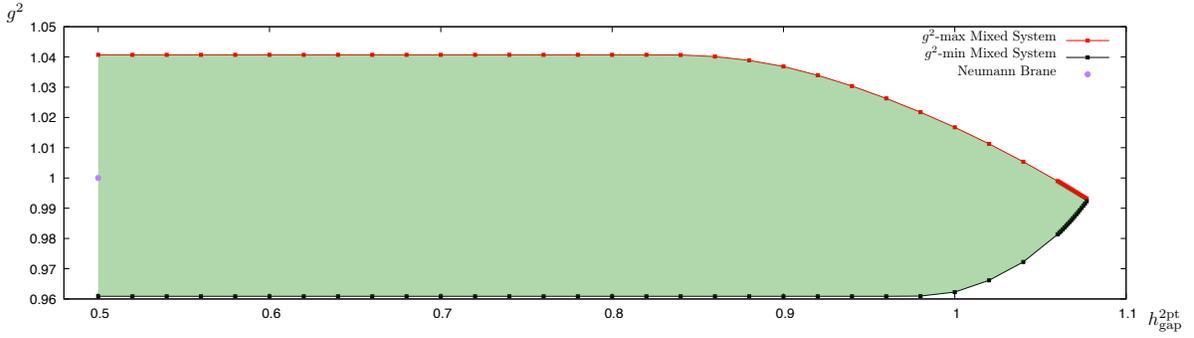
\begin{figure}[!htp]
    \centering
    \begin{tikzpicture}
        \node at (0,0) {\includegraphics[width=\textwidth]{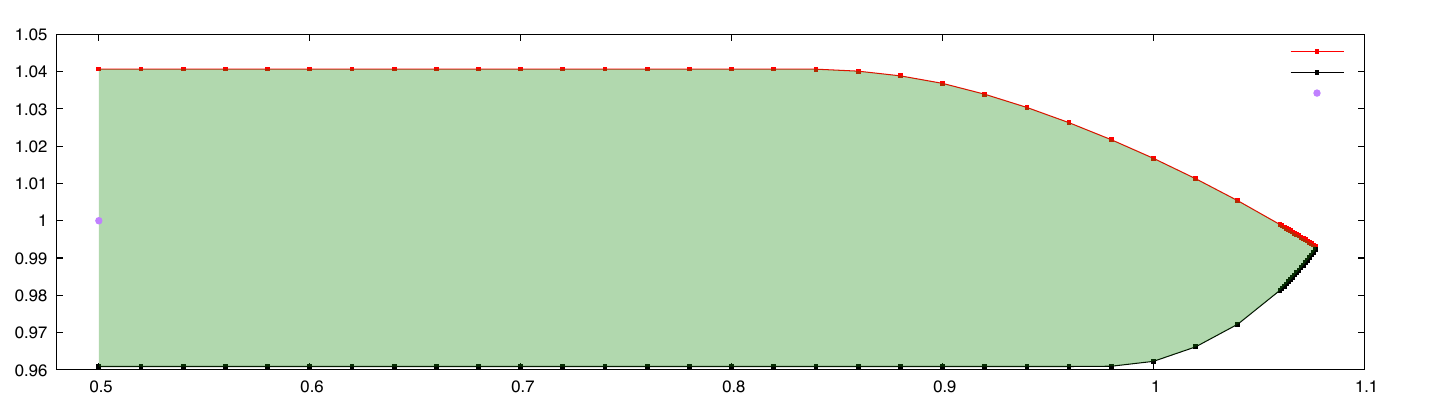}};
    
    
        \node[scale=0.525, anchor=east] at (6.1,1.683) {$g^2$-max Mixed System}; 
        \node[scale=0.525, anchor=east] at (6.1,1.457) {$g^2$-min Mixed System};
        \node[scale=0.525, anchor=east] at (6.1,1.231) {Neumann Brane};
    
        \node[scale=0.7, anchor=east] at (-7.6,2) {$g^2$}; 
        \node[scale=0.7, anchor=east] at (7.8,-2.1) {$\hgaptp$}; 
    \end{tikzpicture}
    \caption{Bounds on the $g$-function as a function of $\hgaptp$ for $c=1$ and $\vec{p}=(\frac{1}{4}|1;\frac{1}{2},\hgaptp|0)$, compatible to Neumann boundary conditions with $R=2$. The red curve denotes the upper bound on $g^2$ and the black curve denotes the lower bound on $g^2$, which were obtained for $N=35$. We would like to highlight that the difference between the upper and the lower bound is of the order of $0.08$.}
    \label{fig:c=1_whale_0.5}
\end{figure}

\begin{figure}[!ht]
    \centering
    \begin{tikzpicture}
        \node at (0,0) {\includegraphics[width=\textwidth]{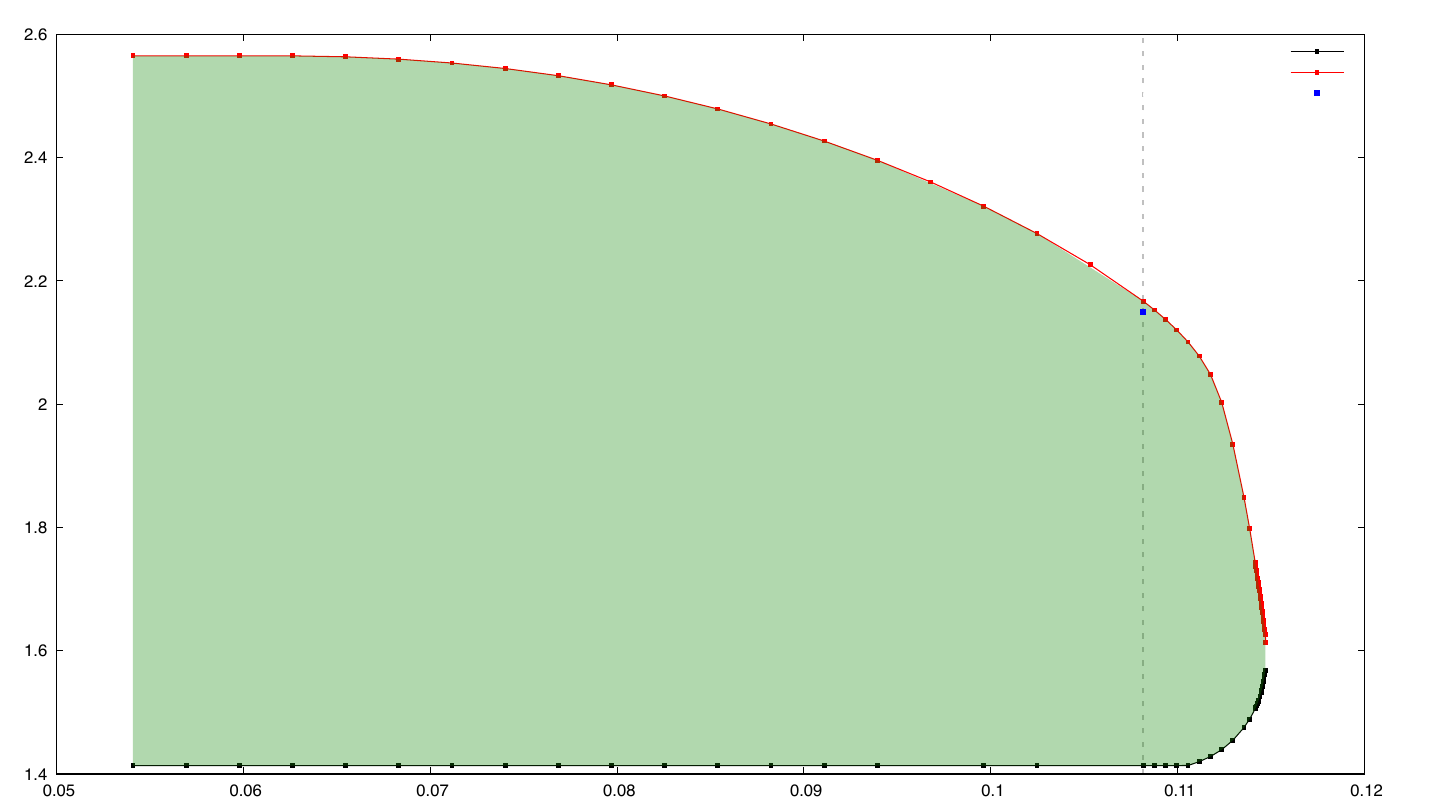}};
        
    
        \node[scale=0.525, anchor=east] at (6,3.83) {$g^2$-min Mixed System}; 
        \node[scale=0.525, anchor=east] at (6,3.6) {$g^2$-max Mixed System}; 
        \node[scale=0.525, anchor=east] at (6,3.37) {Neumann Brane}; 
    
        \node[scale=0.7, anchor=east] at (-7.4,4.25) {$g^2$}; 
        \node[scale=0.7, anchor=east] at (7.8,-4.2) {$h_{\text{gap}}^{\text{2pt}}$}; 
        
    \end{tikzpicture}
    \caption{Bounds on $g^2$ for $c=1$ theories as a function of $h_{\text{gap}}^{\text{2pt}}$, with gap assumptions $\vec{p}=(\Delta_\phi| 2; 2\Delta_\phi, h_{\text{gap}}^{\text{2pt}}|0)$ and $\mathcal{I}_4 = \{ 4\Delta_\phi \}$ at $\Delta_\phi = 0.0540833$ ($R=4.3$), corresponding to the Neumann boundary condition for $R\geq 2\sqrt{2}$. The bounds were obtained for $N=35$ derivatives. The red curves denote the upper bound on $g^2$ and the black curves denote the lower bound. The blue square at $(0.108167, 2.15)$ is the Neumann brane. The grey dotted line is to indicate that while $h_{\text{gap}}^{\text{2pt}} < h_{\text{gap}}^{\text{ann}}$ is allowed by the numerical bootstrap system, it is not physically meaningful.}
    \label{fig:c=1-nice-whale}
\end{figure}
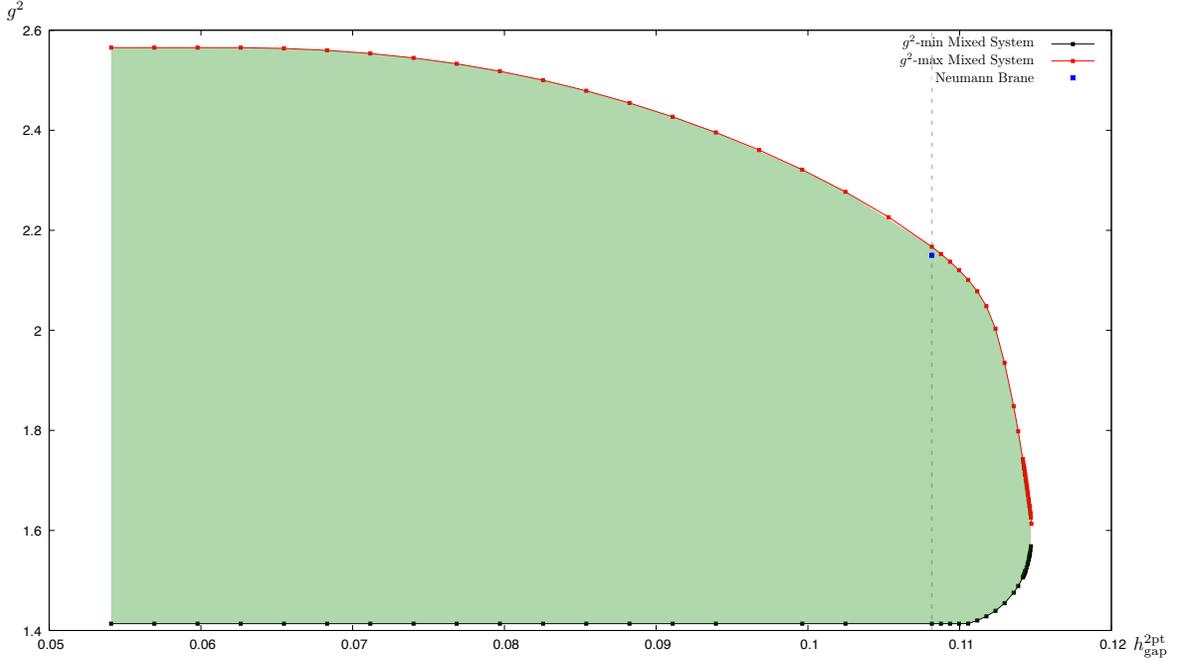

We get a clearer picture of the difference between region \circled{1} and the others by exploring the $(\hgaptp,g^2)$ subspace. Bounding the gap in the two-point function is a natural way to test the constraints arising from crossing of this observable, since it is expected that no solution to crossing exists if the bulk identity is present and $\hgaptp \to \infty$. Contrasting \cref{fig:c=1_whale_0.5} and \cref{fig:c=1-nice-whale}, one immediately notices a striking difference. In \cref{fig:c=1_whale_0.5}, we observe no improvement over the annulus subproblem for $\Delta_\phi=0.25,$ which is at the boundary between regions \circled{2} and \circled{3}. In this case, there remains a very large degree of freedom in $\hgaptp$. More to the point, the bounds on $g^2$ are insensitive to the value of $\hgaptp$ until the latter is much larger than the Neumann brane value. In contrast, the value of $\Delta_\phi$ in \cref{fig:c=1-nice-whale} is well within region \circled{1}. There, the Neumann brane is almost at the boundary of the allowed region, and the upper bound on $g^2$ strongly depends on $\hgaptp$ in its neighbourhood.

A detailed analysis of the extremal functionals reveals that, in the regions where the bounds on $g^2$ are independent of $\hgaptp$, the functionals acting on the four-point and the two-point function vanish. It is only at larger values of $\hgaptp$ where the four-point and two point-functions ``activate'', swiftly bringing the bounds closer to each other. This is of course in accordance with the intuition given above---at large enough $\hgaptp$ there should be no solution to crossing---but these examples show that the bounds on $g^2$ can often be solely dictated by the annulus bootstrap. In the next subsection, we will see that the situation is radically different when we optimize OPE data appearing in the two-point function.

\subsubsection{$U(1)_m$-breaking boundary conditions}
\label{subsec:dbranes}

If the boundary conditions break the shift symmetry of the boson, all momentum modes, including our external operator $\phi$, are allowed to acquire a one-point function. Regarding the boundary spectrum, as discussed above, the current, evaluated at the boundary, always provides a boundary operator of dimension $h=1.$ Because the first momentum mode has dimension $\Delta\leq 1/2$, it is consistent with the bound in \cref{fig:c=1_virasoro_ann_sub} to assume that no lighter boundary operators exist, so we set $\hgapa=\hgaptp=1$.
In order to study Dirichlet branes with our system, the information above is sufficient: we set $\Delta_{\text{gap}} = 4\Delta_{\phi}$ for all SDPB runs involving Dirichlet branes, and we add $\Delta_\phi$ to $\mathcal{I}_a$. Using the notation introduced in \eqref{pdef},
\begin{equation}\label{eq:dirichlet-gaps}
   \textup{Dirichlet:}\qquad \vec{p}=(\Delta_\phi| 4\Delta_\phi;1,1 | a_{\Delta_{\phi}}), ~~ \mathcal{I}_a = \{\Delta_\phi\}~.
\end{equation}

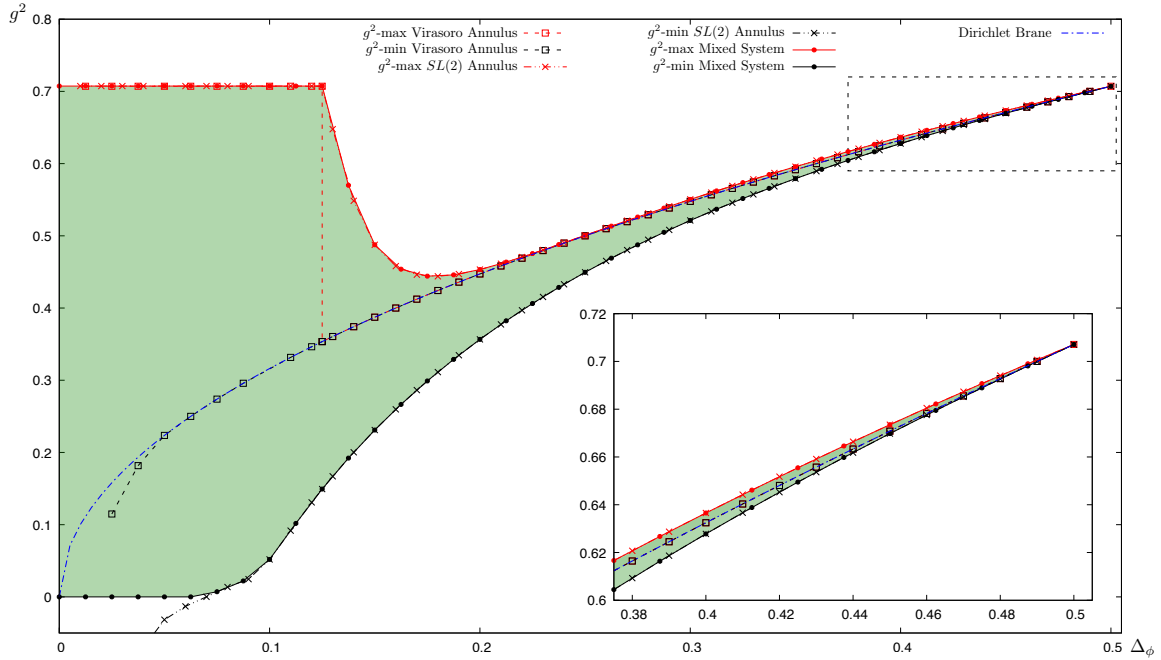
\begin{figure}[ht]
    \centering
    \begin{tikzpicture}
        \node at (0,0) {\includegraphics[width=\textwidth]{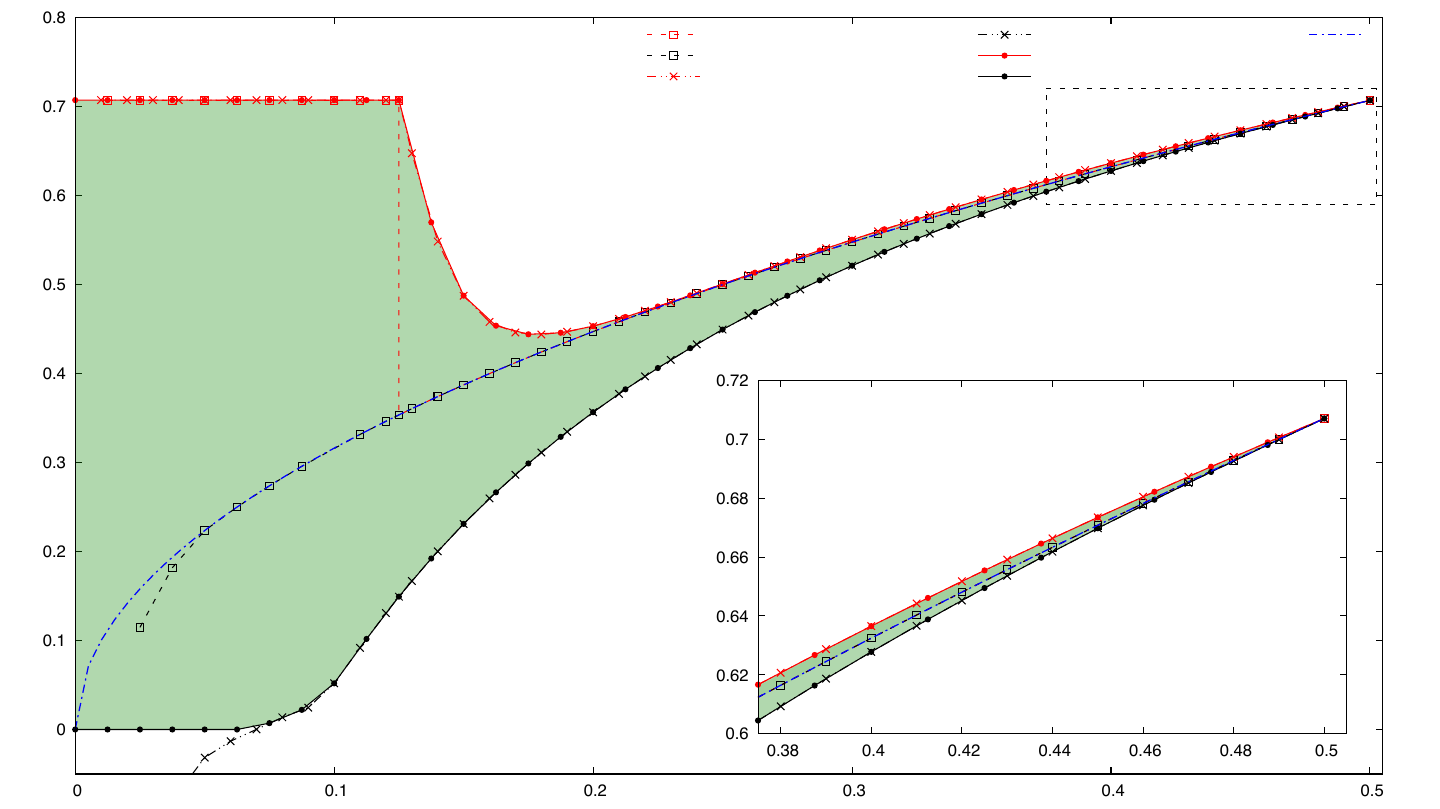}};
        
    
        \node[scale=0.525, anchor=east] at (-0.8,4) {$g^2$-max Virasoro Annulus}; 
        \node[scale=0.525, anchor=east] at (-0.8,3.775) {$g^2$-min Virasoro Annulus};
        \node[scale=0.525, anchor=east] at (-0.8,3.55) {$g^2$-max $SL(2)$ Annulus};
    
        \node[scale=0.525, anchor=east] at (2.75,4) {$g^2$-min $SL(2)$ Annulus}; 
        \node[scale=0.525, anchor=east] at (2.75,3.775) {$g^2$-max Mixed System};
    \node[scale=0.525, anchor=east] at (2.75,3.55) {$g^2$-min Mixed System};
    
        \node[scale=0.525, anchor=east] at (6.3,4) {Dirichlet Brane}; 
    
        \node[scale=0.7, anchor=east] at (-7.3,4.25) {$g^2$}; 
        \node[scale=0.7, anchor=east] at (7.7,-4.2) {$\Delta_{\phi}$}; 
        
    \end{tikzpicture}
    \caption{Bounds on $g^2$ as a function of $\Delta_\phi$ for $c=1$ theories with the choice of gaps described in \eqref{eq:dirichlet-gaps}: $\vec{p}=(\Delta_\phi| 4\Delta_\phi;1,1)$, $\mathcal{I}_a = \{\Delta_\phi\}$. We compare the bounds obtained from the annulus subproblem with Virasoro characters for $N=111$ and $N_{\text{deg}}=10$ degenerate characters (dotted curves), the annulus subproblem with $SL(2)$ characters for $N=35$, and the full system with $N=35$ (solid curves). The red curves denote the upper bound on $g^2$ and the black curves denote the lower bound. The blue dotted curve is the Dirichlet brane.}
    \label{fig:c=1-dirichlet}
\end{figure}

We begin by finding bounds on $g$ as a function of the radius. We plot the result in
\cref{fig:c=1-dirichlet}. Similar to the regions $\circled{2}$ and $\circled{3}$ in \cref{fig:c=1-neumann} for the Neumann branes, we observe no significant improvement of the mixed-correlator bound over the annulus results, again indicating that the correlators involving bulk operators decouple in the full system. Restricting our attention to the annulus subproblem, using Virasoro symmetry yields especially sharp bounds when $\Delta_\phi>1/8$. On the scale of the $sl(2)$ bounds, the Virasoro ones are indistinguishable from the line of Dirichlet branes. We notice that a similar plot already appeared in \cite{Collier:2021ngi}. The difference is that the assumptions on the bulk spectrum \eqref{eq:dirichlet-gaps} here are refined by setting a gap above $\Delta_\phi$ in the bulk spectrum. The sharp rise of the upper bounds at $\Delta_\phi=1/8$ is easy to explain. There is a solution to the annulus crossing equation with an isolated bulk state at $\Delta=1/2$.  This is the $D$-brane for a free boson with $\Delta_\phi=1/2,$ the rightmost tip of the plot in \cref{fig:c=1-dirichlet}. The same solution is allowed by our assumptions when $\Delta_\phi\leq 1/8,$ where this time the state at $\Delta=1/2$ is interpreted as part of the continuum above $\Delta_\text{gap}=1/2,$ while $a_{\Delta_\phi}=0$. Such spurious solution could be excluded by including more bulk states.

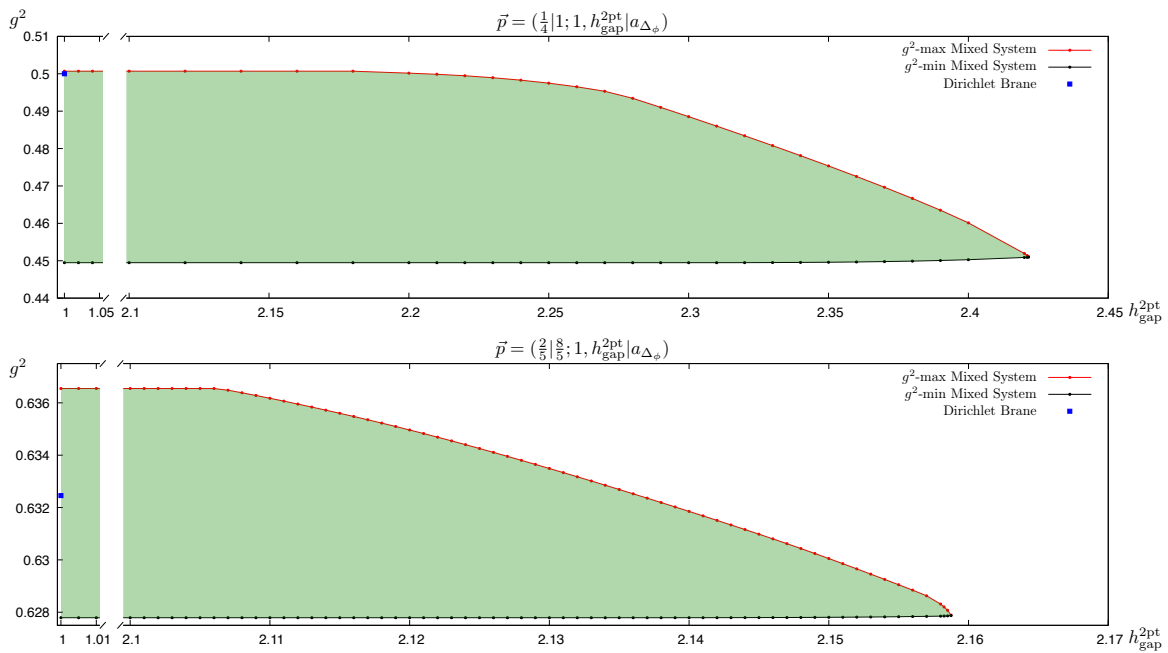
\begin{figure}[ht]
    \centering
    \begin{tikzpicture}
        \node at (0,0) {\includegraphics[width=\textwidth]{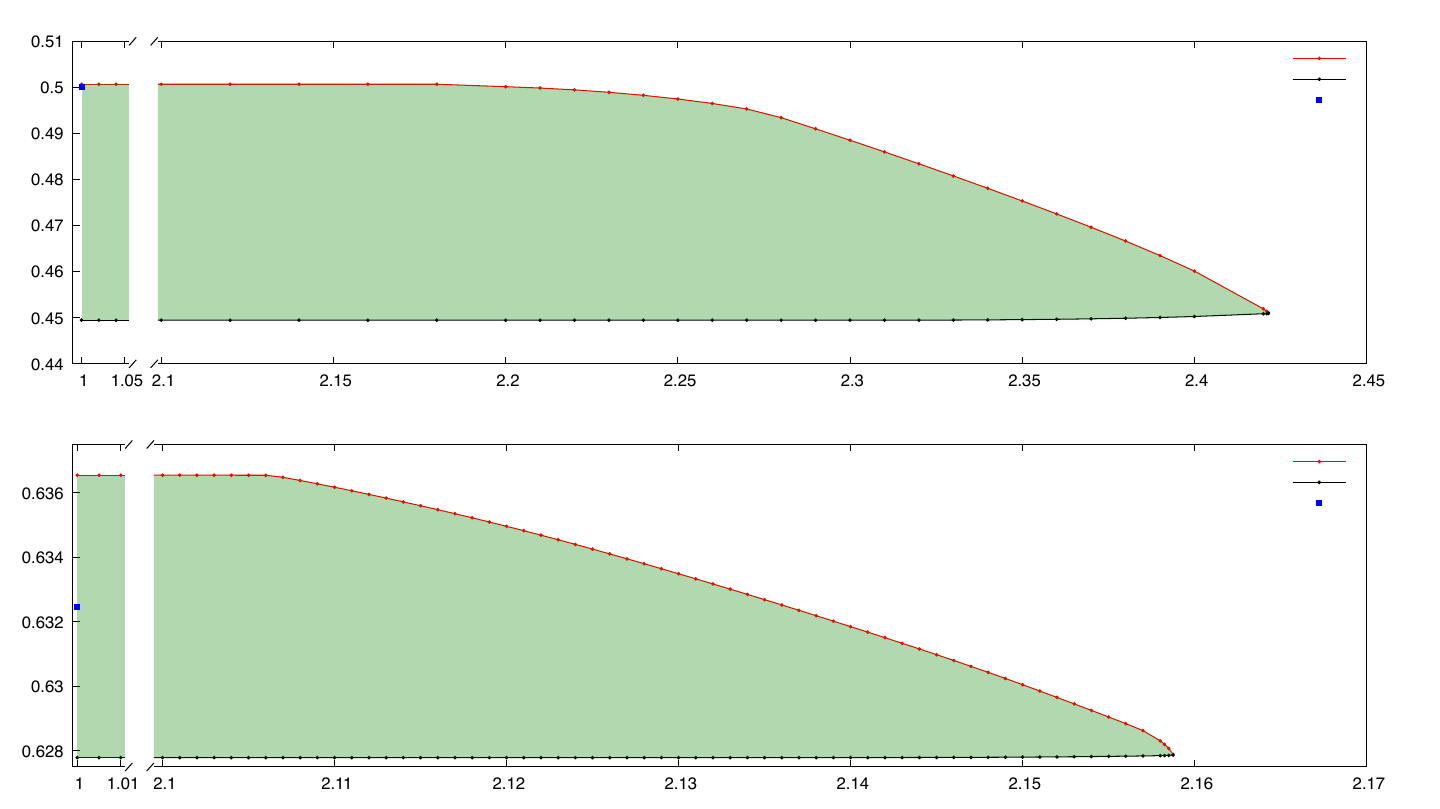}};
    
    
        \node[scale=0.7] at (0,4.1) {$\vec{p}=(\frac{1}{4}|1;1,\hgaptp|a_{\Delta_\phi})$};
        
        \node[scale=0.525, anchor=east] at (6.1,3.75) {$g^2$-max Mixed System}; 
        \node[scale=0.525, anchor=east] at (6.1,3.525) {$g^2$-min Mixed System};
        \node[scale=0.525, anchor=east] at (6.1,3.3) {Dirichlet Brane};
    
        \node[scale=0.7, anchor=east] at (-7.3,4.15) {$g^2$}; 
        \node[scale=0.7, anchor=east] at (7.8,0.25) {$\hgaptp$}; 
    
    
        \node[scale=0.7] at (0,-0.25) {$\vec{p}=(\frac{2}{5}|\frac{8}{5};1,\hgaptp|a_{\Delta_\phi})$};
        
        \node[scale=0.525, anchor=east] at (6.1,-0.61) {$g^2$-max Mixed System}; 
        \node[scale=0.525, anchor=east] at (6.1,-0.835) {$g^2$-min Mixed System};
        \node[scale=0.525, anchor=east] at (6.1,-1.06) {Dirichlet Brane};
    
        \node[scale=0.7, anchor=east] at (-7.3,-0.5) {$g^2$}; 
        \node[scale=0.7, anchor=east] at (7.8,-4.1) {$\hgaptp$}; 
    \end{tikzpicture}
    \caption{Bounds on the $g$-function as a function of $\hgaptp$ for $c=1$ and the choice of gaps $\vec{p}=(\frac{1}{4}|1;1,\hgaptp|a_{\Delta_\phi})$ (top) and $\vec{p}=(\frac{2}{5}|\frac{8}{5};1,\hgaptp|a_{\Delta_\phi})$ (bottom). The region of the plot for $\hgaptp$ between $1$ and $2$ has been omitted for clarity.}
    \label{fig:c=1_whale_dirichlet}
\end{figure}

It is unfortunate that the the full system, which is directly sensitive to $\Delta_\phi$ via the external operator, is unable to improve the bounds on $g$ over the annulus.
The lack of sensitivity of the boundary entropy bound to the bulk operators is again emphasized by varying $\hgaptp$. In \cref{fig:c=1_whale_dirichlet}, one can see that the crossing constraints from the bulk become effective only at much higher gaps than the ones relevant for the boundary conditions of the free-boson.

Plots like the ones in figure \cref{fig:c=1_whale_dirichlet} do show that the mixed-correlator bootstrap of points and lines sets non-perturbative constraints on the data appearing in the two-point function of bulk operators with a defect, specifically on the gap $\hgaptp.$ It is therefore natural to optimise other CFT data which appear in the same correlator, in order to exploit its crossing constraints. When the external operator $\phi$ gets a one-point function, as in the case of a $D$-brane, $a^2_{\Delta_\phi}$ appears in the boundary OPE of $\phi$ via $b_{\Delta_\phi,h=0} = a_{\Delta_\phi}$---see eq. \eqref{eq:bd-id}---and provides a natural objective.

We will make one more assumption, which holds for a $D$-brane but can be stated more generally: \emph{we will be looking for a boundary condition with a sparse spectrum, meaning that the only (low lying) boundary states which are singlets under the symmetry preserved by the defect are the ones dictated by the Ward identities.} This assumption is satisfied in particular by the identity Cardy branes in rational theories, and often by symmetry breaking boundary conditions in higher dimensions as well \cite{AJBray_1977}. In our case, this implies a selection rule for the boundary OPE of the momentum vertex operators $\mathcal{V}_{n,0}$, which are singlets under the $U(1)_w$ preserved by the $D$-brane: in the strongest version of the assumption, only the identity module of the $U(1)_w$ chiral algebra appears in its boundary OPE, with an OPE coefficient given by the one-point function. 
In practice, we will impose a weaker version, that is, we will assume that the \emph{tilt} ($h=1)$ and the \emph{displacement} ($h=2)$ operators are the only operators in the boundary OPE up to $h=2$. 

The displacement operator is part of the Virasoro identity module, and the corresponding boundary OPE coefficient is related to $a^2_{\Delta_\phi}$ as follows:
\begin{equation}
    b_{\Delta_\phi,h=2} = \frac{\Delta_\phi}{\sqrt{2c}} a_{\Delta_\phi}.
    \label{dispVsa}
\end{equation}

Regarding the $h=1$ state, our external operator $\phi$ in eq. \eqref{eq:phi-c-1} is one component of a $U(1)_m$ vector, and its OPE coefficient $b_{\Delta_\phi,h=1}$ is proportional to the \emph{variation} of the one point function $\braket{\phi}$ under the action of $U(1)_m$. Such variation can always be set to zero by the action of a $U(1)_m$ group element. Indeed, defects that break a bulk continuous global symmetry are always a part of a conformal manifold, whose tangent space is spanned by the tilt operator (see \cite{Recknagel:1998ih}, or \emph{e.g.} \cite{Drukker:2022pxk} for a recent discussion). CFT data on the conformal manifold are trivially related by the action of the symmetry. In particular, the defect conformal manifold for a boundary condition breaking $U(1)$ is a circle, whose points are labeled by a coordinate $\theta$ which, in the case of the boundary state \eqref{DCardy}, is the position of the $D$-brane in target space. Therefore, in all generality, $\braket{\phi}=|a_{\Delta_\phi}|\cos \theta$, and we can always choose $\theta=0.$ Then, $\delta_{U(1)}\braket{\phi}=0,$ and so $b_{\Delta_\phi,h=1}=0$.

\begin{figure}[!ht]
    \centering
    \begin{tikzpicture}
        \node at (0,0) {\includegraphics[width=\textwidth]{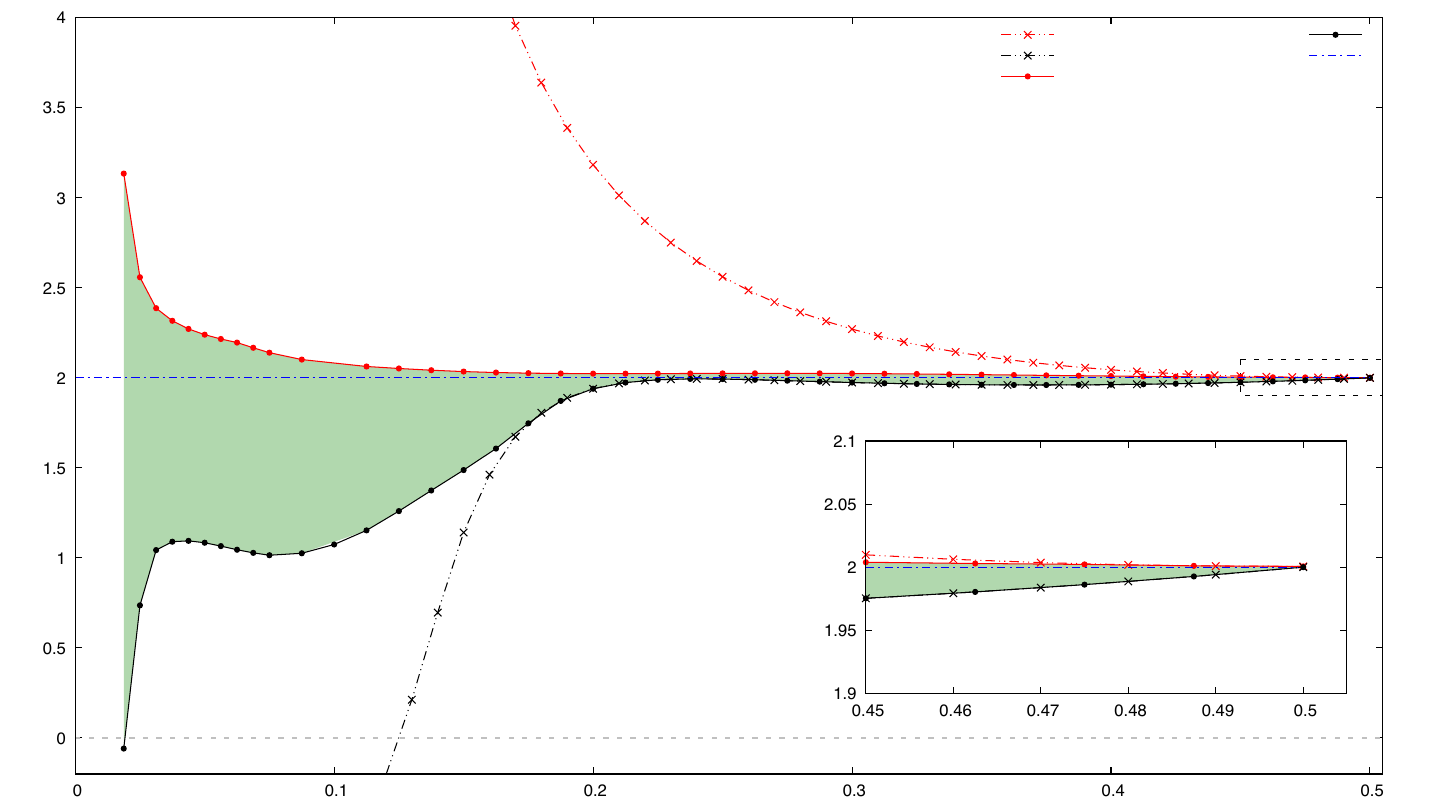}};
    
    
        \node[scale=0.525, anchor=east] at (3,3.985) {$a^2$-max $SL(2)$ Annulus}; 
        \node[scale=0.525, anchor=east] at (3,3.775) {$a^2$-min $SL(2)$ Annulus};
        \node[scale=0.525, anchor=east] at (3,3.565) {$a^2$-max Mixed System};
    
        \node[scale=0.525, anchor=east] at (6.3,3.985) {$a^2$-min Mixed System}; 
        \node[scale=0.525, anchor=east] at (6.3,3.775) {Dirichlet Brane};
    
        \node[scale=0.7, anchor=east] at (-7.175,4.25) {$a^2_{\Delta_\phi}$}; 
        \node[scale=0.7, anchor=east] at (7.75,-4.2) {$\Delta_\phi$}; 
    \end{tikzpicture}
    \caption{Bounds on $a^2_{\Delta_\phi}$ as a function of $\Delta_\phi$ for $c=1$ and $\vec{p}=(\Delta_\phi|4\Delta_\phi;1,2.001|a^2_{\Delta_\phi})$. The data points for the bounds from Virasoro annulus have been excluded, since they are consistently equal to $2$.}
    \label{fig:free-boson-asq}
\end{figure}
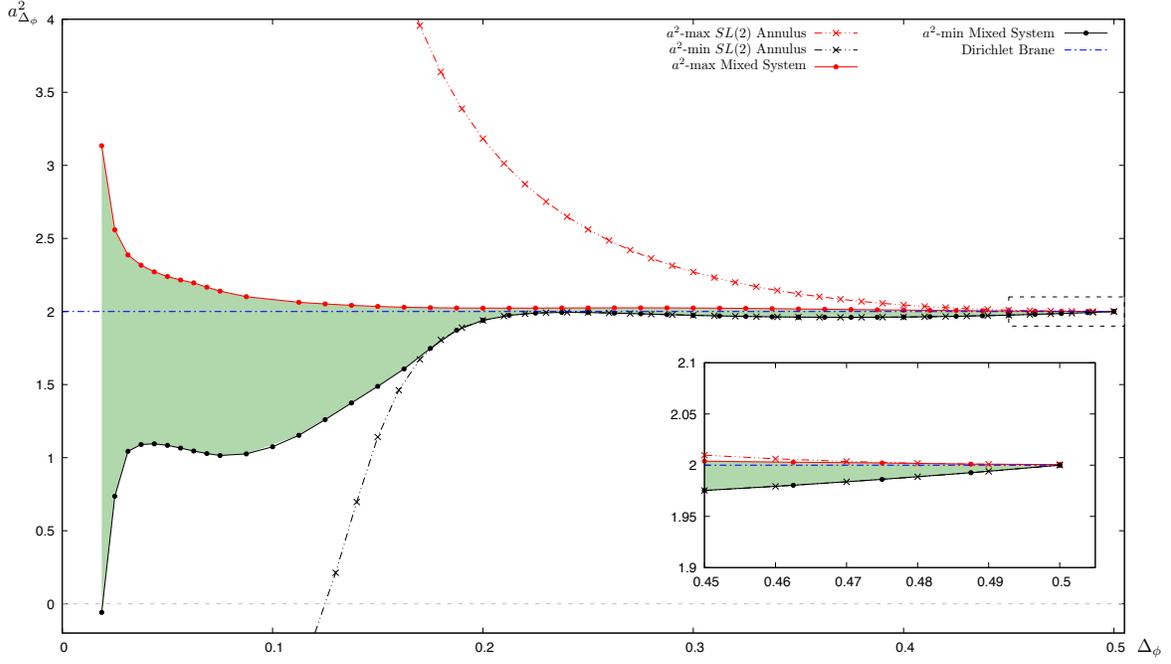

All in all, optimizing $a_{\Delta_\phi}$ corresponds to choosing the objective
\begin{equation}
    \vec{V}_{\Delta_\phi}^a + \vec{V}_{0}^{\text{2pt}} + \frac{\Delta_\phi^2}{2c} \vec{V}_{2}^{\text{2pt}}
    \label{dbranes_obj_a}
\end{equation}
 in the notation of eq. \eqref{blockVectors}, which involves components of the functional that act both on $\tau$ and on $\xi$. Notice that, in selecting the objective \eqref{dbranes_obj_a}, we also imposed the relation $b_0=a_{\Delta_\phi}$. As discussed at the end of \cref{ssec:twopcross}, this equation is valid because the operator $\text{Im} (\mathcal{V}_{(1,0)})$, degenerate with and orthogonal to $\phi,$ does not get a vev. The same constraint also eliminates the non-elementary branes from the primal bootstrap problem.   As for the spectrum, we implement the discussion above by including the identity and the displacement in the boundary channel of the two-point function, and leaving a small gap above the latter: 
\begin{equation}
    \vec{p}=(\Delta_\phi|4\Delta_\phi;1,2.001|a^2_{\Delta_\phi})~.
\end{equation}

The result is shown in \cref{fig:free-boson-asq}, where the advantage of using the mixed system is evident for all values of $\Delta_\phi.$ This concludes our exploration of the boundary conditions of a single free boson via the numerical bootstrap. In the next section, we will use the same strategy to put bounds on a rational CFT whose set of boundary conditions has not been classified.

\subsection{$c=3/2$ and the $\mathfrak{su}(2)_2$ WZW model}
 
We now begin the exploration of the $c=3/2$ moduli space. In this work, we mostly focus on the Wess-Zumino-Witten model \cite{Witten:1983ar}, although some results---\cref{fig:su(2)2_hann_delta} in particular---are valid for any boundary CFT with this value of the central charge.
We begin in \cref{subsec:wzw} with a brief overview of the $\sutwok$ WZW model and of the known conformal boundary conditions. In \cref{subsec:wzw_ann} we compute bounds deriving from the annulus crossing constraints, and in \cref{subsec:wzw_mixed} we move on to the mixed system.

\subsubsection{A brief review}
\label{subsec:wzw}

We are concerned with CFTs describing $SU(2)$ valued fields $g:S^2 \rightarrow SU(2)$ given by the action
\begin{equation}
    S = \frac{k}{4\pi} \int_{S^2} d^2z ~~ \text{tr}(g^{-1}\partial_\mu g g^{-1} \partial^\mu g) - \frac{ik}{12\pi} \int_B d^3y ~~ \epsilon_{\alpha\beta\gamma}\text{tr}\left( g^{-1}\partial^{\alpha}g g^{-1}\partial^{\beta}g g^{-1}\partial^{\gamma}g\right),
    \label{WZWaction}
\end{equation}
where, in the second term (called the Wess-Zumino term), we use the conventional overloading of $g$ to denote the extension of the map $g: S^2 \rightarrow SU(2)$ to $g:B \rightarrow SU(2)$ such that $\partial B = S^2$.
Quantising the fields results in the condition $k \in \mathbb{Z}$, called the level of the theory. The theory has a conserved current $J(z) = -k\partial g g^{-1}$, whose modes form an \emph{affine Kac-Moody algebra}
\begin{equation}\label{eq:su2-chiral-algebra}
    [J^a_m, J^b_n] = km\delta^{ab} \delta_{m+n, 0} + if^{ab}_c J^c_{m+n}, 
\end{equation}
which is a central extension of the $su(2)$ algebra, justifying the nomenclature of $\sutwok$ WZW model. 
There exist many detailed reviews of WZW models, \emph{e.g.} \cite{DiFrancesco:1997nk,LorenzNotes}. Here, and henceforth, $a$, $b$, $\dots$ denote the adjoint indices. One can construct a stress tensor using the Sugawara construction as 
\begin{equation}
    T(z) = \frac{1}{2(k+2)} \left( J^a J^a \right)(z),
    \label{TSugawara}
\end{equation}
whose OPE with itself takes the usual form for a CFT
\begin{equation}
    T(z)T(w) \sim \frac{c/2}{(z-w)^4} + \frac{2T(w)}{(z-w)^2} + \frac{\partial T(w)}{(z-w)}~,
\end{equation}
where $c$ is the central charge 
\begin{equation}
    c = \frac{3k}{k+2}.
\end{equation}
Operators in the model fall into the so-called integrable representations of $\sutwok$, which are labeled by the $SU(2)$ spin of the primary. The spins of the integrable representations take the values
$j=0,\frac{1}{2},1,\dots,\frac{k}{2}$. 
The chiral scaling dimensions of the corresponding primaries are
\begin{equation}
    h = \frac{j(j+1)}{k+2}.
    \label{wzwh}
\end{equation}
Each $\mathfrak{su}(2)_k$ module contains infinitely many Virasoro primaries.
The $\sutwok$ modules satisfy the following fusion rules:
\begin{equation}\label{eq:fusion-rule}
    j_1 \times j_2 = |j_1-j_2| \oplus \dots \oplus \min(j_1+j_2, k-j_1-j_2)~.
\end{equation}
This can be deduced from the the S-matrix
\begin{equation}
    S_{j}^{j'} = \sqrt{\frac{2}{k+2}} \sin\left( \frac{\pi (2j+1)(2j'+1) }{k+2} \right)~,
    \label{su2S}
\end{equation}
via the Verlinde formula \cite{Verlinde:1988sn}:
\begin{equation}\label{eq:degeneracy}
    n_{j j'}^{j''} = \sum_{i} \frac{S_{j}^{i}S_{j'}^{i}S_{j''}^{i}}{S_{0}^{i}},
\end{equation}
where we have used the fact that $S_{i}^{j}$ is real and symmetric.
The $\sutwok$ modules have a level one null descendant for all values of $j$.
The ensuing Knizhnik–Zamolodchikov differential equation \cite{Knizhnik:1984nr} allows in particular to compute the four-point function of local operators in the model. We review the solution of the equation in appendix \cref{ssec:KZ}, applied to a four-point function of primaries in the fundamental of $SU(2)$.

The $\mathfrak{su}(2)_1$ WZW model is isomorphic to the free boson at the self dual radius, and its conformal boundary conditions are classified, as summarized in \cref{ssec:free-boson}. 
In this paper, we work with $k=2$ as the minimal example where the boundary states are not all known, and leave the treatment of $k>2$ to future work. By specializing eq. \eqref{wzwh} to $k=2$, we obtain the chiral spectrum of $\mathfrak{su}(2)_2$ primaries:
\begin{equation}
    h_0=0, ~\quad h_{1/2}=\frac{3}{16}~, \quad
    h_{1}=\frac{1}{2}~,
    \label{hsu22}
\end{equation}
where the lower index denotes the spin label.
In the diagonal modular invariant, which is the only one at level $k=2$, all the integrable representations appear with multiplicity one, and all the $\mathfrak{su}(2)_2$ primaries are scalars, with dimensions
\begin{equation}
    \Delta_0=0, ~\quad \Delta_{1/2}=\frac{3}{8}~, \quad
    \Delta_{1}=1~.
    \label{Deltasu22}
\end{equation}

The moduli space of theories at $c=3/2$ is richer than just the $\mathfrak{su}(2)_2$ WZW model, and in particular it contains the product of a compact free boson and a free fermion. Orbifolds of the latter theory by discrete symmetries produce the full known moduli space of $\mathcal{N}=1$ superconformal field theories \cite{Dixon:1988ac}. Boundary conditions preserving the $\mathcal{N}=1$ supersymmetry of the bulk were constructed in \cite{Gaberdiel:2001zq,Cappelli:2002wq}. Furthermore, orbifolds of the model under discrete symmetries obviously belong to the same moduli space. Since our focus for this work is on the WZW model, we will not explore these boundary conditions thoroughly, but we will compare our bounds to the tensor product of a free boson and a free fermion in \cref{fig:su(2)2_hann_delta} below.

To the best of our knowledge, the branes of the $\mathfrak{su}(2)_2$ WZW model whose boundary state is known exactly and have a gap in the defect spectrum fall into two categories, called the $A$- and $B$-branes \cite{Maldacena:2001ky}. The former preserve one copy of the full $\sutwok$ chiral algebra, while the latter preserve the smaller $(\sutwok/\mathfrak{u}(1)_k)\times \mathfrak{u}(1)_k$ algebra. Beyond these two classes, partial knowledge of other boundary states was obtained in \cite{Kudrna:2021rzd,Kudrna:2024ekn} via numerical solution of the string field theory equations of motion, but only solutions corresponding to the $A$-branes where found for $k=2$. 

The $A$-branes are simply the Cardy boundary conditions for the $\sutwok$ chiral algebra, and are therefore labeled by the same $SU(2)$ index as the bulk primary operators. 
The Cardy states are
\begin{equation}\label{eq:su2-brane}
    \ket{j}_A = \sum_{j'} \frac{S_{j}^{j'}}{\sqrt{S_{0}^{j'}}} \ket{{j'}}\rangle,
\end{equation}
where $\ket{j}\rangle$ is the Ishibashi state for the $\sutwok$ bulk module with spin $j$ and the S-matrix was reported in \eqref{su2S}.
Explicitly, the Cardy states for $k=2$ read
\begin{subequations}
\begin{align}
\Ket{0}_A &= \frac{1}{\sqrt{2}} \ket{0}\rangle + \frac{1}{2^{\frac{1}{4}}} \ket{1/2}\rangle + \frac{1}{\sqrt{2}} \ket{1}\rangle~, \label{CardyZero} \\
\Ket{1/2}_A &= \ket{0}\rangle - \ket{1}\rangle~, \label{Cardyonehalf} \\
\Ket{1}_A &=  \frac{1}{\sqrt{2}} \ket{0}\rangle - \frac{1}{2^{\frac{1}{4}}} \ket{1/2}\rangle + \frac{1}{\sqrt{2}} \ket{1}\rangle~. \label{CardyOne}
\end{align}
\label{AbraneCardy}
\end{subequations}
The $\ket{0}_A$ and $\ket{1}_A$ boundary states are mapped to each other by the $g_{\alpha \dot{\alpha}} \to -g_{\alpha \dot{\alpha}}$ symmetry of the action \eqref{WZWaction}, while the $\Ket{1/2}_A$ boundary condition is invariant under it.\footnote{This symmetry can be identified with a chiral $SU(2)$ transformation, showing that the $\ket{0}_A$ and $\ket{1}_A$ belong to the same conformal manifold parametrized by the gluing condition \eqref{su2gluing}---see \emph{e.g.} \cite{Kudrna:2021rzd}.}

\bgroup
    \def\arraystretch{1.5}
    \begin{table}
        \centering
        \begin{tabular}{|c|c|c|c|}
            \hline
            $\alpha$ & $\Delta_{\text{gap}}^{\text{ann}}$ & $h_{\text{gap}}^{\text{ann}}$ & $g^2$ \\
            \hline
            $\ket{0}_A$ & $\frac{3}{8}$ & 1 & $\frac{1}{2}$\\
            \hline
            $\ket{1/2}_A$ & 1 & $\frac{1}{2}$ & 1\\
            \hline
            $\ket{1}_A$ & $\frac{3}{8}$ & 1 & $\frac{1}{2}$\\
            \hline
        \end{tabular}
        \caption{Gaps and boundary entropy of Cardy states for $k=2$.}
        \label{tab:k=2-gaps}
    \end{table}
\egroup

The Ishibashi states $\ket{j}\rangle$ are fixed once a gluing condition,
\begin{equation}
    J^a(z)-\Omega^a_b\,\bar{J}^b(z)=0~, \qquad z\in \mathbb{R},
\label{su2gluing}
\end{equation}
for the currents is chosen. Therefore each Cardy brane comes in a family parametrized by the elements $\Omega$ in the automorphism group of the algebra. 
Since $SU(2)$ only has inner automorphisms, this means that $A$-branes are parametrized by a group element. One can change the gluing condition by acting on the boundary state with the topological operator associated to the copy of $SU(2)$ broken by the boundary \cite{Recknagel:1998ih}:
\begin{eqnarray}
    U(\theta)=\exp \mathrm{i} \int \theta_a \tilde{J}^a~,
    \label{Utiltsu2}
\end{eqnarray}
where $\tilde{J}$ is a linear combination of $J$ and $\bar{J}$ different from \eqref{su2gluing}, or in other words, a tilt operator. Correlation functions of bulk operators change by conjugating each operator by $U(\theta)$. The spectrum of boundary operators is invariant, and so is the $g$ function---see eq. \eqref{gdef}---as expected for a marginal boundary deformation.\footnote{The spectrum of boundary condition changing operators does depend on the twist in the overlap of two branes with different gluing conditions. These states appear in the spectrum of non-elementary branes, as discussed in \cref{ssec:elementary}. We have not explored how much of the allowed region in \cref{fig:su2-annulus} below is filled by them.}

The boundary spectrum on each $A$-brane, including the degeneracy of each $\mathfrak{su}(2)_2$ primary, is determined by the fusion rules in the same way as for the boundary conditions in the Ising model discussed in \cref{subsec:ising}. 
The spectrum of $\mathfrak{su}(2)_2$ boundary primaries can therefore be deduced from eqs. \eqref{hsu22} and \eqref{AbraneCardy}, and the $su(2)$ representation of the primary corresponds to the spin label, as for the bulk operators.
The summary of relevant gaps for the bootstrap and the values of $g$ for the $A$-branes are reported in  \cref{tab:k=2-gaps}. 

Moving on to the $B$-branes, these are rational boundary conditions that break a part of the bulk chiral algebra \cite{Maldacena:2001ky}, see also \cite{Fuchs:1999zi,Quella:2002ct,Blakeley:2007gu}. The preserved algebra involves a factor of $\mathfrak{u}(1)_k$, which is the chiral algebra of a free boson at radius $R=\sqrt{2k}$, and a factor $\mathfrak{su}(2)_k/\mathfrak{u}(1)_k$, which is the parafermionic coset. The $\mathfrak{u}(1)_k$ chiral algebra is generated by the (appropriately normalized) current $J^3$ and by an extra chiral vertex opertor with spin $k$. At level $k=2$, the parafermionic coset is an ordinary free fermion.

Like for the boundary conditions of the free boson and of the $A$-branes, the global symmetries broken by the defect imply that the latter form conformal manifolds. In the case of the $B$-branes, one has 6 (anti)holomorphic operators in the bulk and in general---but see below for $k=2$---only the gluing condition for $J^3$, so the moduli space is five dimensional. Correlation functions of bulk operators at different points of the moduli space are related by acting on the operators with the appropriate charge. The target space interpretation of the moduli space is discussed in \cite{Maldacena:2001ky}. Of course, $g$ is constant on the moduli space, as is the spectrum of boundary operators supported on each boundary.\footnote{Notice that the gluing conditions for the stress tensor \eqref{TSugawara} require that the deforming charges leave the Killing form invariant, \emph{i.e.} they are unitary operators and cancel out in the annulus partition function. } 

The $B$-branes boundary states are the product of a factor living in the $\mathfrak{u}(1)_k$ sector times a factor living in the parafermionic Hilbert space. These branes are nicely reviewed in \cite{Kudrna:2021rzd}. Here, we focus on the $k=2$ case, where there are four elementary boundary states:
\begin{subequations}
\begin{align}
    \ket{B,\theta_0,+} &=\frac{1}{\sqrt{2}}\ket{N(\theta_0)}_{\mathfrak{u}(1)} \left(\ket{\mathbb{1}}_{Ising}+\ket{\epsilon}_{Ising}\right)~, \\
     \ket{B,\theta_0,-} &=\ket{N(\theta_0)}_{\mathfrak{u}(1)} \ket{\mathbb{\sigma}}_{Ising}~,
\end{align}
\label{bbranes}
\end{subequations}
where $\theta_0$ takes the two values
\begin{equation}
    \theta_0= 0,\ \pi~.
    \label{theta0}
\end{equation}
In eqs. \eqref{bbranes}, the first factor is the Neumann Cardy state from eq. \eqref{DNCardy}, and the value of the Wilson line $\theta_0$ is fixed by the requirement that the brane preserves the extended chiral algebra. The second factor is a linear combination of the Ising boundary states \eqref{IsingCardy}, and specifically, the two linear combinations correspond to the two gluing conditions $\psi=\pm \bar{\psi}$ for the free fermion that generates the chiral algebra of the $c=1/2$ theory. 

Since the the four boundary states \eqref{bbranes}  can be obtained from each other via the shift symmetry of the free boson and the $\mathbb{Z}_2$ chiral symmetry of the free fermion (Kramers-Wannier duality of the Ising model), the $B$-branes have identical annulus partition functions, which, in terms of the bulk channel modular parameter $q$, read:
\begin{equation}
    Z_B(q) = \left( \sum_{n\in\mathbb{Z}^+} \frac{q^{n^2}}{\eta(q)} + \sum_{\omega\in\mathbb{Z}} \frac{q^{\frac{\omega^2}{2}}}{\eta(q)} \right)\left( q^{-\frac{1}{48}} \prod_{n=0}^{\infty} \left( 1+q^{n + \frac{1}{2}} \right) \right).
    \label{ZBbranes}
\end{equation}
Now, a computation shows that 
\begin{equation}
    Z_B=Z_{A,1/2}~,
\end{equation}
where the latter is the partition function associated to the state \eqref{Cardyonehalf}. We conclude that, when $k=2$, the $B$-branes preserve in fact the full $\mathfrak{su}(2)_2$ algebra and belong to the moduli space of $\ket{1/2}_A$.\footnote{The $A$- and $B$-branes also coincide in the purely (para)fermionic theory \cite{Maldacena:2001ky}. In the full WZW model, the representatives \eqref{Cardyonehalf} and \eqref{bbranes} must be related by a change in gluing condition, which we did not work out.} Correspondingly, they have three-fold (rather than five-fold) degeneracy in the boundary channel at $h=1$. In the basis \eqref{bbranes} only one of the direction in the moduli space has a transparent meaning: the deformation generated by the $u(1)$ singlet corresponds to continuously changing $\theta_0$ away from the values \eqref{theta0}. 
In the following, we will sometimes consider non-elementary branes obtained as sum of the $B$-brane states $\eqref{bbranes}$, as examples of boundary states arising from summing over different gluing conditions.  

The space of boundaries that only preserve the Virasoro subalgebra of $\sutwok$ is in principle larger than the $A$- and $B$-branes discussed above. Inside this space, one might find non-rational boundary conditions preserving a smaller sub-algebra of $\sutwok$ than the one preserved by the $B$-brane, for instance a discrete symmetry or the sub-algebra generated by the modes of $J^3(z)$ alone. In the following, we initiate the exploration of this landscape via the numerical bootstrap. 

\subsubsection{The annulus constraint}
\label{subsec:wzw_ann}

\begin{figure}[!ht]
    \centering
    \begin{tikzpicture}
        \node at (0,0) {\includegraphics[width=\textwidth]{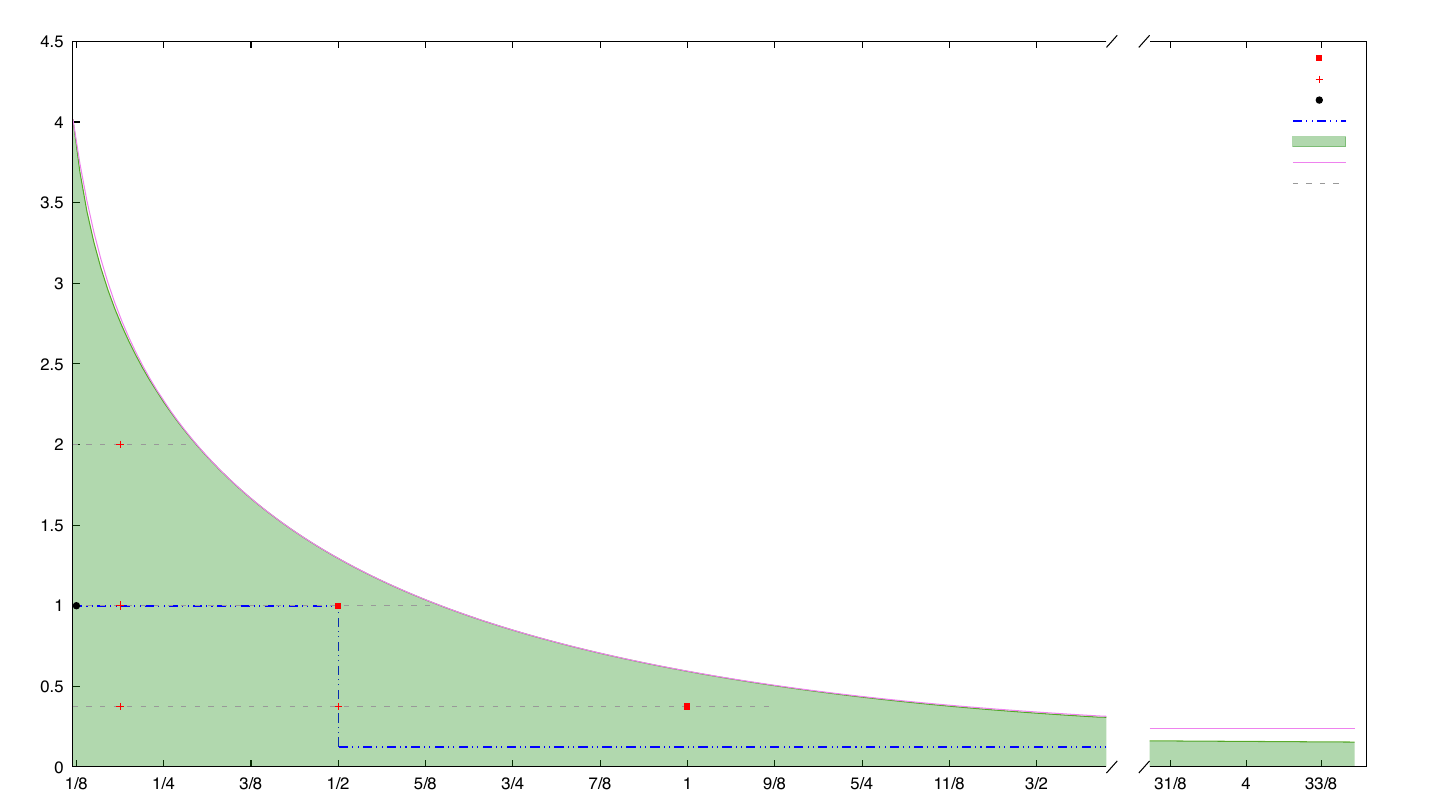}};
    
    
        \node[scale=0.525, anchor=east] at (6.15,3.75) {Elementary A branes};
\node[scale=0.525, anchor=east] at (6.15,3.525) {Linear Combinations of A branes};
\node[scale=0.525, anchor=east] at (6.15,3.300) {Linear Combinations of B branes};
\node[scale=0.525, anchor=east] at (6.15,3.075) {Free Boson and Ising Product States};
\node[scale=0.525, anchor=east] at (6.15,2.850) {Virasoro Annulus Bounds};
\node[scale=0.525, anchor=east] at (6.15,2.625) {$sl(2)$ Annulus Bounds};
\node[scale=0.525, anchor=east] at (6.15,2.400) {Bounds in \cref{tab:su2-ann-2}};
    
        \node[scale=0.7, anchor=east] at (-7.1,4.25) {$\Delta_{\text{gap}}$}; 
        \node[scale=0.7, anchor=east] at (7.6,-4.2) {$h_{\text{gap}}^{\text{ann}}$}; 
    \end{tikzpicture}
    \caption{Bounds on $\Delta_{\text{gap}}$ as a function of $h_{\text{gap}}^{\text{ann}}$ for $c=\frac{3}{2}$ with Virasoro and $sl(2)$ annulus bootstrap at 75 derivatives. Similar to \cref{fig:c=1_virasoro_ann_sub}, the $sl(2)$ bounds for $h_\text{gap}>2$ have been set to the value at $h_\text{gap} = 2$.}
    \label{fig:su(2)2_hann_delta}
\end{figure}
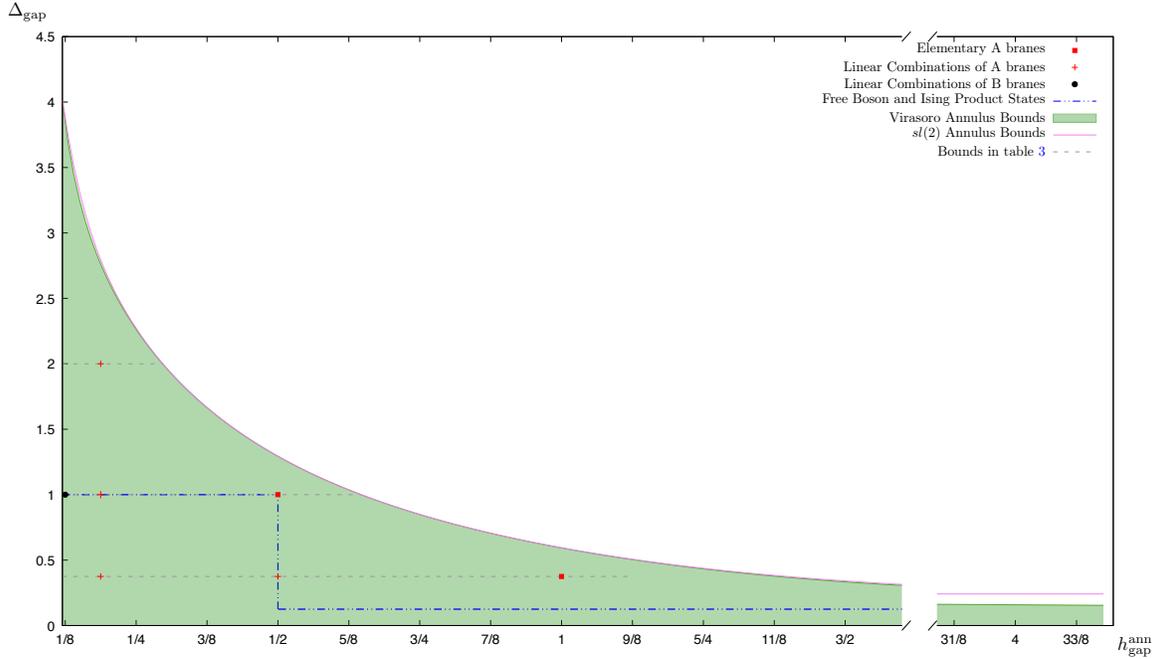

Analogous to the explorations in \cref{ssec:free-boson}, we start with the annulus partition function bootstrap before exploring the space of boundaries allowed by the full system. The first plot we present, in \cref{fig:su(2)2_hann_delta}, maps the allowed space in the $(h_\text{gap},\Delta_\text{gap})$ plane. The only assumptions going into this plot are Virasoro/$sl(2)$ symmetry and the value of the central charge. Therefore, this plot is relevant for the full moduli space of theories with $c=3/2$. The bound has the same rough behavior of the $c=1$ case in \cref{fig:c=1_virasoro_ann_sub}, namely that $h_\text{gap}$ and $\Delta_\text{gap}$ are inversely correlated. While this is generically expected, the curve quantifies the dependence of one quantity on the other. In particular, we see that a boundary condition can be stable only if the first bulk primary with a one-point function has dimension
\begin{equation}
    \Delta_\text{gap}\leq 0.593200 \dots~, \qquad \Leftarrow \quad \textup{stable boundary.}
    \label{stable3over2}
\end{equation}
\Cref{fig:su(2)2_hann_delta} only includes data with $\hgapa\geq 0.12$. Below this point, numerical instabilities seem to arise in the Virasoro bootstrap bound, a feature that we have observed in cases with small gaps. By extrapolating the available points, one finds hints that $\Delta_\text{gap}$ diverges as $\hgapa \to 0$.

Let us discuss how some known theories populate the allowed region.
As already remarked, if the theory in the bulk has global symmetries, the boundary gap can never be larger than 1. Correspondingly, the $A$-branes lie in the $h_\text{gap}\leq 1$ half-space, with the stable branes, \emph{i.e.} $\ket{0}_A$ and $\ket{1}_A$, obeying the inequality \eqref{stable3over2}. In addition to the elementary $A$-branes, an infinite set of valid boundary conditions can be obtained from linear combinations, with integer coefficients to keep the boundary character expansion well-behaved. As we discussed in \cref{ssec:elementary}, they can only have boundary gaps which are lower than the gaps of the branes being combined, while in principle $\Delta_\text{gap}$ can increase. Two subclasses of these linear combinations, given by linear combinations of $A$-branes and linear combinations of $B$-branes,  can be found in \cref{fig:su(2)2_hann_delta}. One can recognise points with $\hgapa=1/8$ as contributions from the boundary changing operators in the free boson (the ground state between boundary conditions with $\theta=0$ and $\theta=\pi$). Similarly, $\hgapa=3/16$ arises from the ground state between the $\ket{0}_A$ and $\ket{1/2}_A$ boundary states. One can reduce $\hgapa$ indefinitely by moving away from the $\theta=0,\,\pi$ points of the moduli space (see eq. \eqref{ZN}) or more generally by summing $A$-branes with different gluing conditions.  
One can also construct a boundary state proportional to the identity Ishibashi state: $1/\sqrt{2}(\ket{0}_A+\ket{1}_A)+\ket{1/2}_A$. Albeit not crucial for the bootstrap, notice that this boundary state has integer degeneracies, since it can be understood as a brane in the $\mathbb{Z}_2$ orbifold under the $g_{\alpha\dot{\alpha}} \to -g_{\alpha\dot{\alpha}}$ symmetry. The bulk gap of this boundary state is $\Delta_\text{gap}=2$, given by the product $J^a \bar{J}^b$, and it is the largest possible among linear combinations of $A$-branes. 

A chunk of the allowed region in \cref{fig:su(2)2_hann_delta} can be covered by the product of boundary states of a free boson and a free fermion, of which the $B$-branes \eqref{bbranes} are but one example compatible with the bulk $\mathfrak{su}(2)_2$ bulk spectrum. In order to understand the envelop of the resulting points, we look for the largest $\Delta_\text{gap}$ as a function of $h_\text{gap}$. Firstly, notice that operators should be classified according to the Virasoro symmetry generated by the sum of the stress tensors, $T_\text{Ising}+T_\text{f.b.}$. Bilinear combinations of Virasoro descendants can be primaries under the diagonal Virasoro symmetry. In the bulk channel, though, if a descendant gets a one-point function, so does the primary, and of course one-point functions of bilinears of the form $O_\text{Ising}O_\text{f.b.}$ factorize. Since we are interested in the gap above the identity, we only need to consider this phenomenon in the identity modulus. There, the lowest lying scalar primary under the diagonal Virasoro algebra has dimension $\Delta=4$.\footnote{This is the product $W\bar{W}$, where $W$ is the linear combination of $T_\text{Ising}$ and $T_\text{f.b.}$ which is a primary of the algebra generated by $T_\text{Ising}+T_\text{f.b.}$, see for instance appendix A of \cite{Quella:2006de}.} An analogous reasoning leads to a Virasoro primary of dimension $h=2$ in the boundary channel, which is always present.

Boundary gaps in the Ising model belong to the allowed Virasoro representations: $h=1/16,\ 1/2,\ 2$, where $2$ is the first state above identity in the identity module. The boundary gap in the free boson can instead vary continuously, see \cref{fig:c=1_virasoro_ann_sub}. Therefore, we split \cref{fig:su(2)2_hann_delta} in 4 subregions. Due to the reasoning above, no product of boundary states can have $h_\text{gap} > 2$. The second strip lies between $1/2\leq h_\text{gap}\leq 2$, and can be populated by the product of $\ket{\mathbb{1}}_{\text{Ising}}$ (or $\ket{\epsilon}_{\text{Ising}}$) times an appropriate free boson boundary state. The bulk gap cannot exceed $\Delta_\text{gap}=1/8$, since $\braket{\sigma}\neq0$. In the strip $1/16\leq h_\text{gap}\leq 1/2$, any of the boundary states in \eqref{IsingCardy} become available. The one with larger bulk gap is $\ket{\sigma}_{\text{Ising}}$, so, comparing with \cref{fig:c=1_virasoro_ann_sub}, we conclude that $\Delta_\text{gap}=1$ in this region. Finally, linear combinations of the three boundary states involving $\ket{\sigma}_{\text{Ising}}$ have $h_\text{gap}\leq 1/16$. One such linear combination is, like the analogous non-elementary $A$-brane discussed above, proportional to the identity Ishibashi state: $1/\sqrt{2}(\ket{\mathbb{1}}+\ket{\epsilon})+\ket{\sigma}$, where the states are understood to be Cardy states in the Ising model. Multiplying this state with a boundary state of the free boson, we find a family of boundary conditions with $\Delta_\text{gap}=2$ and $h_\text{gap}\leq 1/16$. This concludes the analysis of this class of factorized boundaries, and their envelope in the $(h_\text{gap},\Delta_\text{gap})$ plane is shown in \cref{fig:su(2)2_hann_delta}.

It would be interesting to fully explore the space of (non-elementary) analytically computable boundary states in the moduli space of $c=3/2$ CFTs, and see whether any of them are extremal. Conversely, based on the set of boundary states enumerated above, we could not exclude that new boundary conditions are hidden inside the allowed region of \cref{fig:su(2)2_hann_delta}. 

\bgroup
    \def\arraystretch{1.5}
    \begin{table}
        \centering
        \begin{tabular}{|c|c|c|}
            \hline
            $\Delta_\text{gap}$ & $h_{\text{gap}}^{\text{ann}}$ Virasoro & $h_{\text{gap}}^{\text{ann}}$ SL2 \\
            \hline
            $\frac{3}{8}$ & 1.12515(9) & 1.13416(5)\\
            \hline
            $1$ & 0.64606(9) & 0.64612(9)\\
            \hline
            $2$ & 0.29642(6) & 0.29708(4)\\
            \hline
            $\frac{19}{8}$ & 0.23313(2) & 0.23444(3) \\
            \hline
            $3$ & 0.16642(5) & 0.16987(2) \\
            \hline
            $4$ & 0.12023(9) & 0.12038(1)\\
            \hline
            
        \end{tabular}
        \caption{Bounds on $\hgapa$ from the annulus bootstrap, with $\mathfrak{su}(2)_2$ bulk states added up to $\Delta = 20$. $N=75$ derivatives.}
        \label{tab:su2-ann-2}
    \end{table}
\egroup
\smallbreak
One way to focus on the $\mathfrak{su}(2)_2$ WZW model within the moduli space is to input the bulk spectrum into the annulus bootstrap. 
When we do this, we find the stronger bounds shown in \cref{tab:su2-ann-2} and in \cref{fig:su(2)2_hann_delta}.
Now, the allowed values of $\Delta_\text{gap}$ jump discretely according to the Virasoro primaries in the model. We stopped at $\Delta_\text{gap}=4$, although of course more stringent bounds can be found by pushing the gap to higher values. 
None of the $A$-branes or their linear combinations saturate these bounds, leaving room for more exotic boundary conditions.
In particular, since $\Delta=3/8$ is the lightest operator above the identity, the first line of \cref{tab:su2-ann-2} contains the largest possible boundary gap in the $\mathfrak{su}(2)_2$ WZW model. As discussed above, as long as the global symmetries in the bulk are intact, the gap cannot exceed 1. Furthermore, it is unlikely that the boundary states of the orbifolds of the model can saturate the gap in \cref{tab:su2-ann-2}: the regular branes are linear combinations of branes of the original WZW model, and so their gap cannot exceed $\hgapa=1$; the boundary states of the fractional branes have overlaps with the twisted sector, and so they violate the spectrum assumptions of \cref{tab:su2-ann-2}. It would be interesting to check this explicitly. It is possible that no boundary condition with integer degeneracies exists between $\hgapa=1$ and $\hgapa=1.125\dots$, or on the contrary, that a new boundary condition lies there.

One can also focus on the stable branes in the $\mathfrak{su}(2)_2$ WZW model by imposing an isolated operator at $h=1$, and a gap above it. In this case, the largest $\hgapa=2.02$. If we continue imposing isolated operators at integer values of $h$, the gap above them remains close to the next integer: the bootstrap is aware that the sparsest boundary condition should include the boundary limit of the bulk global currents. Interestingly, the bounds on $g$ get tighter, and by imposing that the boundary spectrum is only supported at integer values of $h$ one eventually isolates the unique value $g^2=1/2$, giving evidence that the $\ket{0}_A$ boundary state, together with its moduli space, is the unique boundary condition with the sparsest possible boundary spectrum in the $\mathfrak{su}(2)_2$ WZW model.\footnote{In particular, with a spectrum composed of $h=0,\,1,\,2,\dots,10$ and a gap at $h_\text{gap}=10.99$, one gets $0.499999999996 \leq g^2\leq 0.50000000000026$.} It might be possible to prove this statement analytically, using the known crossing kernel for the Virasoro characters.

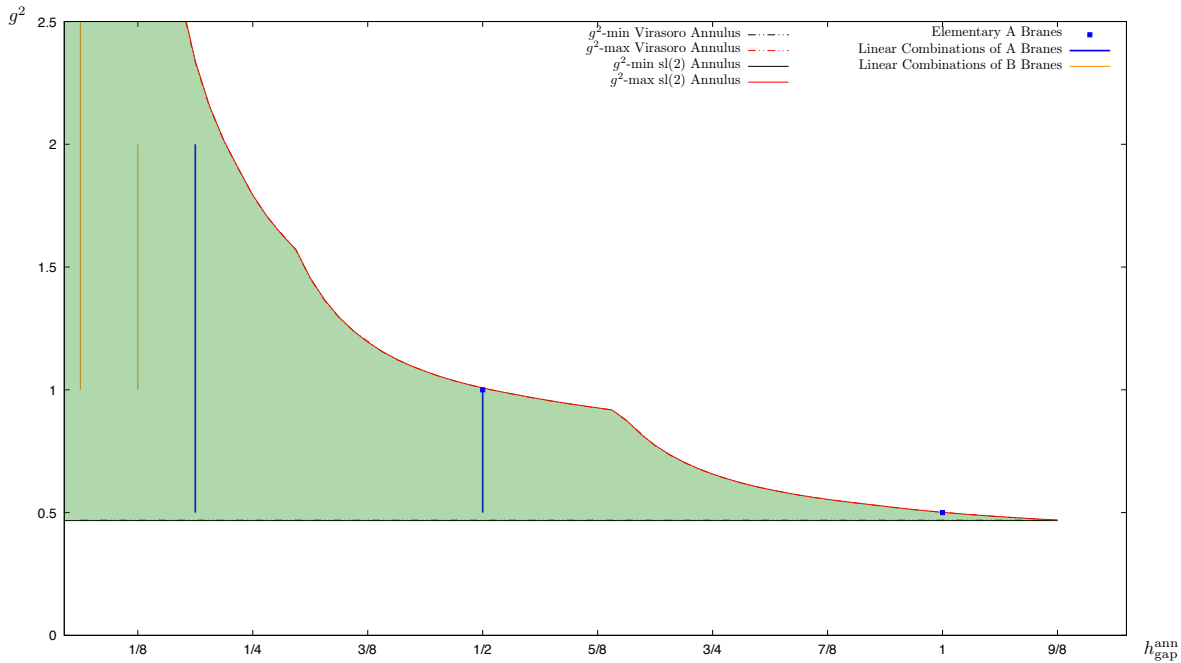
\begin{figure}[!ht]
    \centering
    \begin{tikzpicture}
        \node at (0,0) {\includegraphics[width=\textwidth]{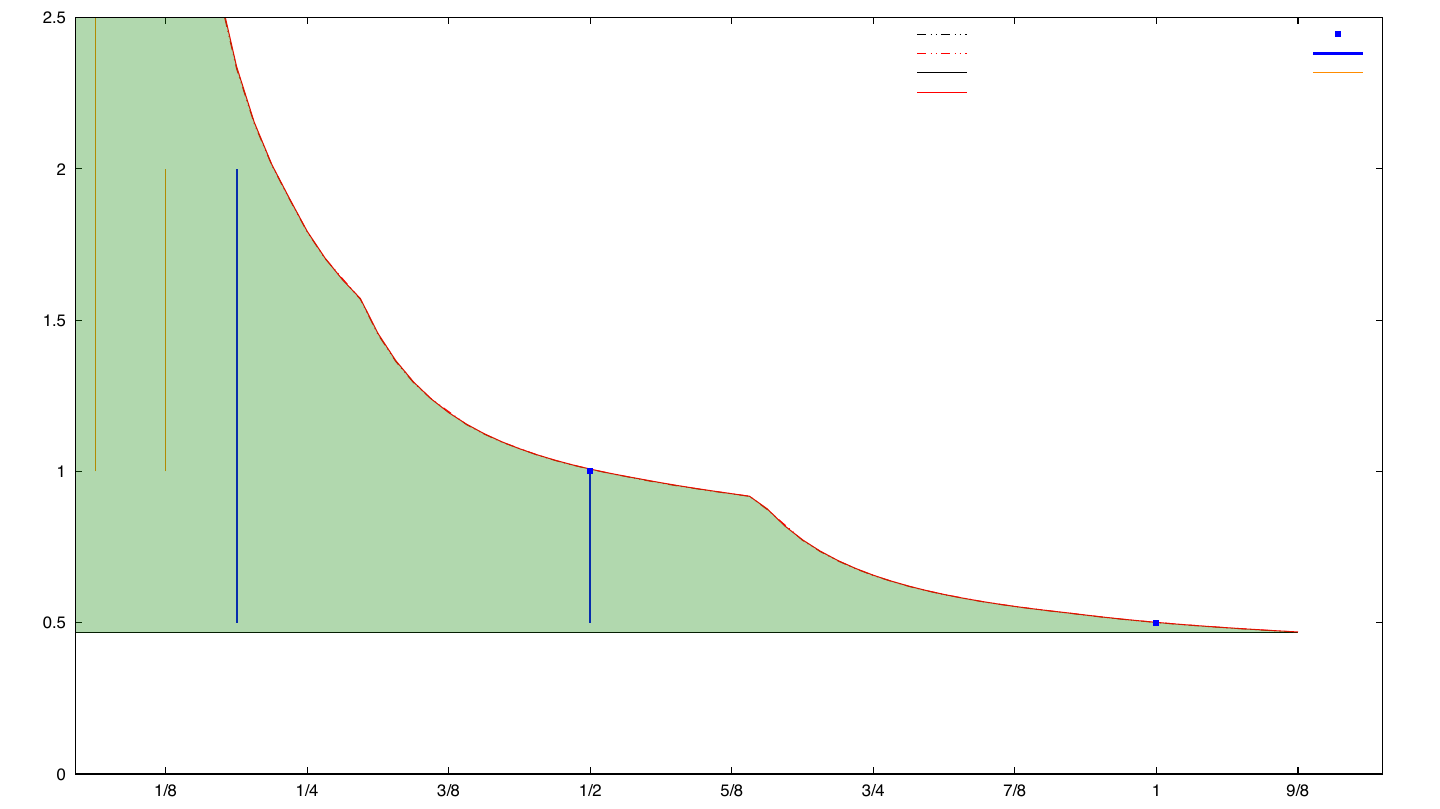}};
    
    
        \node[scale=0.525, anchor=east] at (2.1,4.03) {$g^2$-min Virasoro Annulus}; 
        \node[scale=0.525, anchor=east] at (2.1,3.82) {$g^2$-max Virasoro Annulus}; 
        \node[scale=0.525, anchor=east] at (2.1,3.61) {$g^2$-min sl(2) Annulus}; 
        \node[scale=0.525, anchor=east] at (2.1,3.4) {$g^2$-max sl(2) Annulus}; 
        
        \node[scale=0.525, anchor=east] at (6.375,4.03) {Elementary A Branes}; 
        \node[scale=0.525, anchor=east] at (6.375,3.82) {Linear Combinations of A Branes}; 
        \node[scale=0.525, anchor=east] at (6.375,3.61) {Linear Combinations of B Branes}; 
    
        \node[scale=0.7, anchor=east] at (-7.375,4.25) {$g^2$}; 
        \node[scale=0.7, anchor=east] at (7.95,-4.2) {$\hgapa$}; 
    \end{tikzpicture}
    \caption{Bounds on $g^2$ as a function of $h_{\text{gap}}$ for $c=3/2$ at $N=75$ derivatives. We input the first 20 bulk states belonging to the $\mathfrak{su}(2)_2$ WZW model. The linear combinations of $A$- and $B$-branes produce the following segments: $(\hgapa,g^2_\text{min},g^2_\text{max} )=\{(1/16,1,4),(1/8, 1, 2), (3/16, 1/2, 2), (1/2, 1/2, 1)\}$. }
    \label{fig:su2-annulus}
\end{figure}
\Cref{fig:su2-annulus}, the final plot of this subsection, shows the dependence of the ground state degeneracy on the boundary gap. In this case as well, we input the spectrum of bulk Virasoro primaries of the $\mathfrak{su}(2)_2$ WZW model. The first feature to be noticed is the independence of the lower bound of $\hgapa$, which confirms what we already saw in \cref{fig:c=1-neumann-h2pt,fig:c=1-neumann}.  

The second interesting feature of this plot is provided by the two kinks along the upper bound. They have an easy technical origin, similar to the phenomenon observed when discussing \cref{fig:c=1_virasoro_ann_sub}. Comparing the position of the kinks with \cref{tab:su2-ann-2}, we see that they appear for values of $h_\text{gap}$ where a certain spectrum of bulk operators becomes extremal. As we increase $h_\text{gap}$, at the values in the second column of \cref{tab:su2-ann-2} a new bulk operator must get a one-point function, and correspondingly the upper bound of \cref{fig:su2-annulus} develops a kink. It would be nice to understand these kinks along the lines of \cite{El-Showk:2016mxr}. Finally, $g_\text{min}$ and $g_\text{max}$ coincide at the largest allowed $h_\text{gap}$ from \cref{tab:su2-ann-2}, as expected. In the $c=1$ case, the kinks in \cref{fig:c=1_virasoro_ann_sub} roughly correspond to existing branes: it would be nice to inquire about the kinks in \cref{fig:su2-annulus} as well.

As for the known solutions to crossing, the $A$-branes are seen to be extremal. We also included in \cref{fig:su2-annulus} a few linear combinations of $A$- and $B$-branes, since they help understand some features of the bounds. These branes form vertical segments in the $(\hgapa,g^2)$ plane, whose extensions are reported in the caption and match the rules explained in \cref{ssec:elementary}.

\subsubsection{The mixed system}
\label{subsec:wzw_mixed}

We now move on to the bootstrap of mixed correlators. As usual, we use $sl(2)$ blocks and characters throughout this subsection. As external operator, we choose the lowest-lying primary of the model, which transforms into the $(j,\bar{j})=(1/2,1/2)$ representation of $\mathfrak{su}(2)_2\times\mathfrak{su}(2)_2$. This multiplet has dimension
\begin{equation}
    \Delta_\phi=\frac{3}{8}~,
\end{equation}
and can be identified with the fundamental field in the action \eqref{WZWaction}, $g_{\alpha \dot{\alpha}}$. The correspondence between matrix elements $g_{\alpha \dot{\alpha}}$ and eigenfunctions of dilatations and the Cartan of the global symmetry are as follows:
\begin{equation}
    g_{\alpha \dot{\alpha}}= 2^{\frac{1}{4}}
    \begin{pmatrix}
        \phi_{\frac{1}{2},\frac{1}{2}} & \phi_{\frac{1}{2},-\frac{1}{2}} \\
        \phi_{-\frac{1}{2},\frac{1}{2}} & \phi_{-\frac{1}{2},-\frac{1}{2}} \\
    \end{pmatrix}
    \label{gtophi}
\end{equation}
where $\phi_{ij}$ is the component of the $\mathfrak{su}(2)$ primary with eigenvalues $(i,j)$ under $(J^3,\bar{J}^3)$.\footnote{We normalize $\phi_{ij}$ so that in the OPE limit $\phi_{\frac{1}{2},-\frac{1}{2}}(z,\bar{z})\phi_{-\frac{1}{2},\frac{1}{2}}(w,\bar{w})\sim 1/|z-w|^{3/4}$ and $\phi_{\frac{1}{2},\frac{1}{2}}(z,\bar{z})\phi_{-\frac{1}{2},-\frac{1}{2}}(w,\bar{w})\sim 1/|z-w|^{3/4}$.  \label{footNormalization}} 
We will first consider boundary conditions where $\phi$ is not given a one-point function. This condition is protected by the $g_{\alpha \dot{\alpha}} \to -g_{\alpha \dot{\alpha}}$ symmetry of the action \eqref{WZWaction}, and is compatible with the $\ket{\ket{1/2}}_A$ $A$-brane.

\begin{figure}[!b]
    \centering
    \begin{tikzpicture}
        \node at (0,0) {\includegraphics[width=\textwidth]{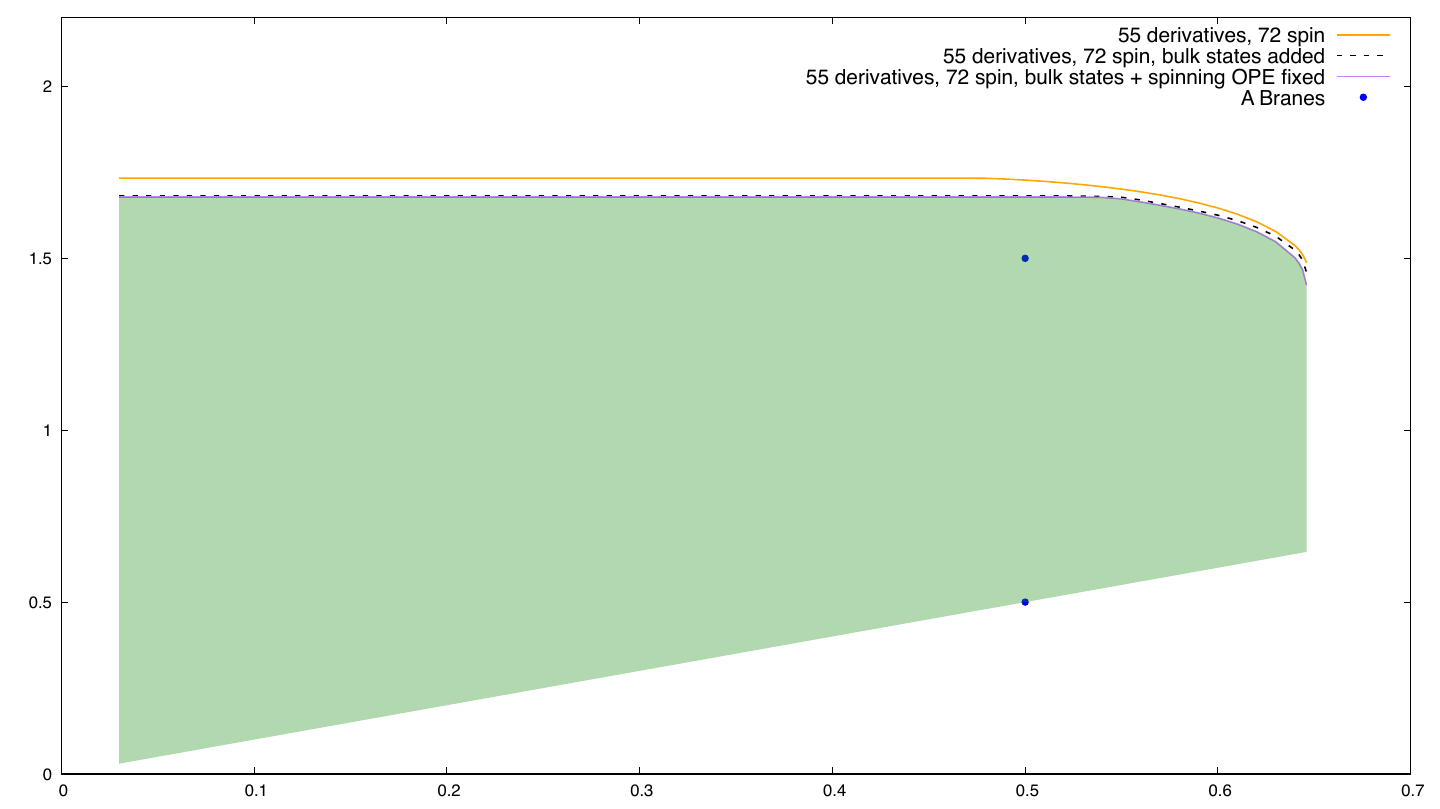}};
    
    
    
    
        \node[scale=0.7, anchor=east] at (-7.175,4.25) {$h_{\text{gap}}^{\text{2pt}}$}; 
        \node[scale=0.7, anchor=east] at (8.25,-4.2) {$h_{\text{gap}}^{\text{ann}}$}; 
    \end{tikzpicture}
    \caption{Bounds on the boundary gap in the two-point correlator as a function of the boundary gap in the partition function, for the $\mathfrak{su}(2)_2$ WZW model. The looser bound is obtained with $\vec{p} = (\frac{3}{8}|1, h_{\text{gap}}^{\text{ann}},h_{\text{gap}}^{\text{2pt}}|0)$. More stringent spectral assumptions lead to the other bounds, and are detailed in the legend and the text. In all runs, the functional includes $N=55$ derivatives and operators with spin up to 72 are included in the spectrum. }
    \label{fig:su(2)2_hann_h2pt}
\end{figure}

This information, together with the fusion rules \eqref{eq:fusion-rule}, is sufficient to obtain the first mixed-correlator bound, which we present in \cref{fig:su(2)2_hann_h2pt}. In the $(\hgapa,\hgaptp)$ plane, the green region is allowed for a boundary CFT with $c=3/2$ and a spectrum compatible with the following vector of gaps---see eq. \eqref{pdef}
\begin{equation}
    \vec{p} = \left(\frac{3}{8}\Bigg|1, h_{\text{gap}}^{\text{ann}},h_{\text{gap}}^{\text{2pt}}\Bigg|0\right)~.
    \label{pvector_vanilla_su2}
\end{equation}
As already emphasized when discussing \cref{fig:ising_hann_h2pt}, the presence of an upper bound on $\hgaptp$ is a consequence of the crossing constraint provided by the two-point function. On the other hand, the vertical boundary on the right of the plot is determined by the annulus crossing, and coincides with the second row of \cref{tab:su2-ann-2}. The lower bound at 45 degrees is imposed by hand and simply reflects the fact that all boundary states should appear in the annulus partition function.

In attempting to zero in on the $\mathfrak{su}(2)_2$ WZW model, we can do better in at least two ways. Firstly, we can input bulk quasi-primaries of the model beyond $\Delta=3/8,\,1$. The resulting bound is the dotted line in \cref{fig:su(2)2_hann_h2pt}. The $\vec{p}$ vector is modified with respect to \eqref{pvector_vanilla_su2} as follows:
\begin{equation}
    \vec{p} = \left(\frac{3}{8}\Bigg|20+\eps, h_{\text{gap}}^{\text{ann}},h_{\text{gap}}^{\text{2pt}}\Bigg|0\right)~,
    \quad \mathcal{I}_{a} = \left\{\frac{3}{8}+2,\frac{3}{8}+3,\dots, \frac{3}{8}+19\right\},
    \quad \mathcal{I}_{2\times 2} =  \left\{1,2,3,\dots, 20\right\}~,
    \label{pvector_refined_su2}
\end{equation}
where $\eps$ is a small positive number. It is worth noticing that we are assuming none of the $SU(2)$ components of the primary of dimension $\Delta=3/8$ acquires an expectation value, while its $\mathfrak{su}(2)_2$ descendants can have a vev. This is a strictly weaker condition than the one imposed by the $\mathbb{Z}_2$ symmetry of the action. From the $\mathfrak{su}(2)_2$ character, one can check that there is at least one Virasoro primary for every operator dimension in $\mathcal{I}_{2\times 2}$: hence, the spectrum is consistent with Virasoro symmetry.  Different setups are of course possible: one could allow a subset of components of the $\mathfrak{su}(2)_2$ primary to acquire a one-point function, while choosing as the external operator $\phi$ a component that does not. This would describe, for instance, a boundary state preserving a continuous or discrete subgroup of $SU(2)$. However, in that case, it is not hard to argue that $\phi$ must couple to a tilt operator, and hence the $\hgaptp\leq 1$.\footnote{The argument goes as follows. Since the spin $1/2$ representation is transitive, there is an element of the $su(2)$ algebra that connects any two components of $g_{\alpha\dot{\alpha}}$. The transformation connecting a component whose vev vanishes ($\phi$) to one whose vev is non-zero must belong to the symmetries broken by the boundary condition. Ward identities then impose that the corresponding tilt operator couples to $\phi$.} We comment more on this scenario in the last part of this section. 

We can go further in isolating our model of interest, by feeding the bootstrap the OPE coefficients $c_{\D \ell}$ in the fusion of $\phi$ with itself---see eq. \eqref{bulkOPE}. These are known in closed form \cite{Kudrna:2021rzd}. In fact, the four-point function itself in the WZW model is computable, as mentioned in the introduction, and the procedure is reviewed in appendix \ref{ssec:KZ}. The values of $c_{\D \ell}$ depend on which component in the $(1/2,1/2)$ multiplet is chosen as the external operator. Since $\phi$ must be real, we are left with a linear combination of the following options:
\begin{subequations}
\begin{align}
    O^1_+ &=\frac{1}{\sqrt{2}}\left(\phi_{\frac{1}{2}\frac{1}{2}}+\phi_{-\frac{1}{2}-\frac{1}{2}}\right)~, \\
    O^1_- &=\frac{i}{\sqrt{2}} \left(\phi_{\frac{1}{2}\frac{1}{2}}-\phi_{-\frac{1}{2}-\frac{1}{2}}\right)~, \\
    O^1_0 &=\frac{1}{\sqrt{2}}\left(\phi_{\frac{1}{2}-\frac{1}{2}}+\phi_{-\frac{1}{2}\frac{1}{2}}\right)~, \\
    O^0_0 &=\frac{i}{\sqrt{2}} \left(\phi_{\frac{1}{2}-\frac{1}{2}}-\phi_{-\frac{1}{2}\frac{1}{2}} \right)~. \label{O00}
\end{align}
\label{bulkRealSu2}
\end{subequations}
The upper label of the operators $O$ refers to the representation under the diagonal $SU(2)$ inside $SU(2)_L\times SU(2)_R$. For definiteness, we choose
\begin{equation}
    \phi=  O^1_+~.
\end{equation}
The purple line in \cref{fig:su(2)2_hann_h2pt} is obtained by inputting the value of the OPE coefficients $c_{\D \ell}$ of the bulk operators with spin larger than 0 and dimension and spin bounded from above ($0<\ell<\ell_\text{max}$, $\Delta<\Delta_\text{max}$). In practice, this is done by including the contribution of these operators in the normalization of the functional, eq. \eqref{eq:SDP-bulk-id}, and requiring positivity only above $\ell_\text{max}$ and $\Delta_\text{max}$. We see that this additional constraint only improves the bound for sufficiently large values of $\hgaptp$. It is possible to impose further constraints: in particular, it would be interesting to input the OPE coefficients of scalar operators as well. We leave this attempt, which requires scanning over the one-point functions \cite{Kos:2016ysd}, to future work.\footnote{We would like to thank Denis Karateev and Slava Rychkov for discussions on this point.}

The spectral assumptions \eqref{pvector_refined_su2} are compatible with the $\ket{1/2}_A$ boundary condition, including the linear combinations of representatives of its moduli space. The elementary brane contributes two points, which are marked in blue in \cref{fig:su(2)2_hann_h2pt}. In particular, the point at $(\hgapa,\hgaptp)=(1/2,3/2)$ exists because, by choosing as external operator a linear combination which is neutral under the copy of $su(2)$ preserved by the brane---for instance $O_0^0$ in eq. \eqref{O00}, with trivial gluing conditions for $su(2)$---then $\phi$ does not couple to the primary of dimension $h=1/2$, which transforms in the adjoint. Rather, it couples to the first singlet in the module, which lies at level 1. As discussed in \cref{ssec:elementary}, linear combinations of boundary states have decreasing values of $\hgapa$ due to the boundary-changing operators, but the latter do not couple to the two-point function. Generically, a single bulk operator $\phi$ is not in the singlet of the preserved $su(2)$ of a superposition of $A$-branes, therefore most corresponding points would lie on the segment $0<\hgapa<1/2$, $\hgaptp=1/2$. We did not draw them, since they only showcase the convexity of bounds like the ones in \cref{fig:su(2)2_hann_h2pt}.

\begin{figure}[!ht]
    \centering
    \begin{tikzpicture}
        \node at (0,0) {\includegraphics[width=\textwidth]{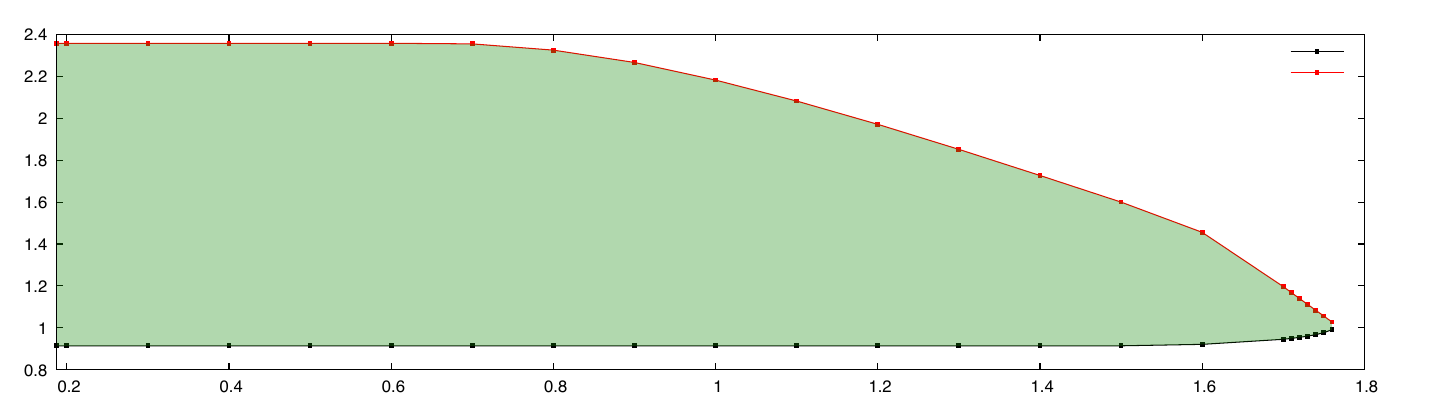}};
    
    
        \node[scale=0.525, anchor=east] at (6.1,1.65) {$g^2$-min Mixed System}; 
        \node[scale=0.525, anchor=east] at (6.1,1.43) {$g^2$-max Mixed System}; 
    
        \node[scale=0.7, anchor=east] at (-7.5,2.125) {$g^2$}; 
        \node[scale=0.7, anchor=east] at (7.9,-2.1) {$h_{\text{gap}}^{\text{2pt}}$}; 
    \end{tikzpicture}
    \caption{Bounds on the boundary entropy as a function of $\hgaptp$ for $c=3/2$ and $\vec{p}=(\frac{3}{8}|1;\frac{3}{16},\hgaptp|0)$, at $N=35$ derivatives. }
    \label{fig:su2-whale-0.1875}
\end{figure}
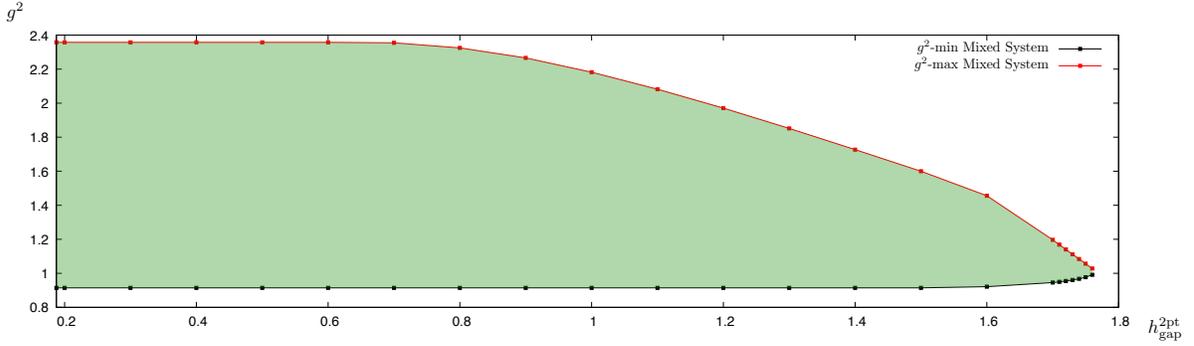

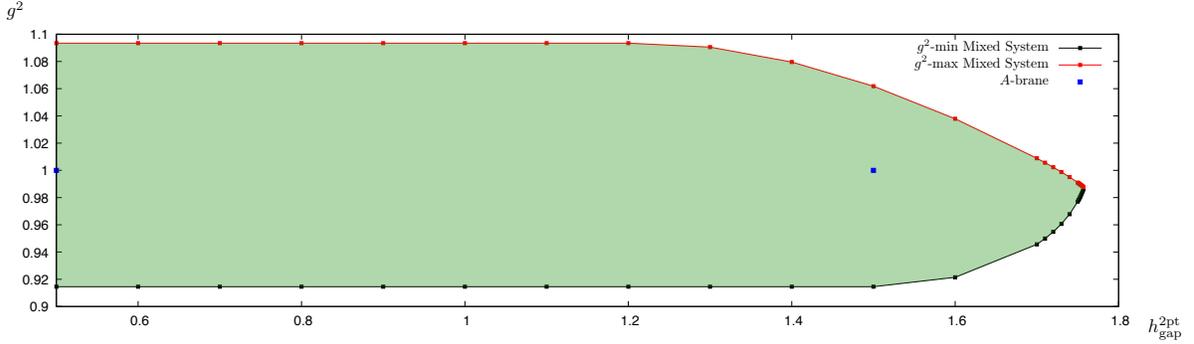
\begin{figure}[!ht]
    \centering
    \begin{tikzpicture}
        \node at (0,0) {\includegraphics[width=\textwidth]{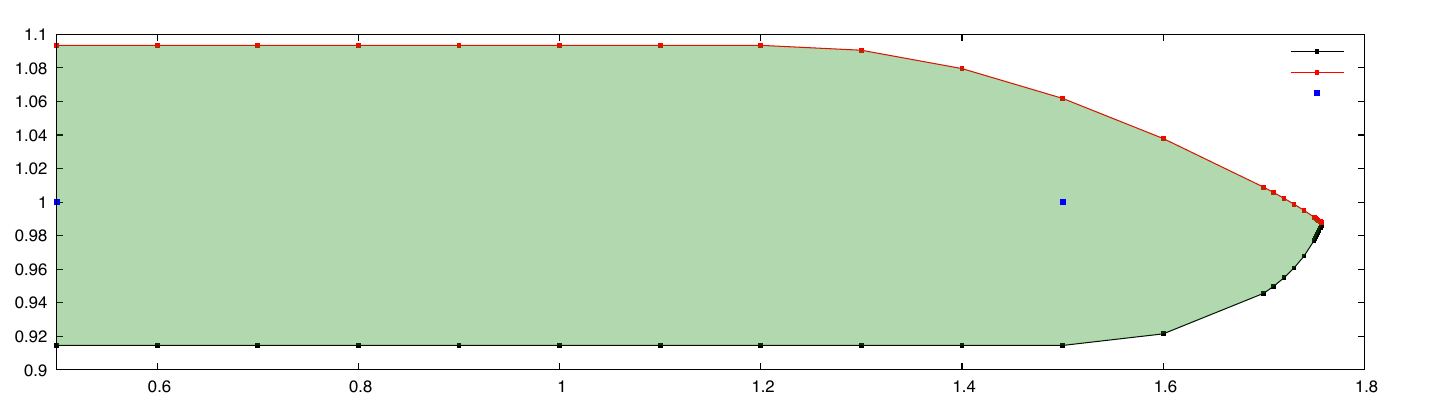}};
    
        \node[scale=0.525, anchor=east] at (6.1,1.65) {$g^2$-min Mixed System}; 
        \node[scale=0.525, anchor=east] at (6.1,1.43) {$g^2$-max Mixed System}; 
        \node[scale=0.525, anchor=east] at (6.1,1.21) {$A$-brane}; 
    
        \node[scale=0.7, anchor=east] at (-7.5,2.125) {$g^2$}; 
        \node[scale=0.7, anchor=east] at (7.9,-2.1) {$h_{\text{gap}}^{\text{2pt}}$}; 
    \end{tikzpicture}
    \caption{Bounds on the boundary entropy as a function of $\hgaptp$ for $c=3/2$ and $\vec{p}=(\frac{3}{8}|1;\frac{1}{2},\hgaptp|0)$, at $N=35$ derivatives.}
    \label{fig:su2-whale-0.5}
\end{figure}

Instead, we move on to discussing bounds on the ground state degeneracy $g$ in the same setup as above. They are presented in \cref{fig:su2-whale-0.1875} and \cref{fig:su2-whale-0.5}, as a function of $\hgaptp$, for two sample values of $\hgapa$. As observed in various other examples, the constraint from the mixed system kicks in for large enough values of $\hgaptp$. Interestingly, such a region is of concrete physical interest, because it is populated (at least) by the $\ket{1/2}_A$ boundary state in \cref{fig:su2-whale-0.5}. The two figures meet the sanity check that the allowed region in \cref{fig:su2-whale-0.5} is contained in the one of \cref{fig:su2-whale-0.1875}: indeed, the only difference between the two runs is a stricter value of $\hgapa$ in the former.  Figures \ref{fig:su(2)2_hann_h2pt},  \ref{fig:su2-whale-0.1875} and  \ref{fig:su2-whale-0.5} are intersections of a three-dimensional shape with three different planes. It would be nice to fully map out this solid in the $(\hgapa,\hgaptp,g^2)$ space, which contains all possible boundary conditions of the $\mathfrak{su}(2)_2$ WZW model whose boundary state does not include the $j=1/2$ primary.

The opposite case, where the $j=1/2$ operators acquire a one-point function, is the object of our last exploration. There are two scenarios one can tackle with the setup of this subsection. Either $\phi$ acquires a one-point function, or it does not. The first scenario might arise because all the components $g_{\alpha \dot{\alpha}}$ have an expectation value, or because one chooses a combination of components that do, among the \eqref{bulkRealSu2}. Furthermore, if all symmetries are broken by the boundary condition, the vevs $\braket{g_{\alpha \dot{\alpha}}}$ are independent, while a preserved subgroup would impose constraints. For concreteness, we focus on a boundary condition which produces the following pattern:
\begin{equation}
    \braket{\phi_{\frac{1}{2},-\frac{1}{2}}}=-\braket{\phi_{-\frac{1}{2},\frac{1}{2}}}= \frac{i\,a_\frac{1}{2}}{(2\text{Im} z)^{\Delta_\phi}}~, \qquad \braket{\phi_{\frac{1}{2},\frac{1}{2}}}=\braket{\phi_{-\frac{1}{2},-\frac{1}{2}}}=0~,
    \label{su2vevs}
\end{equation}
 where the  $a_\frac{1}{2} \in \mathbb{R}$.\footnote{The fact that $a_\frac{1}{2}$ must be real can be seen as follows. $O_0^0$ acquires an expectation value which must be real, because $O_0^0$ is Hermitian in our conventions and reflection positivity forces the square of the one-point function to be positive.} As we shall see in a moment, the expectation values in \eqref{su2vevs} are always a part of a larger conformal manifold of boundary conditions. One way to realise eq. \eqref{su2vevs} is by requiring the boundary conditions to preserve a $\mathbb{Z}_2 \ltimes U(1)$ subgroup of $SU(2)_L\times SU(2)_R$. Both the $\mathbb{Z}_2$ and the $U(1)$ factors are embedded in the diagonal $SU(2)$. Specifically, the gluing condition for the $U(1)$ factor can be deduced from eq. \eqref{su2vevs}, 
while $\mathbb{Z}_2$ is generated by the element  $\exp i \pi (J^1)_\text{diag}$. This element sends $J^3 \to - J^3$ and $\phi_{\frac{1}{2},-\frac{1}{2}}\to -\phi_{-\frac{1}{2},\frac{1}{2}}$. Since $\mathbb{Z}_2 \ltimes U(1) \in SU(2)_\text{diag}$, eq. \eqref{su2vevs} is compatible with the $A$-branes, and in particular $a_\phi$ is different from zero with $\ket{0}_A$ and $\ket{1}_A$ boundary states: $a_\frac{1}{2}=\pm 2^{1/4}$. The fact that only $O_0^0$ acquires a one-point function among the \eqref{bulkRealSu2} is of course in keeping with the fact that it lies in the singlet of $SU(2)_\text{diag}$.

As usual, eq. \eqref{su2vevs} describes a point in a moduli space, generated by the action of (the exponentials of) the tilt operators. An interesting subspace of this manifold is generated by acting with the tilt operator in the boundary OPE of $J^3$. The operators $\phi_{\frac{1}{2},-\frac{1}{2}}$ and $\phi_{-\frac{1}{2},\frac{1}{2}}$ have opposite charges under this transformation, and we get the more general one-point functions
\begin{equation}
    \braket{\phi_{\frac{1}{2},-\frac{1}{2}}}=\braket{\phi_{-\frac{1}{2},\frac{1}{2}}}^*\propto i e^{i \theta}\,a_\frac{1}{2}~, \quad \theta \in [0,2\pi)~.
    \label{su2vevsTheta}
\end{equation}
The parameter $\theta$ has the same role as the position of the $D$-brane in target space in \cref{subsec:dbranes}: it can be tuned to make the defect spectrum in the two-point function sparser. Specifically, by plugging eq. \eqref{su2vevsTheta} in \eqref{O00}, we find
\begin{equation}
    a_\phi=\sqrt{2} a_{\frac{1}{2}} \cos \theta~,
\end{equation}
with the choice
\begin{equation}\label{eq:phi-asq}
 \phi=O_0^0~.  
\end{equation}
Since the coupling to the $J^3$-tilt is proportional to $\partial_\theta\, a_\phi,$ we have two situations where the boundary OPE of $\phi$ is sparse: either $\theta=0$, and the tilt decouples, or $\theta=\pi/2$, and the identity decouples. Notice that, although the $h=1$ subspace also includes 4 other tilt operators---$J^1,\,\bar{J}^1,\,J^2,\,\bar{J}^2$---they are all charged under $U(1)_\text{diag}$ and therefore do not couple to $\phi.$

\begin{figure}[!ht]
    \centering
    \begin{tikzpicture}
        \node at (0,0) {\includegraphics[width=\textwidth]{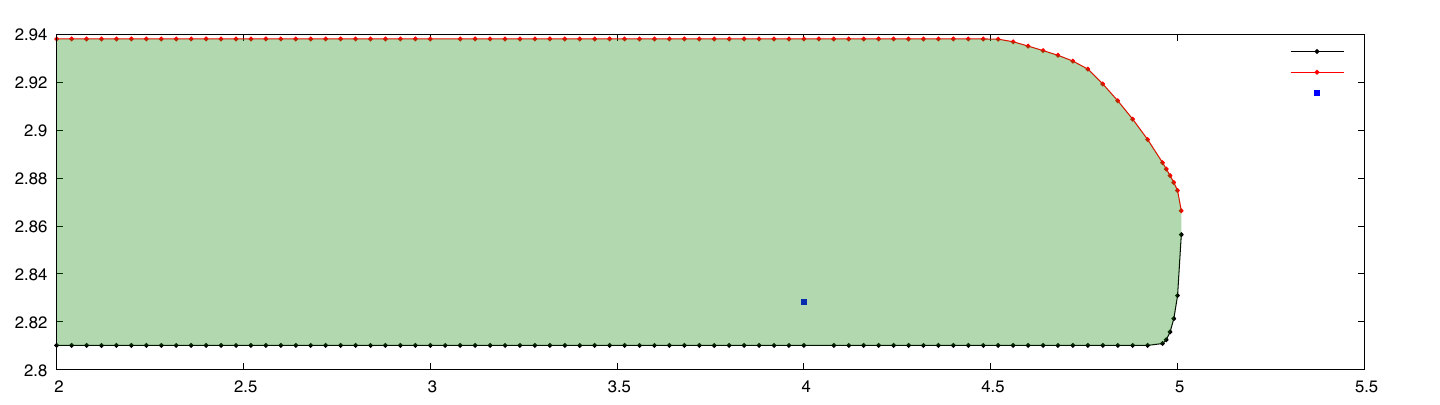}};
    
    
        \node[scale=0.525, anchor=east] at (6.1,1.63) {$a^2_\phi$-min Mixed System}; 
        \node[scale=0.525, anchor=east] at (6.1,1.415) {$a^2_\phi$-max Mixed System}; 
        \node[scale=0.525, anchor=east] at (6.1,1.2) {$A$-brane}; 
    
        \node[scale=0.7, anchor=east] at (-7.6,2.125) {$a^2_{\phi}$}; 
        \node[scale=0.7, anchor=east] at (7.7,-2.1) {$h_{\text{gap}}^{\text{2pt}}$}; 
    \end{tikzpicture}
    \caption{Bounds on $a^2_{\Delta_\phi}$, for the choice of $\phi$ in \eqref{eq:phi-asq}, as a function of $\hgaptp$ for $c=3/2$ and $\vec{p}=(\frac{3}{8}|20+\eps;1,\hgaptp|a^2_{\Delta_\phi})$, with $\mathcal{I}_{2\times 2} = \{1, 2, 19/8, 3, 4, 35/8, \dots, 20\}$, $\mathcal{I}_a = \{\frac{3}{8}\}$, $\mathcal{I}_{2} = \{2\}$, at $N=35$ derivatives.}
    \label{fig:su2-asq}
\end{figure}
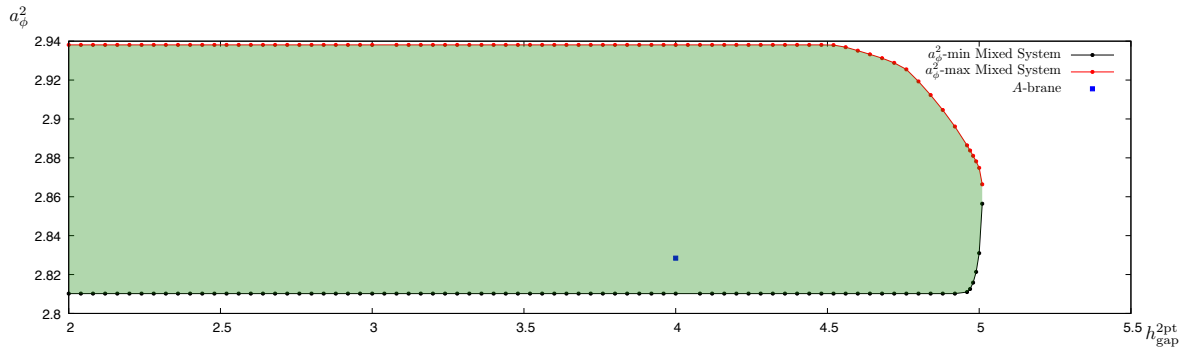

A natural question for the numerics is then to bound $a^2_\phi$, similar to \cref{fig:free-boson-asq} for the $D$-brane. Since $\Delta_\phi$ is fixed, and the annulus boundary gap cannot be larger than $\hgapa=1$ it makes sense again to plot $a^2_\phi$ as a function of $\hgaptp$, which also allows to explore how sparse the spectrum can be in the two-point function. The simplest setup is then the following:
\begin{equation}
    \vec{p}=\left(\frac{3}{8}\Bigg|20+\eps;1,\hgaptp\Bigg|a^2_{\Delta_\phi}\right),\quad  \mathcal{I}_{2\times 2} = \{1, 2, 19/8, 3, 4, 35/8, \dots, 20\},\quad \mathcal{I}_a = \left\{\frac{3}{8} \right\}, \quad\mathcal{I}_{2} = \{ 2\}~,
    \label{pforaphi2u2}
\end{equation}
where we input the first few bulk operators, we decouple the tilt operator, and we set the annulus gap to the largest allowed value. In order to optimize over the one-point function, we choose the same objective as in \eqref{dbranes_obj_a},
\begin{equation}
    \vec{V}_{\Delta_\phi}^a + \vec{V}_{0}^{\text{2pt}} + \frac{\Delta_\phi^2}{2c} \vec{V}_{2}^{\text{2pt}},
\end{equation}
which also includes the contribution from the displacement operator. Notice that, since the displacement is not the only $h=2$ operator in the boundary OPE of the $su(2)$ currents, we are equivalently setting a lower bound on the total OPE coefficient in that subspace. The result is shown in \cref{fig:su2-asq}. We see that the full system does not affect the bounds on $a_\phi^2$ until $\hgaptp>4.5$, then the allowed region quickly shrinks away.

Various refined assumptions could be made, \emph{e.g.} imposing the presence of a $u(1)$ current algebra at the level of the blocks. However, many of the expected features of the optimal bound can be argued for analytically, including the value of $\hgaptp=4$, hence we rather turn to explaining this, and leave further comments about the numerics for the end.

Firstly, the value $\hgaptp=4$ is the one corresponding to the $A$-branes \eqref{CardyZero} and \eqref{CardyOne}. The two-point function in the presence of this boundary state can be computed using the fact that it is proportional to the identity $su(2)$ block in the boundary channel. The specific expression can be found, for instance, by applying the method of images to the four-point function computed in \cref{ssec:KZ}, and reads 
\begin{equation}
    \braket{g_{\alpha\dot{\alpha}}(z_1,\bar{z}_1) g_{\beta\dot{\beta}}(z_2,\bar{z}_2)} = \frac{1}{|z_1-z_2|^{3/4}}\left[(\epsilon_{\alpha\dot{\beta}}\epsilon_{\dot{\alpha}\beta} - \epsilon_{\alpha\dot{\alpha}}\epsilon_{\beta\dot{\beta}})\,f_s(-1/\xi)+\epsilon_{\alpha\beta}\epsilon_{\dot{\alpha}\dot{\beta}}\,f_t(-1/\xi)\right]~,
\end{equation}
where the functions $f_s,\,f_t,\,$ are equal to the functions $\mathcal{F}_{s,\text{id}},\,\mathcal{F}_{t,\text{id}},\,$ defined in (\ref{Fsid},\ref{Ftid}), up to a phase chosen such that the former are real when $\xi>0$. When $k=2$, these two functions become elementary:\footnote{One easily sees that these functions have the requested reality properties, and in particular the explicitly real alternative expression
\begin{align}
f_s(-1/\xi) &=\frac{\sqrt{\xi } \cos \left(\frac{1}{2} \arctan \left(\frac{1}{\sqrt{\xi
   }}\right)\right)}{\sqrt[8]{\xi +1}}~, \\
f_t(-1/\xi) &= \frac{\sqrt{\xi } \sin \left(\frac{1}{2} \arctan\left(\frac{1}{\sqrt{\xi
   }}\right)\right)}{(\xi +1)^{5/8}}-\frac{(\xi +2) \cos \left(\frac{1}{2} \arctan\left(\frac{1}{\sqrt{\xi }}\right)\right)}{(\xi +1)^{5/8}}~.
\end{align}
}
\begin{align}
f_s(-1/\xi) &=\frac{\sqrt{\xi }\left(\left(\sqrt{\xi }+i\right)^{1/2}+\left(\sqrt{\xi }-i\right)^{1/2}\right) }{2 (\xi +1)^{3/8}}~, \\
f_t(-1/\xi) &=-\frac{i \sqrt{\xi } \left(\left(\sqrt{\xi }+i\right)^{1/2}-\left(\sqrt{\xi }-i\right)^{1/2}\right)}{2 (\xi
   +1)^{7/8}}-\frac{(\xi +2)\left(\left(\sqrt{\xi }+i\right)^{1/2}+\left(\sqrt{\xi }-i\right)^{1/2}\right) }{2 (\xi
   +1)^{7/8}}~.
\end{align}
Using eq. \eqref{gtophi}, one sees that the gluing conditions are compatible with \eqref{su2vevs}---\emph{i.e.} $\theta=0$ in \eqref{su2vevsTheta}---and that $a_\frac{1}{2}=\pm 2^{1/4}$ as expected. In particular,
\begin{equation}
\braket{\phi_{\frac{1}{2},-\frac{1}{2}}(z_1,\bar{z}_1)\phi_{-\frac{1}{2},\frac{1}{2}}(z_2,\bar{z}_2)}=\frac{\sqrt{2} (\xi +1)^{3/8} \left(\sqrt{\xi }+\sqrt{\xi +1}\right)}{\sqrt{\xi +\sqrt{\xi 
   (\xi +1)}+1}}~.    
\end{equation}
The decomposition in $sl(2)$ blocks of the two-point function of $\phi=O_0^0$ shows that, beyond the tilts, all the odd dimensional quasi-primaries also decouple from the boundary channel, leaving $\hgaptp=4$, in the notation \eqref{pforaphi2u2}, as claimed.

Let us now argue that this value is optimal among boundary CFTs of the $\mathfrak{su}(2)_2$ WZW model obeying eq. \eqref{su2vevs}. As discussed in \cref{subsec:wzw_ann}, the spectrum in the annulus cannot be sparser than the one of the $A$-brane, which has the largest set of gluing conditions. One may wonder whether boundary conditions might exist with the same boundary spectrum but different bulk-to-boundary couplings. On one hand, in \cref{subsec:wzw_ann} we gave some evidence that the $A$-brane is the unique boundary state with boundary spectrum supported only on the integers. More to the point, the spectrum appearing in the two-point function cannot be sparser than that of the $A$-brane. This is due to the fact that the bulk-to-boundary couplings of a bulk scalar primary $O$ with any operator in the boundary OPE of the currents are fully determined by the bulk OPE and by the one-point function of the primaries in the same multiplet of $O$. 

Since the previous statement holds for any operator appearing in the boundary OPE of a holomorphic operator, we now describe in general how to compute the corresponding OPE coefficients. In the WZW model, all such boundary operators can be expressed in radial quantization as products of modes of the bulk current. Consequently, the claim follows straightforwardly by commuting the currents through correlators of the form
\begin{equation}
\braket{0|O\, (J^1_{-1})^{n}\dots (J^3_{-m})^{k}|\hat{0}},
\end{equation}
where $\bra{0}$ denotes the bulk vacuum and $\ket{\hat{0}}$ the boundary vacuum.

More generally, consider a holomorphic quasi-primary $Y(z)$ of dimension $h_Y$. The two-point function with $O$ in the presence of a boundary is fixed by holomorphy and conformal symmetry to take the form:
\begin{equation}
\braket{O(w,\bar{w}) Y(z)} = \frac{b_{OY}}{(2\text{Im}\, w)^{\Delta_O - h_Y} [(z-w)(z-\bar{w})]^{h_Y}}~.
\end{equation}
Integrating $w$ around $z$, and using the fact that the result is entirely determined by the bulk OPE, one obtains an equation for the boundary OPE coefficient $b_{OY}$:
\begin{equation}
b_{OY} = - a_{\delta O}\, \frac{4^{1-h_Y}(h_Y-1)!}{(1/2)_{h_Y-1}}~,
\end{equation}
where $a_{\delta O}$ is the one-point function coefficient of the operator appearing with a simple pole in the bulk OPE of $O$ and $Y$:
\begin{equation}
Y(z) O(w,\bar{w}) \sim \dots + \frac{\delta O(w)}{z - w} + \dots~.
\end{equation}
Here, $\delta O$ has conformal weights
\begin{equation}
(h, \bar{h}) = \left(\frac{\Delta_O}{2} + h_Y - 1, \frac{\Delta_O}{2} \right)~.
\end{equation}
Thus, unless $h_Y = 1$ (corresponding to the case where $Y$ is a current and the associated operator is the tilt), we must have $b_{OY} = 0$ unless $\delta O$ is an $sl(2)$ descendant of a scalar quasi-primary. In other words,
\begin{equation}
\delta O = \lambda\, \partial_z^{h_Y - 1} \tilde{O},
\end{equation}
where $\tilde{O}$ is an operator of the same scaling dimension as $O$.

If $Y$ is a descendant of a current with $h = 1$, then $\tilde{O}$ lies in the same multiplet as $O$ under the global symmetry, and the coefficient $\lambda$ can be computed algebraically. This confirms that the boundary OPE coefficient $b_{OY}$ is fixed by bulk data and one-point functions, as claimed.

Let us conclude with a remark on the numerics. It appears that the more natural question regarding the gap in the two-point function of an operator whose multiplet acquires a vacuum expectation value can be addressed analytically. Nonetheless, many other directions for numerical exploration remain, both within the $\sutwok$ model and its orbifolds. Among the simplest are questions about boundary conditions whose spectrum includes light operators beyond the boundary values of the currents. For example, one could isolate a boundary operator with a gap and perform a scan over its scaling dimension.  We leave this and other avenues for future work. 

\section{Conclusion and Outlook}
\label{sec:outlook}

In this work, we initiated the study of crossing equations involving both local operators and conformal boundaries via convex optimization methods. This effort is a natural step forward in the program aimed at applying the conformal bootstrap to extended operators. The mixed-correlators system involves the partition function on the annulus, the four-point function on the infinite plane, and the two-point function on the upper half-plane. Along the way, we also derived new bounds from the annulus partition function alone, extending the work done in \cite{Friedan:2012jk,Collier:2021ngi}. 

We have found non-perturbative bounds in a space involving five parameters: the dimension of the lightest operator with a one-point function ($\Delta_\text{gap}$), the dimensions of the lightest boundary operator ($\hgapa$) and of the first operator in the boundary OPE of a specific bulk operator ($\hgaptp$), the ground-state degeneracy ($g$), and the one-point function of the same specific bulk operator ($a_\phi$). These bounds carve the space of theories with $c=1/2$, $c=1$ and $c=3/2$. At the latter value of the central charge, the bounds in \cref{fig:su(2)2_hann_delta,fig:su2-annulus,fig:su(2)2_hann_h2pt,fig:su2-whale-0.1875,fig:su2-whale-0.5,fig:su2-asq} provide new information on a space of boundary conditions of which we know but a handful of examples.

Let us emphasise a few upshots of the bootstrap of points and lines: firstly, it gives access to new CFT data; secondly, it often strengthens the bounds on the data accessible via the annulus partition function, in regions of parameter space where physical theories are known to live. This is visible in most of the plots presented above---see \cref{fig:c=1-neumann-h2pt}, \cref{fig:free-boson-asq} and \cref{fig:su2-whale-0.5} for some noteworthy examples. 

The setup also allows us to input knowledge about the bulk CFT data in the bootstrap, in order to zoom in on the boundary conditions of known theories. Beyond the spectrum, we have also sometimes included the OPE coefficients of the spinning operators which contribute to the four-point function, finding in \cref{fig:su(2)2_hann_h2pt} that the bounds are strengthened, albeit slightly. Including yet more data is possible, in particular the three-point OPE coefficients of scalar operators, which appear in the matrix part of eq. \eqref{full-system} and so require doing OPE scans over the related one-point functions \cite{Kos:2016ysd}. Recent examples have shown that tight bounds can be obtained when precise knowledge of a large part of the CFT data is exploited \cite{Cavaglia:2021bnz}.

The improvement of the mixed-correlator system over the annulus bootstrap is not always realised. For low values of $\hgaptp$, the extremal functional typically suppresses the contribution of the two-point function and of the four-point function, when optimising data which appears in the annulus partition function. This effect is visible, for instance, in \cref{fig:c=1_whale_0.5}. Of course, the issue is absent when looking for bounds on scaling dimensions or OPE coefficients which only appear in the two-point function: $\hgaptp$ is itself an example considered in this work. 

This ``decoupling'' problem sometimes presents us with a tension. On one hand, two-point function constraints are especially consequential when $\hgaptp>\hgapa$. This selection rule is most naturally enforced by requiring the boundary to preserve a group $H$ of symmetries, with $H \subset G$, the symmetry group on the infinite plane. On the other hand, if  $G$ is continuous, then the boundary always supports operators with integer dimensions starting from $h=1$. If the operator $\phi$ inserted in the correlator is part of a $G$-multiplet some of whose components get a vev, then it always couples to (at least some of) the integer-dimensional operators, so $\hgaptp$ cannot be too large. In this regime, numerical bootstrap bounds are more likely to be fully determined by the annulus crossing constraint---see \emph{e.g.} \cref{fig:su2-asq}. This issue is particularly challenging because the natural objectives of the bootstrap of points and lines are rational theories with $c>1$. These theories always have continuous extended symmetries, and their Cardy branes have the sparsest spectrum of protected operators, since they have the largest number of gluing conditions.

Various lessons can be drawn from this tension. When studying boundary conditions of models with a continuous (possibly higher-spin) symmetry, it is easier to get constraints on the spectrum charged under $H$, if there is a bulk operator $\phi$ in a representation of $G$ that contains no singlet under $H$. Examples in this work are momentum vertex operators with a Neumann brane in the free boson theory, or the $(1/2,1/2)$ representation of $\mathfrak{su}(2)_2$, which is charged under the discrete symmetry $g_{\alpha\dot{\alpha}} \to -g_{\alpha\dot{\alpha}}$ of the WZW action \eqref{WZWaction}.\footnote{It would be interesting to compare the bounds obtained this way with bounds on charged boundary operators which might arise from appropriately twisted annulus partition functions.} Then, $\braket{\phi}=0$, and the boundary OPE of $\phi$ does not contain any operator belonging to the protected spectrum---see the discussion at the end of \cref{subsec:wzw_mixed}.

Another promising avenue is the study of boundary conditions in theories obtained as orbifolds of rational CFTs under discrete groups. If all currents are charged, the gauged theory has no holomorphic operators left and the boundary spectrum can be sparse---see \emph{e.g.} \cref{fig:c=1_virasoro_ann_sub}. 

A different lesson that can be drawn from our explorations is that interesting possibilities open up when bootstrapping a CFT with moduli. Bounds obtained by scanning over the scaling dimensions of bulk operators often have more features, and regions where the mixed-correlator system is especially powerful. Notice in particular that bootstrap bounds are not necessarily convex along directions labeled by the scaling dimension of an external operator, or of an exchanged operator isolated from the others by a gap---see \emph{e.g.} \cref{fig:c=1-dirichlet} and \cref{fig:free-boson-asq}. Similarly, it is possible in these setups that bootstrap bounds isolate compact allowed regions, \emph{i.e.} islands \cite{Kos:2014bka}. In this respect, an obvious target is to extend the results of \cref{subsec:wzw_mixed} to the entire known moduli space of theories with $c=3/2$. The space of Narain CFTs is another natural playground.

In the same spirit, it would be interesting to include correlators involving boundary operators \cite{Liendo:2018ukf,Gimenez-Grau:2019hez,Behan:2020nsf} in the mixed system considered here, which would entail challenges similar to the higher-point bootstrap in infinite space \cite{Poland:2023vpn,Antunes:2023kyz}.  A different mixed-correlator system where boundary operators are included has been proposed in the context of QFT in Anti-de Sitter in \cite{Meineri:2023mps}, and implemented in \cite{Ghosh:2025sic}. Similar ideas should work for $2d$ boundary CFTs, which are special cases of QFTs in AdS$_2$.\footnote{Another situation where defect operators can be included in the bootstrap without further complication arises when studying defects that can end \cite{Lanzetta:2024fmy,LanzettaLiuMetlitski}. The boundary-changing operators which interpolate between an opaque interface and a topological one provide an interesting target for this setup.} 

At a more basic level, an obvious step forward for the bootstrap of points and lines in two dimensions is to take into account from the start the largest symmetry preserved by the boundary: the Virasoro algebra in the generic case, and extended algebras when needed. This would in particular mitigate the issue arising from the protected defect spectrum, by grouping many of the boundary operators in the appropriate blocks.

Let us conclude by highlighting a different challenge, which arises if one wants to extend this program to higher dimensions. There, the analogue of the open string channel in the annulus partition function is not under sufficient control. While simple principles govern the fusion of conformal defects \cite{Bachas:2007td,Diatlyk:2024zkk,Cuomo:2024psk}, the defect product expansion involves summing over lower dimensional quantum field theories, and it is unclear whether unitarity imposes useful positivity constraints, and whether the kinematics can be brought under sufficient control. Besides explicit examples, one might hope to make progress by divising sum rules, or by combining the bootstrap with other methods to obtain estimates for the correlator of two defects---\emph{e.g.} the lattice or the fuzzy sphere \cite{Zhu:2022gjc}.

\vspace{2cm}


\acknowledgments{We thank Matthijs Hogervorst for collaborating during the early stages of the project. The authors would like to thank Costas Bachas, Connor Behan, Marco Billò, Gabriele Di Ubaldo, Denis Karateev, Madalena Lemos, Carlo Maccaferri, Jay Padayasi, Ingo Runkel, Slava Rychkov, Ning Su, Alessandro Tomasiello, and Alesandro Vichi for fruitful and enlightening discussions, and especially Joao Penedones, for the conversation which sparked the idea for the project. BR would particularly like to thank Aike Liu, David Simmons-Duffin, and Balt van Rees for aid with diagnosing numerical instabilities, and Vasiliy Dommes for help with technical issues with SDPB. BR and MM would like to thank the organisers of Bootstrap 2022 and Bootstrap 2024, and MM would like to thank the organisers of Bootstrap 2023, where parts of this work were completed. All numerical work featured in this paper were run on the Baobab cluster at the University of Geneva, and we would like to thank the admin team, especially Adrien Albert and Yann Sagon, for their prompt support. MM and BR are supported by the Swiss National Science Foundation through the Ambizione grant number 193472. MM is also is funded by the Italian Ministry of University and Research through the Rita Levi Montalcini program, and by the European Union through the MSCA Staff Exchange program (project HORIZON-MSCA-2023-SE-01-101182937-HeI).

\appendix


\section{Details on Block Implementation and SDPB Runs}\label{sec:block-imp}

In appendix, we provide details on the implementation of the code used to calculate derivatives of the blocks, along with the choices of cross ratio and OPE normalisation alluded to in \cref{ssec:SDPs}. 
We also provide the SDPB parameters that were used for our numerics. 

\subsection{Annulus partition function}

We first focus on the blocks appearing in \eqref{eq:annulus-unnormal}, which contain the terms 
\begin{align}
    & \sum_{h \geq h_{\text{min}}} n_h \, \chi_h(1/t), \\
  &\sum_{\Delta \geq \dm} a_{\Delta}^2 \, \chi_{\frac{\Delta}{2}}(t).
\end{align}
We redefine $a_{\Delta}^2$ using
\begin{equation}
    \sum_{\Delta \geq \dm} a_{\Delta}^2 \, \chi_{\frac{\Delta}{2}}(t) \rightarrow \sum_{\Delta \geq \dm} a_{\Delta}^2 \, 
  \nu^{2(\Delta-\dm)}  \, \chi_{\frac{\Delta}{2}}(t) \\
\end{equation}
the importance of which will be made clear in \cref{ssec:cross-ratio-ope}. Let
\begin{equation}
Z_y(q) \ldef \frac{q^y}{1-q},
\end{equation}
where the coordinate $q$ is defined in \eqref{eq:t-defn}, and $\tilde{q}(t)=q(1/t)$. Derivatives w.r.t. $q$ of $Z$ are of the form
\begin{equation}
\frac{d^n}{dq^n} \, Z_y(q) = q^y \, \mfr{z}_n(y,q)
\end{equation}
for some computable function $\mfr{z}_n(y,q)$ that combines the derivatives of $q^y$ and $(1-q)^{-1}$ using the Leibniz rule, giving a polynomial of degree $n$ in $x$.
Thus the derivatives 
\begin{equation}
    \frac{d^n}{dt^n} Z_{y}(q)
\end{equation}
can be computed as a simple chain rule. On the boundary side, we have
\begin{equation}
\chi_{h = 0}(1/t) = \tilde{q}(t)^{-\frac{c}{24}},
\quad
\chi_{h > 0}(1/t) = Z_{h-\frac{c}{24}}(\tilde{q}(t)).
\end{equation}
When the block is evaluated at a point $t_\star$, the leading scaling in the operator dimensions is
\begin{equation}
\tilde{q}(t_\star)^{\hm - c/24} \cdot \beta^x,
\quad
\beta = \tilde{q}(t_\star),
\quad
x = h - \hm \geq 0 .
\end{equation}
The quantity $n_h$ is required to be a positive integer in order to satisfy unitarity. However, we do not impose the integrality of $n_h$ in our numerics. Similar to the boundary channel, we have on the bulk side
\begin{equation}
\chi_{\Delta = 0}(t) = q(t)^{-\frac{c}{24}},
\quad
\chi_{\Delta > 0}(t) = Z_{\frac{\Delta}{2}-\frac{c}{24}}(q(t)).
\end{equation}
Including the $\nu$-dependence, the leading scaling is
\begin{equation}
\label{eq:ZLS}
\nu^{2(\Delta-\dm)} \, \chi_{\frac{\Delta}{2}}(\tau) \sim q(t_\star)^{\dm/2- c/24} \cdot \beta^x, \quad
\beta = \nu^2 \sqrt{q(t_\star)},
\quad
x = \Delta - \dm \geq 0.
\end{equation}


\subsection{Two-point function}\label{ssec:two-pt}

The contribution of the two-point function to the bootstrap equation consists of the terms 
\begin{align}
    & \sum_{\Delta \geq \dm} c_{\Delta} a_\Delta \, \mathcal{G}^\text{bulk}_\Delta(\xi), \\
    &\sum_{h \geq \hm} b_h^2 \, \mathcal{G}_h^\text{bry}(\xi).
\end{align}
We define the functions
\begin{equation}
\kappa_h(z) \ldef {}_2F_1(h,h,2h,z)
\qaq
F(x,\rho) \ldef (4\rho)^x \, {}_2F_1\left(x,\frac{1}{2},x+\frac{1}{2},\rho^2\right).
\end{equation}
For the two-point function, we introduce the coordinates \cite{Lauria:2017wav}
\begin{equation}
\rho_\text{bry}(\xi) = 1 +2\xi - 2\sqrt{\xi(1+\xi)},
\quad
\rho_\text{bulk}(\xi) = \rho_\text{bry}(1/\xi).
\end{equation}
In the bulk channel, we have 
\begin{equation}
\mathcal{G}^\text{bulk}_{\Delta}(\xi) \ldef \xi^{\frac{\Delta}{2}} \, \kappa_{\frac{\Delta}{2}}(-\xi)
\end{equation}
and in particular
\begin{equation}
\mathcal{G}^\text{bulk}_{\Delta = 0}(\xi) = 1
\end{equation}
automatically. This means we can write
\begin{equation}
\mathcal{G}^\text{bulk}_{\Delta}(\xi) = F(\tfrac{\Delta}{2},\rho_\text{bulk}(\xi)).
\end{equation}
For values of $\xi \in [0,\infty)$, the cross ratio $\rho_\text{bulk}$ is bounded as $|\rho_\text{bulk}|<1$, and thus we can use a series expansion of the hypergeometric function in $F(\tfrac{\Delta}{2},\rho_\text{bulk}(\xi))$. 
In this work, we always take the truncation order $N_\text{trunc}$ for the series to be $21, 27, 36$ for $35, 45, 55$ number of derivatives respectively. This is the lowest number of terms observed to give us a stable SDP. Including $\nu$ from the redefinition of $a_{\Delta}$, the leading scaling will be
\begin{equation}
\label{eq:2ptLS}
\nu^{\Delta-\dm} \, \mathcal{G}_\Delta^\text{bulk}(\xi) \sim \frac{1}{\mca{P}(x)} \, (4\rho_\text{bulk}(\xi_\star))^{\dm/2} \cdot b^x,
\quad
b = \nu \sqrt{4\rho_\text{bulk}(\xi_\star)}\, ,
\quad
x = \Delta - \dm \geq 0
\end{equation}
where
\begin{equation}
\mca{P}(x) = \prod_{\pi \in \mrm{P}} x-\pi,\quad
\mrm{P} = \{-2N_\text{trunc}+1,\, -2N_\text{trunc}+3,\ldots,-1\} - \dm.
\end{equation}
On the boundary side we have
\begin{equation}
\mathcal{G}^{\text{bry}}_h(\xi) \ldef \xi^{\Delta_\phi - h} \, \kappa_h\!\left(-\frac{1}{\xi} \right),
\quad
\mathcal{G}^{\text{bry}}_{h=0}(\xi) = \xi^{\Delta_\phi}~,
\end{equation}
whence
\begin{equation}
\mathcal{G}^\text{bry}_h(\xi) = \xi^{\Delta_\phi} \, F(h,\rho_\text{bry}(\xi)).
\end{equation}
Thus the leading scaling, including a rescaling of $b^2_h$ as $\omega^{-(h-h_{\text{min}})} b^2_h$ is
\begin{equation}\label{eq:two-pt-bry-scaling}
\mathcal{G}^\text{bry}_h(\xi) \sim \frac{1}{\mca{Q}(x)}\, (4\rho_\text{bry}(\xi_\star))^{\hm} \cdot b^x,
\quad
b = 4 \omega \rho_\text{bry}(\xi_\star), \quad x = h - h_\text{min} \geq 0
\end{equation}
where
\begin{equation}
\mca{Q}(x) = \prod_{\pi \in \mrm{Q}} x - \pi,
\quad
\mrm{Q} = \{-N_\text{trunc}+\frac{1}{2},\, -N_\text{trunc}+\tfrac{3}{2},\ldots,-\frac{1}{2}\} - \hm.
\end{equation}
We note that the other normalisation, 
\begin{equation}
    \braket{\phi(x_1)\phi(x_2)}_{\textup{UHP}} = \frac{1}{|z_1 - z_2|^{2\Delta_\phi}} \tilde{H}(\xi)~.
\end{equation}
is also a valid choice theoretically. However, we noticed that \eqref{eq:2pt-fn} is the only numerically stable choice. This is due to the fact that stability in SDPB seems to require that most entries of the normalisation vector be non-zero. Thus if the identity block in the bulk expansion of the two-point function, which is used in the normalisation in \cref{ssec:SDPs}, is a constant, all except one entry of the $\alpha_2$ directions in the normalisation vector will be zeros.


\subsection{Four-point blocks}
The two terms in the bootstrap equation from the four-point function crossing relation are 
\begin{align}
&\sum_{\Delta \geq \dm} c_{\Delta}^2\, F_{\Delta,0}(z,\zb)\\
&\sum_{\Delta \geq \ell} c_{\Delta,\ell}^2\, F_{\Delta,\ell}(z,\zb),
\end{align}
where we have separated out the contribution from spinning states and scalars. The general global blocks for the four-point function in two dimensions take the form \cite{Dolan:2000ut}
\begin{equation}
G_{\Delta,\ell}(z,\zb) = \frac{1}{1+\delta_{\ell,0}} \left[ z^h \zb^{\hb} \kappa_h(z) \kappa_{\hb}(\zb) + z \lra \zb \right]
\quad
h,\hb = \frac{\Delta \pm \ell}{2}.
\end{equation}
It is convenient to rewrite the blocks using the quantities \cite{Hogervorst:2013sma}
\begin{equation}
z^h \kappa_h(z) = F(h,\rho(z)),
\quad
\rho(z) = \frac{z}{(1+\sqrt{1-z})^2}.
\end{equation}
The scalar blocks and the spinning blocks need a slightly different treatment, which is as follows. 

\subsubsection{Scalar Blocks}
The scalar blocks ($\ell = 0$) take the form
\begin{equation}
G_{\Delta,\ell=0}(z,\zb) = F(\tfrac{\dm+x}{2},\rho(z))F(\tfrac{\dm+x}{2},\rho(\zb)).
\end{equation}
Similar to the case with the two-point function, $|\rho| < 1$ for physical values of $z$ and $\bar{z}$. Thus, we use the series expansion of the hypergeometric in $F(h,\rho(z))$ up to order $N$ described in \cref{ssec:two-pt}. The leading scaling of this approximation therefore is
\begin{equation}
\label{eq:4pt0LS}
\frac{1}{\mca{P}(x)^2}\,  (4\rho(\th))^{\dm} \cdot (4\rho(\th))^x, \quad x = \Delta - \dm \geq 0
\end{equation}
where $\theta$ is the value for $z$ and $\bar{z}$ around which we evaluate the derivatives. We note that there is no need to take a special limit for the unit block,
\begin{equation}
G_{0,0}(z,\zb) = 1.
\end{equation}

\subsubsection{Spinning blocks}

For the spinning block we have
\begin{equation}
G_{\Delta,\ell}(z,\zb) = F(\tfrac{x}{2} + \ell,\rho(z))F(\tfrac{x}{2},\rho(\zb)) + (z \lra \zb),
\quad
\Delta = \ell + x.
\end{equation}
The leading scaling is therefore
\begin{equation}
\frac{1}{\prod_{\pi \in \mrm{R}} x - \pi}\,  (4\rho(\th))^\ell \cdot (4\rho(\th))^x
\end{equation}
where
\begin{equation}
\mrm{R} = \{ -2N_\text{trunc}+1,\, -2N_\text{trunc}+3,\ldots,-1 \} \cup \{ -2N_\text{trunc} -2\ell + 1,\, -2N_\text{trunc} -2\ell +3,\ldots,-2\ell-1 \}.
\end{equation}

The relevant combination for the bootstrap is
\begin{equation}
G^{-}_{\Delta,\ell}(z,\zb) = \left[(1-z)(1-\zb)\right]^{\Delta_\phi} G_{\Delta,\ell}(z,\zb) - (z,\zb \mapsto 1-z,1-\zb).
\end{equation}
By construction, this is symmetric under $z \lra \zb$ but antisymmetric under crossing $z,\zb \mapsto 1-z,1-\zb$. In addition
\begin{equation}
G^{-}_{0,0}(z,\zb) =  \left[(1-z)(1-\zb)\right]^{\Delta_\phi} - (z\zb)^{\Delta_\phi}.
\end{equation}

\subsection{Cross Ratio and OPE Normalisation}\label{ssec:cross-ratio-ope}
To recap, the full bootstrap system consists of six terms---three in the bulk scalar sector, one in the bulk spinning sector, and two in the boundary sector. The three scalar bulk blocks combine to form the two-by-two matrix in \eqref{full-system}. We further introduced a rescaling of the one-point function as 
\begin{equation}
    \sum_{\Delta \geq \dm} a_{\Delta}^2 \, \chi_{\frac{\Delta}{2}}(t) \rightarrow \sum_{\Delta \geq \dm} a_{\Delta}^2 \, 
  \nu^{2(\Delta-\dm)}  \, \chi_{\frac{\Delta}{2}}(t), \\
\end{equation}
leading to the redefinition of the bulk scalar contributions as
\label{eq:allBlocks}
\begin{align}
  &\sum_{\Delta \geq \dm} a_{\Delta}^2 \, 
  \nu^{2(\Delta-\dm)}  \, \chi_{\frac{\Delta}{2}}(t) \\
 & \sum_{\Delta \geq \dm} c_{\Delta} a_\Delta \,  \nu^{\Delta-\dm} \, B^\text{bulk}_\Delta(\xi)\\
  &\sum_{\Delta \geq \dm} c_{\Delta}^2\, G^{-}_{\Delta,0}(z,\zb).
\end{align}
The rescaling is important to make the system of crossing equation compatible with SDPB. This is due to the fact that in the formulation of the bootstrap problem as a convex optimisation problem, the elements of the polynomial matrix are required to have a common positive prefactor. 

In practice, this leads to a crucial limitation---the choice of $\tau_\star$, $\xi_\star$, and $\theta_\star$ is not completely free. This can be seen by comparing the leading scalings in~\reef{eq:ZLS}, \reef{eq:2ptLS}
and \reef{eq:4pt0LS}. The requirement of having the same leading behaviour in $x$ implies
\begin{equation} \label{eq:reln-b/w-crossratio}
4\rho(\th) = \nu \sqrt{4\rho_\text{bulk}(\xi_\star)} = \nu^2
\sqrt{q(t_\star)}
\quad
\Rightarrow
\quad
\nu = \frac{2\rho(\th)}{\sqrt{\rho_\text{bulk}(\xi_\star)}}
\qaq
\exp(-\pi t_\star) = \frac{\rho_\text{bulk}(\xi_\star)}{\rho(\th)}.
\end{equation}
We follow the standard convention of evaluating the derivatives in $z$ and $\bar{z}$ at the crossing symmetric point,
\begin{equation}
\theta = (z, \bar{z}) = \left( \frac{1}{2}, \frac{1}{2} \right),
\end{equation}
leaving us the option to either fix $\xi_\star$ or $\tau_\star$. Based on our experiments, setting $\xi_\star$ to be at the crossing symmetric point ($\xi_\star = 1$) leads to numerical instabilities, while setting $\tau_\star = 1$ does not lead to stability issues. From \eqref{eq:reln-b/w-crossratio}, this leads to the numerical value for $\xi_\star$ and $\nu$ to be
\begin{equation}
    \xi_\star \simeq 0.03010 \qaq \nu \simeq 3.98513.
\end{equation}

Finally, we need to make sure that the positive prefactor in \eqref{eq:two-pt-bry-scaling} is not divergent for all $x \geq 0$. For the choice of $\xi_\star$ above, we obtain $4\rho_{\text{bry}} \simeq 2.83212 $. In order to have $b < 1$ we introduce a rescaling of $b_h^2$ as
\begin{equation}
b^2_h \rightarrow 4^h b^2_h
\end{equation}
and choose $\omega = 1/4$.

\subsection{SDPB parameters}

For the purpose of reproducibility, we provide the parameters used in SDPB runs in \cref{tab:sdpb-params}. In the runs where we used $N = 35$ derivatives, the spins included in the four point function were
\begin{equation}
    S_{N=35} = \{ 2, 4, 6, \dots, 42 \},
\end{equation}
while for \cref{fig:su(2)2_hann_h2pt} where we used $N=55$ derivatives, the spins included were 
\begin{equation}
    S_{N=55} = \{ 2, 4, 6, \dots, 72 \}.
\end{equation}

\bgroup
    \def\arraystretch{1.5}
    \begin{table}
        \centering
        \begin{tabular}{|c|c|c|}
            \hline
            Parameters & OPE problems & Gap problems \\
            \hline
            \texttt{findPrimalFeasible}           & false & true \\
            \texttt{findDualFeasible}             & false & true \\
            \texttt{detectPrimalFeasibleJump}     & false & true \\
            \texttt{detectDualFeasibleJump}       & false & true \\
            \texttt{precision}(actual)            & 2500(2560) & 2500(2560)\\
            \texttt{dualityGapThreshold}          & 1e-80 & 1e-80 \\
            \texttt{primalErrorThreshold}         & 1e-30 & 1e-30\\
            \texttt{dualErrorThreshold}           & 1e-30 & 1e-30\\
            \texttt{initialMatrixScalePrimal}     & 1e+50 & 1e+50\\
            \texttt{initialMatrixScaleDual}       & 1e+50 & 1e+50\\
            \texttt{feasibleCenteringParameter}   & 0.1 & 0.1 \\
            \texttt{infeasibleCenteringParameter} & 0.3 & 0.3 \\
            \texttt{stepLengthReduction}          & 0.7 & 0.7 \\
            \texttt{maxComplementarity}           & 1e+300 & 1e+300 \\
            \hline
            
        \end{tabular}
        \caption{Parameters used in SDPs}
        \label{tab:sdpb-params}
    \end{table}
\egroup

\section{Four-point functions in $\sutwok$ WZW models}
\label{ssec:KZ}

In this appendix, we review the derivation of the Knizhnik-Zamalodchikov equation \cite{Knizhnik:1984nr}, and its solution in the $\mathfrak{su}(2)_k$ case. In doing so, we correct a few typos found found in other accounts in the literature.

We begin by considering the standard action for $SU(2)$ WZW model 
\begin{equation}
    S = \frac{k}{4\pi} \int d^2 x \text{Tr} \left( \partial^\mu g^{-1} \partial_\mu g \right) + k \Gamma,
\end{equation}
where $\Gamma$ is
\begin{equation}
    \Gamma = - \frac{i}{12 \pi} \int d^3 y \epsilon_{\alpha\beta\gamma} \text{Tr}\left( g^{-1}\partial^\alpha g g^{-1}\partial^\beta g g^{-1}\partial^\gamma g \right),
\end{equation}
and the trace is defined over the generators of the Lie algebra in the fundamental representation as
\begin{equation}
    Tr(t^a t^b) = \frac{1}{2} \delta^{ab}.
\end{equation}
We are interested in finding the four-point function of $g$ in closed form, but in order to be able to do so, we need a little more machinery in the form of OPEs. We start with how the action varies as we vary the field $g$ as $\delta g = \omega g - g \overline{\omega}$,
\begin{equation}
    \delta S = \frac{i}{2\pi} \int dz d\overline{z} \left( \partial_{\overline{z}} \text{Tr} (\omega J) + \partial_{z} \text{Tr} (\overline{\omega} \overline{J}) \right),
\end{equation}
where we defined 
\begin{equation}
    J = -k \partial_z g g^{-1}, \qquad \overline{J} = k g^{-1} \partial_{\overline{z}} g.
\end{equation}
Using the above, one can derive the OPE of $J^a$ with $g$, giving
\begin{equation}
    J^a(z) g(w,\bar{w}) \sim -2\frac{t^ag(w,\bar{w})}{z-w}.
    \label{JgOPE}
\end{equation}
We now look at the four-point function
\begin{equation}\label{eq:WZW-4pt}
    \mathcal{G}(z_i, \bar{z}_i) = \langle g_{\alpha  \dot{\alpha }} (z_1, \bar{z}_1) g_{\beta  \dot{\beta }} (z_2, \bar{z}_2) g_{\gamma  \dot{\gamma }} (z_3, \bar{z}_3) g_{\delta  \dot{\delta }} (z_4, \bar{z}_4) \rangle= \frac{1}{(z_{14}z_{23})^{2h}(\bar{z}_{14}\bar{z}_{23})^{2h}}\, \tilde{\mathcal{G}}(x,\bar{x})~,
\end{equation}
where 
\begin{equation}
h= \frac{3}{4(k+2)}
\end{equation}
and 
\begin{equation}\label{eq:kz-cross-ratio}
    x = \frac{z_{12}z_{34}}{z_{14}z_{32}}~,
\end{equation}
and similarly for $\bar{x}$.
The correlator can be decomposed in $\mathfrak{su}(2)_k$ blocks, which factorize holomorphically. We denote as $\mathcal{F}(x)$ the generic block, leaving for later the specification of the indices specifying the $\mathfrak{su}(2)_k$ representation:
\begin{equation}
    \left.\tilde{\mathcal{G}}(x,\bar{x})\right|_\textup{single block} = \mathcal{F}(x) \mathcal{F}(\bar{x})~.
\end{equation}
The function $\mathcal{F}(x)$ can be expanded in three channels by utilising $SU(2)$ invariant tensors as 
\begin{equation}\label{eq:channel-expansion}
    \mathcal{F}(x) = \epsilon_{\alpha\beta}\epsilon_{\gamma\delta} \mathcal{F}_{s}(x) + \epsilon_{\alpha\gamma}\epsilon_{\beta\delta} \mathcal{F}_{u}(x) + \epsilon_{\alpha\delta}\epsilon_{\beta\gamma} \mathcal{F}_{t}(x).
\end{equation}
The three terms are not independent since 
\begin{equation}\label{eq:eps-reln}
    \epsilon_{\alpha\beta}\epsilon_{\gamma\delta} + \epsilon_{\alpha\delta}\epsilon_{\beta\gamma} = \epsilon_{\alpha\gamma}\epsilon_{\beta\delta}, 
\end{equation}
however we keep the form of $\mathcal{F}(x)$ as is for a moment. 
One can derive a consistency condition on $\mathcal{G}(z_i, \bar{z}_i)$ by inserting the identity
\begin{equation}
    L_{-1} - \frac{1}{2(k + 2)} (J^a J^a)_{-1} = 0,
\end{equation}
leading to the Knizhnik-Zamalodchikov equations
\begin{equation}
    \left( \partial_{i} - \frac{2}{k + 2} \sum_{i \neq j}\frac{t_i^a \otimes t_j^a}{z_i - z_j} \right) \mathcal{G}(z_i, \bar{z}_i) = 0,
\end{equation}
where we used eq. \eqref{JgOPE}.
Setting $i=1$ and using the fact that the KZ equation is valid block by block, one finds
\begin{equation} \label{eq:KZ}
    \left( \frac{1}{2} \partial_x - \frac{1}{k + 2} \frac{t_1^a \otimes t_2^a}{x} - \frac{1}{k + 2} \frac{t_1^a \otimes t_3^a}{x - 1} \right) \mathcal{F}(x)= 0.
\end{equation}
The generators act only on the $SU(2)$ invariant tensors in \eqref{eq:channel-expansion}, and their action is given as 
\begin{align}
    t_1^a \otimes t_2^a ~ \epsilon_{\alpha\beta}\epsilon_{\gamma\delta} &= -\frac{3}{4} \epsilon_{\alpha\beta}\epsilon_{\gamma\delta} \\
    t_1^a \otimes t_2^a ~ \epsilon_{\alpha\gamma}\epsilon_{\beta\delta} &= \frac{1}{2}\left( \frac{1}{2}\epsilon_{\alpha\beta}\epsilon_{\gamma\delta} -  \epsilon_{\alpha\gamma}\epsilon_{\beta\delta} \right)\\
    t_1^a \otimes t_2^a ~ \epsilon_{\alpha\delta}\epsilon_{\beta\gamma} &= \frac{1}{2} \left( \epsilon_{\alpha\gamma}\epsilon_{\beta\delta} -\frac{1}{2} \epsilon_{\alpha\delta}\epsilon_{\beta\gamma} \right) \\
    t_1^a \otimes t_3^a ~ \epsilon_{\alpha\beta}\epsilon_{\gamma\delta} &= -\frac{1}{4}\left( \epsilon_{\alpha\delta}\epsilon_{\beta\gamma} + \epsilon_{\alpha\gamma}\epsilon_{\beta\delta}\right) \\
    t_1^a \otimes t_3^a ~ \epsilon_{\alpha\gamma}\epsilon_{\beta\delta} &= - \frac{3}{4} \epsilon_{\alpha\gamma}\epsilon_{\beta\delta} \\
    t_1^a \otimes t_3^a ~ \epsilon_{\alpha\delta}\epsilon_{\beta\gamma} &= - \frac{1}{4}\left( \epsilon_{\alpha\beta}\epsilon_{\gamma\delta} + \epsilon_{\alpha\gamma}\epsilon_{\beta\delta} \right).
\end{align}
Thus the KZ equation \eqref{eq:KZ} simplifies to 
\begin{multline}
    \epsilon_{\alpha\beta}\epsilon_{\gamma\delta} \left( \frac{\kappa}{2} \partial_x \mathcal{F}_{s}(x) + \frac{1}{x} \left( \frac{3}{4} \mathcal{F}_{s}(x) + \frac{1}{2} \mathcal{F}_{u}(x) \right) + \frac{1}{x-1} \left(\frac{1}{4}\mathcal{F}_{t}(x)\right) \right) \\
    + \epsilon_{\alpha\gamma}\epsilon_{\beta\delta} \left( \frac{\kappa}{2} \partial_x \mathcal{F}_{u}(x) - \frac{1}{x} \left( \frac{1}{4} \mathcal{F}_{u}(x) + \frac{1}{2} \mathcal{F}_{t}(x) \right) + \frac{1}{x - 1}\left( \frac{1}{4} \mathcal{F}_{s}(x) + \frac{3}{4} \mathcal{F}_{u}(x) + \frac{1}{4} \mathcal{F}_{t}(x) \right) \right) \\
    + \epsilon_{\alpha\delta}\epsilon_{\beta\gamma} \left( \frac{\kappa}{2} \partial_x \mathcal{F}_{t}(x) + \frac{1}{x} \left(\frac{1}{4} \mathcal{F}_{t}(x) \right) + \frac{1}{x-1} \left( \frac{1}{4} \mathcal{F}_{s}(x) \right) \right) = 0,
\end{multline}
where we have defined $\kappa = k + 2$. Thus we obtain three differential equations. However, as we alluded to before, the three terms are not independent. Using \eqref{eq:eps-reln}, the two unique coupled differential equations are
\begin{eqnarray}
    \left(-\frac{3}{2} + 2x \right) \mathcal{F}_{s}(x) + 
 \mathcal{F}_{t}(x) + \left(-\frac{1}{2} + 2 x\right) \mathcal{F}_{u}(x) + (x - 1) x \kappa \left(\partial_x \mathcal{F}_{s}(x) + 
    \partial_x \mathcal{F}_{u}(x) \right) &= 0 \\
\frac{1}{2} \left(\mathcal{F}_{t}(x) + \mathcal{F}_{u}(x) \right) + 
 x \left(\mathcal{F}_{s}(x) + 
    \mathcal{F}_{u}(x) \right) + (x - 1) \kappa \left(\partial_x\mathcal{F}_{t}(x) + 
       \partial_x\mathcal{F}_{u}(x) \right) &= 0
\end{eqnarray}
We have two equations for three unknown functions, but the equations inherit an invariance from the relation \eqref{eq:eps-reln}. Indeed, under the following redefinitions,
\begin{equation}
    \mathcal{F}_{s}(x) \rightarrow \mathcal{F}_{s}(x) - g(x),  \qquad \mathcal{F}_{t}(x) \rightarrow \mathcal{F}_{t}(x) - g(x), \qquad \mathcal{F}_{u}(x) \rightarrow \mathcal{F}_{u}(x) + g(x), 
\end{equation}
the differential equations remain unchanged. Thus we are free to make a choice, and we set $\mathcal{F}_{s}(x) = \mathcal{F}_{u}(x)$. Solving the differential equations, we get two solutions, for the identity exchange,
\begin{equation}
    \mathcal{F}_{s, \text{id}}(x) = (1-x)^{-\frac{3}{2\kappa}} x^{-\frac{3}{2\kappa}} ~_2F_1\left( -\frac{1}{\kappa}, \frac{\kappa - 3}{\kappa}, \frac{\kappa - 2}{\kappa}, x \right),
    \label{Fsid}
\end{equation}
\begin{multline}
    \mathcal{F}_{t, \text{id}}(x) = \frac{1}{\kappa -2}(1-x)^{-\frac{3}{2\kappa}} x^{-\frac{3}{2\kappa}} \left(2 (\kappa -3) (x-1) x ~_2F_1\left(2-\frac{3}{\kappa },\frac{\kappa -1}{\kappa }, 2-\frac{2}{\kappa }, x\right) \right.\\
    \left. +(\kappa -2) (2 x-1) ~_2F_1\left(-\frac{1}{\kappa},\frac{\kappa -3}{\kappa },\frac{\kappa -2}{\kappa },x\right)\right),
    \label{Ftid}
\end{multline}
and for the adjoint exchange
\begin{equation}
    \mathcal{F}_{s, \text{adj}}(x) = (1-x)^{\frac{1}{2\kappa}} x^{\frac{1}{2\kappa}} ~_2F_1\left(1+\frac{1}{\kappa },\frac{3}{\kappa },\frac{\kappa +2}{\kappa },x\right),
\end{equation}
\begin{multline}
    \mathcal{F}_{t, \text{adj}}(x) = -\frac{1}{\kappa +2} 3 (1-x)^{\frac{1}{2\kappa}} x^{\frac{1}{2\kappa}} \left((\kappa +2) (2 x-1) \, _2F_1\left(1+\frac{1}{\kappa },\frac{3}{\kappa };\frac{\kappa +2}{\kappa };x\right) \right.\\ 
    \left. +2 (\kappa +1) (x-1) x \, _2F_1\left(2+\frac{1}{\kappa },\frac{\kappa +3}{\kappa };2+\frac{2}{\kappa };x\right)\right).
\end{multline}
Now that we have the blocks, we can write $\mathcal{G}(z_i, \bar{z}_i)$ in a block expansion as 
\begin{multline}
    \mathcal{G}(z_i, \bar{z}_i) = \frac{1}{(z_{14}z_{23})^{2h}(\bar{z}_{14}\bar{z}_{23})^{2h}}\sum_{i,j} X_{ij} \left( \epsilon_{\alpha\beta}\epsilon_{\gamma\delta} \mathcal{F}_{s, i}(x) + \epsilon_{\alpha\gamma}\epsilon_{\beta\delta} \mathcal{F}_{u, i}(x) + \epsilon_{\alpha\delta}\epsilon_{\beta\gamma} \mathcal{F}_{t, i}(x) \right) \\
    \left( \epsilon_{\dot{\alpha}\dot{\beta}}\epsilon_{\dot{\gamma}\dot{\delta}} \mathcal{F}_{s, j}(\bar{x}) + \epsilon_{\dot{\alpha}\dot{\gamma}}\epsilon_{\dot{\beta}\dot{\delta}} \mathcal{F}_{u, j}(\bar{x}) + \epsilon_{\dot{\alpha}\dot{\delta}}\epsilon_{\dot{\beta}\dot{\gamma}} \mathcal{F}_{t, j}(\bar{x}) \right), \qquad i, j \in \{\text{id}, \text{adj}\}.
\end{multline}
Working in the diagonal modular invariant implies $X_{12} = X_{21} = 0$. The remaining $X_i \cong X_{ii}$ can be fixed using crossing from $x$ to $1-x$, which gives
\begin{equation}
    X_1 = 1, \qquad X_2 = \frac{\Gamma\left(\frac{1}{\kappa}\right)\Gamma\left(\frac{3}{\kappa}\right)\Gamma\left(\frac{\kappa - 2}{\kappa}\right)^2}{3\Gamma\left(-\frac{3}{\kappa}\right)\Gamma\left(-\frac{1}{\kappa}\right)\Gamma\left(\frac{\kappa+2}{\kappa}\right)^2}.
\end{equation}

Now that we have the exact $\sutwok$ four-point function of $g_{\alpha \dot{\alpha}}$, we provide a more explicit formula for a combination of components useful for the main text. We first define the real operator
\begin{equation}
    \mathcal{O} = \frac{1}{2} \left( g + g^{-1} \right).
\end{equation}
The $\mathcal{O}_{11}$ element of the matrix is 
\begin{equation}
    \mathcal{O}_{11} = \frac{1}{2} \left( g_{11} + g_{22} \right),
\end{equation}
and its four-point correlator is of the form
\begin{equation}
    \langle \mathcal{O}_{11}(z_1) \mathcal{O}_{11}(z_2) \mathcal{O}_{11}(z_3) \mathcal{O}_{11}(z_4) \rangle = \frac{\tilde{\mathcal{G}}_{\mathcal{O}}(x, \bar{x})}{|z_{12}|^{\frac{3}{(k+2)}}|z_{34}|^{\frac{3}{(k+2)}}}, 
\end{equation}
with the following expression for $\tilde{\mathcal{G}}_{\mathcal{O}}(x, \bar{x})$ when $k=2$:
{\small
\begin{equation}
\tilde{\mathcal{G}}_{\mathcal{O}}(x, \bar{x}) = \frac{ \sqrt{(x-1) (\bar{x}-1)}(2 +\sqrt{x} + \sqrt{\bar{x}} +\sqrt{x\bar{x}}) + 2 \sqrt{x \bar{x}}-\bar{x}+\sqrt{\bar{x}}+2 + x \left(2 \bar{x}+\sqrt{\bar{x}}-1\right) +\sqrt{x} (\bar{x}+1)}{4 (x-1)^{3/8} (\bar{x}-1)^{3/8}\sqrt{\left(\sqrt{x}+1\right) \left(\sqrt{\bar{x}}+1\right)} }~.
\end{equation}
}

\bibliography{./auxi/bibliography}
\bibliographystyle{./auxi/JHEP}

\end{document}